# Chaotic Dynamics of Polyatomic Systems with an Emphasis on DNA Models

---

Malcolm Hillebrand

*A thesis presented for the degree of Doctor of Philosophy in the Department of Mathematics and Applied Mathematics*

University of Cape Town

December 2020

# Declaration

I, Malcolm Hillebrand, hereby declare that the work on which this thesis is based is my original work (except where acknowledgements indicate otherwise) and that neither the whole work nor any part of it has been, is being, or is to be submitted for another degree in this or any other university. I authorise the University to reproduce for the purpose of research either the whole or any portion of the contents in any manner whatsoever.

Signature: 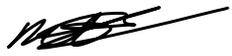  Date: 23rd of December 2020.

# Acknowledgements

Firstly, thanks go to those who have worked with me on different projects – including fellow students Guy Paterson-Jones, Adrian Schwellnus, Bertin Many Manda, and to Professor Alan Bishop (Los Alamos National Laboratory), who has added tremendous insight and experience to our work. Particular thanks go to Professor George Kalosakas (University of Patras), who has been extremely patient and helpful in the course of our collaboration, constantly providing invaluable guidance and expertise even when I have been woefully ignorant about biological details. I also thank Dr Enrico Gerlach (Technische Universität Dresden) for both his contribution through computational skill, as well as his assistance and enormous kindness as my research host during my visit to the TU Dresden. I am indebted to him for selflessly acting as a guide, translator and rock climbing coach during my time in Dresden in February-April 2019. Many thanks also to the other members of the Professur für Astronomie in the Institut für Planetare Geodäsie for their scientific and personal impact: Robin, Hagen, Anke, Lutz, Sven, Thomas, Irina, and Professors Klioner and Soffel. Thanks also go to Matthias Werner for sharing his computational expertise and CUDA wizardry, as well as the computational facilities of the ZIH of the TU Dresden. The financial assistance of the Deutsche Akademische Austauschdienst towards this visit is greatly appreciated, as well as their help with organisation. All those who have contributed by valuable conversations, in various contexts, thank you – particularly Professors Thierry Dauxois, Efthimios Kaxiras and Tassos Bountis, whose suggestions have proved very useful, as well as Tomas Bruce-Chwatt for sharing his biochemistry knowledge. My sincere thanks also go to Morgan Vandeyar for amazing TeXnical assistance and combing my thesis for missing commas. To my family, I thank you all for the combination of emotional support, proof-reading, scientific conversation and technical assistance. Grant, Jenny, Charlotte, Pippa, Darièn, Brandon, many, many thanks. To Claire, for sticking with me through the ups and downs of the PhD life, my eternal gratitude. You are amazing. All the members of our Nonlinear Dynamics and Chaos research group at UCT, who have made this process a joy, shared so many unforgettable experiences, and helped overcome countless obstacles, I cannot thank you enough. Bob, Arnold, Chinenye, Adrian, Many, Henok – my heartfelt thanks. Finally, I would like to thank my supervisor, Professor Haris Skokos. The meeting four years ago that started with "I must warn you, with me you will work hard" kicked off a postgraduate journey that lived up to those words, but has been a truly amazing experience. It is hard for me to imagine a better supervisor. At every step, Professor Skokos has provided more support and time than I could ever have expected, from last-minute motivation letters (sorry) to arranging conference opportunities and painstakingly working through all kinds of documents, using up multiple red pens in his untiring efforts to improve my work. My sincere thanks, for all the hard work that has gone into this process. This work is based on the research supported by the National Research Foundation of South Africa (Grant Number: 112332), supported additionally by the University of Cape Town, with conference assistance from the University of Crete. I also thank the Center for High Performance Computing and the University of Cape Town's ICTS High Performance Computing team for providing computational resources to the project.

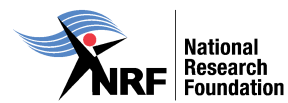

*Soli Deo Gloria*

# Abstract

In this work we investigate the chaotic behaviour of multiparticle systems, and in particular DNA and graphene models, by applying various numerical methods of nonlinear dynamics. Through the use of symplectic integration techniques—efficient routines for the numerical integration of Hamiltonian systems—we present an extensive analysis of the chaotic behaviour of the Peyrard-Bishop-Dauxois (PBD) model of DNA. The chaoticity of the system is quantified by computing the maximum Lyapunov exponent (mLE) across a spectrum of temperatures, and the effect of base pair disorder on the dynamics is investigated. In addition to the inherent heterogeneity due to the proportion of adenine-thymine (AT) and guanine-cytosine (GC) base pairs, the distribution of these base pairs in the sequence is analysed through the introduction of the alternation index. An exact probability distribution for arrangements of base pairs and their alternation index is derived through the use of Pólya counting theory. We find that the value of the mLE depends on both the base pair composition of the DNA strand and the arrangement of base pairs, with a changing behaviour depending on the temperature. Regions of strong chaoticity are probed using the deviation vector distribution, and links between strongly nonlinear behaviour and the formation of bubbles (thermally induced openings) in the DNA strand are studied. Investigations are performed for a wide variety of randomly generated sequences as well as biological promoters. Furthermore, the properties of these thermally induced bubbles are studied through large-scale molecular dynamics simulations. The distributions of bubble lifetimes and lengths in DNA are obtained and discussed in detail, fitted with simple analytical expressions, and a physically justified threshold distance for considering a base pair to be open is proposed and successfully implemented. In addition to DNA, we present an analysis of the dynamical stability of a planar model of graphene, studying the behaviour of the mLE in bulk graphene sheets as well as in finite width graphene nanoribbons (GNRs). The well-attested stability of the material manifests in a very small mLE, with chaos being a slow process in graphene. For both possible kinds of GNR, armchair and zigzag edges, the mLE decreases with increasing width, asymptotically reaching the bulk behaviour. This dependence of the mLE on both energy density and ribbon width is fitted accurately with empirical expressions.

# Publications and Presentations of Results

From this PhD research, results have been presented through publications in international peer reviewed journals:

1. **M. Hillebrand**, G. Paterson-Jones, G. Kalosakas and Ch. Skokos, *Distribution of Base Pair Alternations in a Periodic DNA Chain: Application of Pólya Counting to a Physical System*, Regular and Chaotic Dynamics, **23** No. 2, 135 (2018), doi.org/10.1134/S1560354718020016.

2. **M. Hillebrand**, A. Schwellnus, G. Kalosakas and Ch. Skokos, *Heterogeneity and Chaos in the Peyrard-Bishop-Dauxois Model of DNA*, Physical Review E, **99**, 022213 (2019), doi.org/10.1103/PhysRevE.99.022213.

3. **M. Hillebrand**, B. Many Manda, G. Kalosakas, E. Gerlach and Ch. Skokos, *Chaotic Dynamics of Graphene and Graphene Nanoribbons*, Chaos **30**, 063150 (2020), doi.org/10.1063/5.0007761.

4. **M. Hillebrand**, G. Kalosakas, Ch. Skokos and A.R. Bishop, *Distributions of Bubble Lifetimes and Bubble Lengths in DNA*, Physical Review E, **102**, 062114 (2020), doi.org/10.1103/PhysRevE.102.062114.

Parts of the results of this thesis have been presented, at conferences and seminars, as follows:

1. Seminar Presentation: *Chaotic Dynamics in DNA Chains*, Centre for Theoretical and Mathematical Physics, University of Cape Town (Cape Town, South Africa, March 2018).

2. The International Conference on Nonlinear Localization in Lattices – NLL 2018, Poster Presentation: *Chaotic behaviour of the Peyrard-Bishop-Dauxois model of DNA* (Spetses, Greece, June 2018).

3. Seminar Presentation: *Chaotic Dynamics of DNA and Graphene*, Professur für Astronomie und Lohrmann-Observatorium, Technische Universität Dresden (Dresden, Germany, February 2019).

4. Postgraduate Science Symposium, Stellenbosch University: *Applying Dynamics to Biology: Understanding the Chaos of our DNA* (Stellenbosch, South Africa, September 2019).

5. The 62nd Annual Congress of the South African Mathematical Society, Oral Presentation: *Chaotic Dynamics of DNA* (Cape Town, South Africa, December 2019).

# Contents

# Chapter 1

# Introduction and Numerical Methods

This thesis is a foray into the wonderful world of nonlinear dynamics, being applied to perhaps surprising places. We will explore some of the possibilities of implementing numerical methods of nonlinear dynamics, especially chaotic dynamics, to polyatomic systems. The two systems explored are both fascinating in their own right: DNA is entrenched in the public eye as a cornerstone of modern biology (as a cursory glance through any textbook will reveal, e.g. [1]), and graphene has made waves after its introduction as a supermaterial for the future [2]. There have been many ways of studying these systems, as will be discussed in detail in the relevant chapters, but it is not particularly common to approach this problem from a mathematical physics standpoint, let alone from a chaotic dynamics point of view. This gives us the opportunity here to push forward the understanding of the physical systems through applying these techniques, but also to develop the understanding of how we can use chaotic dynamics in a variety of situations. With regard to the first aim, we hope to bring out fresh information about the systems that can contribute to further studies. The significance of DNA and graphene (as well as the physical robustness of the models used to study them) provides ample motivation for us to use any tools we have at our disposal to try and better understand the functioning and potential uses of these substances. The particular well-documented biophysical importance of openings in the DNA double helix (a fundamentally dynamic process) in the context of transcriptional activity adds further physical motives to the study. In general, it is advantageous to implement different approaches to studying a problem, as they are likely to provide some new understanding and insight. In an increasingly interdisciplinary age, the second aim is an important one – exploring the nonlinear dynamics side of physical systems, and identifying what the possibilities are for the application of chaos theory to these real substances. This approach could give answers to some important questions, such as: How can we use chaos indicators to understand the phenomenal structural strength of graphene? Is there a way to probe the transcriptional properties of DNA through nonlinear dynamics signatures? Furthering the state of knowledge in this direction is a promising avenue to pursue.

So here we will give outlines of the structure of DNA and graphene, as well as their biochemical and physical significance, and also of the various numerical and theoretical methods we have used in studying these polyatomic systems. These methods include chaos detection techniques, mathematical aspects (as Chapter 3 will reveal, sometimes *very* mathematical), high performance computing ideas – including multi-thread parallelism on central processing units (CPUs) and graphical processing units (GPUs), as well as adjustments of these numerical techniques and tricks specific to our application of chaotic dynamics.

In what follows in this chapter, we will briefly introduce the notions of Hamiltonian dynamics and chaos, as well as the main players in our study: The maximum Lyaponuv exponent (mLE) as the primary chaos indicator we have used, the deviation vector distribution (DVD) as a means of identifying regions of strong chaoticity, and the essential numerical techniques of symplectic integration and parallel computing for integrating systems of differential equations.

The remainder of this thesis is structured as follows. In Chapter 2 we present some background to the dynamical study of DNA, with a particular emphasis on the models used in our study. Chapter 3 describes the mathematical basis for our quantification of heterogeneity in DNA sequences, using an extension of Pólya's enumeration theorem. Results from the investigation of the chaotic dynamics of DNA are given in Chapter 4, with emphasis on the effect of heterogeneity on the system's chaoticity. Taking a different approach to studying DNA, Chapter 5 presents results from an investigation into the physical properties of DNA, with a particular focus on thermally induced openings in the double strand called bubbles. Chapter 6 then presents a dynamical model of graphene, some computational details for the implementation of parallelisation, and a study of the chaotic dynamics of 2D graphene models. Finally, in Chapter 7 we conclude the study, summarise our findings and discuss future outlooks.

## 1.1 Hamiltonian Dynamics and Chaos

Within the broad scope of dynamical systems, we are particularly focussing here on Hamiltonian systems. In any Hamiltonian system, the dynamics are governed by a Hamiltonian function $H$. This function generally serves as the energy of the system as well as a generator of the time evolution [3]. If we have generalised positions $\boldsymbol{q}$, and conjugate momenta $\boldsymbol{p}$ which define the state of our dynamical system, then we can write the Hamiltonian expressed in terms of these variables, and possibly also time as

$$H(\boldsymbol{q}, \boldsymbol{p}, t).$$

In a closed or isolated system, the Hamiltonian is simply the sum of the system's kinetic ($T$) and potential ($V$) energies, giving

$$H(\boldsymbol{q}, \boldsymbol{p}, t) = T + V, \tag{1.1}$$

where it is quite straightforward to understand the Hamiltonian as the total energy of the system. If in addition to this the Hamiltonian is not explicitly time dependent, so that

$$\frac{\partial H}{\partial t} = 0, \tag{1.2}$$

then we call the dynamical system autonomous. The Hamiltonian will usually of course still depend on time implicitly, through the time-dependent coordinates and momenta. So we would have an autonomous Hamiltonian in the form

$$H(\boldsymbol{q}(t), \boldsymbol{p}(t)),$$

where the explicit $t$-dependence has fallen away.

Significantly, time independent Hamiltonians conserve the total energy of the system (the value of the Hamiltonian function itself) [3]. This is extremely useful not only for simplifying analytical calculations, but also makes a lot of computational analysis much easier. For instance, it immediately gives us a conserved quantity to check our numerical integration schemes with – if our computation does not conserve the initial value of the Hamiltonian (up to numerical accuracy), then we know we have a problem in the machinery somewhere. We will typically measure this energy conservation through the relative energy error,

$$E_{Rel} = \frac{|H(t) - H(0)|}{|H(0)|},$$
(1.3)

which tells us how close we are staying to the initial energy value.

Quite often in autonomous systems, the Hamiltonian is "separable", i.e. (1.1) can be expressed as

$$H(\boldsymbol{q}, \boldsymbol{p}) = T(\boldsymbol{p}) + V(\boldsymbol{q}),$$
(1.4)

where $T(\boldsymbol{p})$ is the kinetic energy depending only on the momentum of the system, and $V(\boldsymbol{q})$ is the potential energy, a function of only the generalised coordinates.

A further useful aspect of Hamiltonian mechanics is the existence of the Poisson bracket. The Poisson bracket of two functions $f$ and $g$ of the generalised positions and momenta in an $N$-dimensional system with coordinates $(\boldsymbol{q}, \boldsymbol{p})$ is defined as [3]

$$\{f, g\} = \sum_{i=1}^{N} \left( \frac{\partial f}{\partial q_i} \frac{\partial g}{\partial p_i} - \frac{\partial f}{\partial p_i} \frac{\partial g}{\partial q_i} \right).$$
(1.5)

As our main use case for the Poisson bracket, the time derivative of a function with no explicit time dependence in a Hamiltonian system is given by

$$\frac{df}{dt} = \{f, H\},$$
(1.6)

which provides a route to Hamilton's equations of motion, among other things.

In general, much of the significance of Hamiltonian systems consists in the analytical and numerical methods that it is possible to construct for them. Especially for conservative Hamiltonian systems, the beautiful properties of the conserved volume of the phase space – the space spanned by the set of coordinates $\boldsymbol{q}$ and $\boldsymbol{p}$ – lend themselves to some very neat computations, which in our case enable the efficient study of the chaotic dynamics of these systems through symplectic integration (see Section 1.2.1), and the variational equations (Section 1.3.2).

Relevant details of chaotic behaviour will be given in the discussion of the maximum Lyapunov exponent (Section 1.3), but as a brief introduction we will present the basic framework for chaotic motion. An excellent introduction to chaotic dynamics is given in [4], for a range of dynamical systems. For now, we are concerned with chaotic dynamics of Hamiltonian systems, but of course the basic characteristics of fully chaotic systems hold here as well [5]:

1. Sensitivity to initial conditions.

2. Topological mixing.

3. Density of periodic orbits.

The sensitivity to initial conditions is by and large well understood. For any given trajectory in a chaotic region of the phase space an initially nearby orbit will eventually diverge significantly from it. This will be our crucial point in identifying and quantifying chaos later. The topological mixing means that a chaotic orbit will eventually visit every region in the phase space, to an arbitrarily small closeness. The result of this is that the chaotic phase space cannot be separated into different subsystems, but rather has to be treated as a single whole. Finally, the density of unstable periodic orbits means that in the phase space each point will have an unstable periodic orbit in a small neighbourhood around it, yielding some sense of regularity in the system.

Consequently, much of the study of chaotic or mixed (partly chaotic and partly regular phase space) systems involves trying to identify chaotic regions or trajectories based on these criteria. For low-dimensional systems it is sometimes possible to directly integrate a collection of initial conditions and identify chaotic regions through observing the dynamical behaviour, such as in the archetypical Lorenz system for atmospheric flow [6]. The phase space of these low dimensional systems can also be studied through taking surfaces of section, such as the Poincaré surface of section (PSS) [7], which can reveal chaotic and regular regions. This has for example successfully been applied to astronomical models, such as the well known Hénon-Heiles system [8].

Higher dimensional systems (for instance of hundreds of degrees of freedom) make this direct approach challenging to say the least. While methods exist for visualising four dimensional phase spaces such as can arise in the study of three dimensional Hamiltonian models [9, 10], visualising five hundred dimensions is a task beyond most human minds. Fortunately, there are direct methods available for identifying chaotic motion, and distinguishing it from regular behaviour. The method of chaos identification and quantification used in our work is the mLE, described in detail in Section 1.3. Further options include the smaller (SALI) and generalised (GALI) alignment index methods [11–13], which are very efficient ways of identifying chaotic trajectories. Recent reviews of various modern chaos detection techniques can be found in [14].

## 1.2 Numerical Methods

In any research project based largely on computational simulations, it is paramount to use effective and efficient numerical methods. In light of this, there was a significant effort in the course of this study to continuously improve the state of the computer code used for integration and analysis, and throughout the process new ideas were considered for optimisation. The two major tools used in the optimisation process were symplectic integrators, and code parallelisation. Here we briefly discuss these two aspects, with a focus on the practical side of how they were actually implemented in the project.

### 1.2.1 Symplectic Integrators

For the numerical integration of the equations of motion of Hamiltonian systems, a specific class of methods known as *symplectic integrators* have been introduced [15] with great success [16]. These are integration routines that were derived bearing in mind the specific properties of Hamiltonian phase space, and particularly of conservative Hamiltonian systems, in order to provide accurate long-time simulations. A great resource for a thorough overview of the topic is [16], but here we will present a practical outline of what these methods are and how they are implemented in our case.

The basic idea of a symplectic integration method is instead of approximately integrating the exact Hamiltonian (as would happen with a traditional Runge-Kutta integrator for instance), to rather exactly integrate an approximate Hamiltonian, by using a sequence of exact symplectic mappings [17]. The benefit of this approach is that by conserving the symplectic structure of the phase space, we in fact conserve the energy precisely. This is a result of the symplectic preservation, which corresponds to conserving the volume of the phase space and consequently the energy of the system. The magic of this particular approach lies in the fact that it is possible to design these integration methods based on a desired Hamiltonian, and end up with a routine that integrates a very similar Hamiltonian with complete accuracy, allowing us to integrate this slightly perturbed Hamiltonian for arbitrarily long times, and with arbitrary closeness to the original Hamiltonian set by the time step.

As outlined very well in [18], a basic approach to constructing these algorithms is to approximate the exact canonical transformation $(\boldsymbol{q}_0, \boldsymbol{p}_0) \to (\boldsymbol{q}(t), \boldsymbol{p}(t))$, which is usually impossible to write down, with a series of simpler canonical transformations. So let us take a separable Hamiltonian, of the form (1.4)

$$H = T(\boldsymbol{p}) + V(\boldsymbol{q}).$$

With this Hamiltonian there is an associated time evolution operator of the system's state $\exp(\tilde{H}t)$, where $\tilde{H}$ represents an operator that takes the Poisson bracket of its argument with the Hamiltonian: $\tilde{H}(X) = \{X, H\}$. Thanks to the separability of the Hamiltonian, this time evolution operator can be written in terms of two independent operators $A(X) = \{X, T\}$ and $B(X) = \{X, V\}$ corresponding to the kinetic and potential terms respectively as [16]

$$\exp(\tilde{H}t) = \exp[(A + B)t]. \tag{1.7}$$

The trick now is to try to approximate this exponential operator as accurately as possible, by only using simple operations of $\exp(At)$ and $\exp(Bt)$ which can be obtained analytically in the case of Hamiltonians in the form of (1.4) [18]. Of course for most cases the operators do not commute, and the exponential does not simply split out into $\exp[(A + B)t] = \exp(At)\exp(Bt)$, but rather a sequence of exponentials with carefully chosen coefficients $c_i, d_i$ is necessary to approximate the true operator. This sequence of exponentials is necessarily finite, and for an approximation with $n$ applications of each operator we have

$$\exp[(A + B)t] \approx \prod_{i=1}^{n} \exp(c_i A t)\exp(d_i B t). \tag{1.8}$$

How to choose these coefficients, and in general the art of constructing symplectic integrators is a subject that has attracted much attention (see e.g. [16–23]).

The accuracy of a given symplectic integration scheme is defined by the order of the error in the approximation of the true time evolution operator. In the approximation of (1.8), the error term will be of some fixed order in the time step. So with an integration time step of $\tau$, we would have more precisely

$$\exp[(A + B)\tau] = \prod_{i=1}^{n} \exp(\tau c_i A)\exp(\tau d_i B) + \mathcal{O}(\tau^{m+1}), \tag{1.9}$$

where the integration scheme is said to be of order $m \in \mathbb{N}$. Exactly what $m$ is for a given method is dependant on the number of terms in the product and the particular choice of coefficients $c_i$ and $d_i$. The aim with designing these methods is of course to reduce the order of the error as far as possible, while having to perform as few operations at each time step as possible.

We will now discuss some practical implementation details of these methods. In terms of the computations, a single application of an $A$ or $B$ operator is the same as updating the positions $q$ or momenta $p$ respectively. As described in [17], this means that for a single time step we have a sequence of updates to the positions and momenta using the usual Hamiltonian equations of motion:

$$p_{i+1} = p_i - c_i \tau \frac{\partial V(q_i)}{\partial q}, \tag{1.10}$$

$$q_{i+1} = q_i + d_i \tau \frac{\partial T(p_{i+1})}{\partial p}. \tag{1.11}$$

The index $i$ refers to the substeps within one integration time step of the process; so the system has only been evolved by $\tau$ time units after all these substeps have been computed.

The simplest form of symplectic integration is to simply take a first order approximation and set $c = d = 1$, and use only a single substep. This is commonly referred to as the symplectic Euler method for numerical integration [16]. For more involved multistep integrators, these substeps are applied consecutively with the appropriate coefficients, and produce accurate evolutions of the system with a varying degree of precision depending on the method and the system under study.

As a specific example of symplectic integration in action, we can consider the fourth-order, six-stage Symplectic Runge-Kutta-Nyström integration scheme $SRKN_6^b$ [24]. This is the integrator that is used in obtaining the results of Chapters 4 and 5, to integrate the equations for the DNA model. The integrator takes the form

$$\exp[(A+B)t] \approx \exp(c_1 Bt)\exp(d_1 At)\dots\exp(c_6 Bt)\exp(d_6 At)\exp(c_7 Bt). \tag{1.12}$$

Using the iterative procedure of Eqs. (1.10) and (1.11), this routine requires six update stages for the $A$ operator, and seven stages for the $B$ operator. The coefficients for this integration scheme are given in Table 3 of [24], showing the optimised values found to to minimise errors. With these values, we are able to apply the iterative procedure directly, and implement this integrator as a computational algorithm.

As an illustration of the benefit of symplectic integration, let us consider the simulation of the DNA model studied in this thesis. The details of the model and its equations are discussed in Chapter 2, but here we will simply look at one aspect of the system – the conservation of energy, as this is the primary advantage of SIs in our application. Figure 1.1 shows the relative energy error (1.3) throughout the integration process for the symplectic scheme for a particular initial condition, compared to that of the commonly-used fourth order Runge-Kutta integrator (RK4). Here the issue with the non-symplectic scheme quickly becomes apparent – with the same integration time step, the RK4 integrator and $SRKN_6^b$ have a very similar total CPU time (as seen in the inset), but even from the beginning the accuracy of the RK4 method is much worse than the symplectic scheme. Even using a time step ten times smaller, resulting in very slow integration, the energy is poorly conserved at long times with the RK4 integrator. Given the preeminent importance of accuracy in computations of chaotic systems, this is a strong argument in favour of SIs: They provide a guarantee of long-time accuracy, without requiring extremely small time steps.

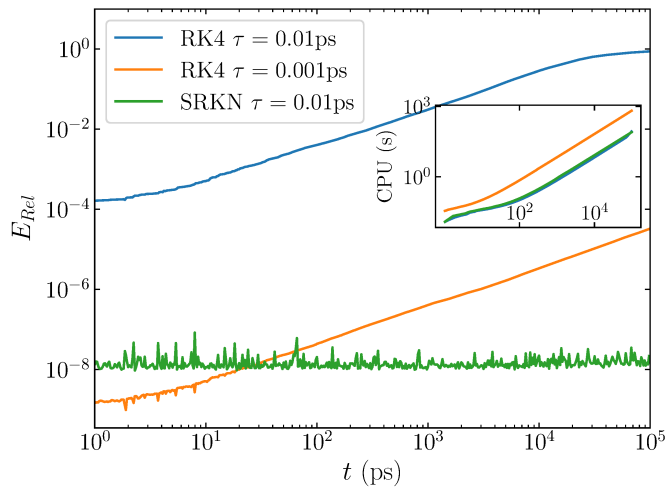

FIGURE 1.1: The relative energy error $E_{Rel}$ (1.3) as a function of time $t$ for the same initial condition of the DNA model (2.4), integrated using the $SRKN_6^b$ method as well as the standard non-symplectic fourth-order Runge Kutta (RK4) method. Two different time steps $\tau$ have been used for the RK4 method, but in both cases the energy error becomes much greater at long times than for the symplectic integrator. Inset: The CPU time in seconds required to reach the final integration time of $10^5$ps, for the three cases. Note that the blue and green curves practically overlap in this plot.

## 1.2.2 Computational Process and Code Parallelism

Having outlined the mathematically important aspect of our integration procedure, here we will briefly discuss some of the technical concerns and the parallel computation methods used in this thesis. Since the main computational expense arising from our work is incurred by the numerical integration of equations of motion (and variational equations), this is the main focus of the parallelisation work. Due to the different models and code platforms used in the course of the investigation, different optimisation approaches have been implemented.

All computations were performed on the Lengau cluster of the South African Centre for High Performance Computing (CHPC) [25]. This allowed access to GPU as well as CPU computing facilities. Each CPU node used on the cluster had a total of 24 available cores, which meant that in all our simulations the largest number of cores used at any one time was 24, as this obviated the need for inter-node communication techniques.

First though we will address the simplest form of parallelism used – running independent programs simultaneously on a single node with many CPU cores available.

**GNU Parallel**

The tool we have used for the simple running in parallel of multiple independent programs is GNU Parallel [26]. The process here is quite straightforward. Say that we have a compute node with 24 CPUs, as was our case on the CHPC cluster, and plenty of memory for our application. We note that in most of our numerical integrations the computer code is strongly compute bound, i.e. never coming close to running out of memory even when parallelised. If we also have a large number of individual programs to run, each using only one CPU core, then it would be convenient to be able to run these simultaneously (24 at a time) rather than having to run them inefficiently in series. This is precisely what GNU parallel enables us to do – arrange our multiple individual files to be fed through the processors in batches.

In the simplest case, we will take serial programs, where each program uses only a single core. This situation can arise in several ways; the main way which it was used in our application was running many cases of the DNA model (see Section 4.1.1), where the same code needs to be run with a large number of different parameters and initial condition. These cases can be split into separate executable files, each running the code with the required conditions, completely independently of the other programs, and run using a single CPU core. Whenever this splitting is possible, it is inherently the most efficient form of parallelisation possible, guaranteeing 100% efficiency as the number of cores is increased (assuming no memory-related slowdown, since there is no need for communication between the cores). The other situation where this parallelisation occurs in our work is the analysis of large numbers of data files, which can be processed in parallel independently. Then again it is highly efficient to analyse the data files in different blocks, each block being processed by a separate program.

This is the first choice for all the parallelisation we have used. If individual runs do not take extremely long, as is the case in all DNA results (see Chapters 4 and 5), then we have used only this form of independent parallelism.

GNU Parallel can also be used when each individual program is not purely serial. So for instance if each individual run requires 4 cores, and we still have access to the 24 core node, then we can make use of GNU Parallel to run 6 instances of the code on each node. This allows for a balancing between the speedup of an individual program due to internal parallelisation, and the efficiency that comes from being able to run many programs at once.

**OpenMP and CUDA**

Turning our attention to this internal parallelisation, we will roughly outline the methods used in our computations. Primarily, our codes have been parallelised using CPU multithreading, and particularly the OpenMP framework (see e.g. [27]). This multithreading has been used in the relatively complex two dimensional graphene model discussed in Chapter 6.

Briefly, in applying multithreading instead of having one CPU core perform all the computations one after the other, we will divide the work between multiple cores and have them perform the operations simultaneously. In the context of lattice integration, this means when the integration step occurs we split the integration of the sites between multiple cores. So if we assign 4 cores to the integration, that means we will give each core one quarter of the atoms to integrate, and these cores will execute simulta-neously. During each application of the integration operators $\exp(\tau c_i A)$ and $\exp(\tau d_i B)$, the equations of motion for each atom are computed and the position/momentum updated. Since these equations of motion depend strictly on the positions and momenta at the *previous* time step, there is no problem with computing them simultaneously. Each atom can be updated seperately, independently of the changes to the other atoms. At the end of each step, the code ensures that all threads (or cores) have finished all their assigned sites, in case some operations take slightly longer or shorter. This then avoids complications of trying to continue the integration with the next step while some sites have not in fact been updated yet.

When parallelising code it is always necessary to take precautions to ensure that there are no race conditions, such as problems arising from different threads attempting to access data before it has been updated or trying to modify memory that is being used by another thread. This demands frequent synchronisation checkpoints to be sure that all threads have completed their assigned tasks. Because of this, as well as other inefficiencies in the parallelisation process (e.g. cache usage, necessarily serial parts, slower individual processors), there are limits to how much speedup can be achieved by this kind of multithreading [27]. In Section 6.2, we will present some results discussing this imperfect speedup, and how we identified a reasonably optimal number of threads to use for each individual integration.

A second possibility for multithreading is using GPU parallelisation. For this, we wrote the integration code in CUDA (for an excellent introduction, see [28]), as the most efficient available way of making use of the NVIDIA GPU architecture, the most common today. The primary distinction between GPUs and CPUs is that GPUs tend to have many more processing cores (for instance, the Tesla V100 GPUs deployed in the CHPC cluster have an impressive 5120 CUDA cores, compared to the 24 CPU processors available on a single node), but with lower processing speeds (the V100 has a peak of 1455MHz, while the CHPC Intel Xeon processors have a peak of 2600MHz). This results in GPUs being much more powerful for computations involving significant parallelisation, and the integration of very large numbers of atoms. The design of CUDA programming also lends itself to very efficient ways of handling enormous lattices (tens or hundreds of thousands of atoms), through various techniques [28].

In CUDA programming, the same ideas as for CPU parallelisation are followed: We assign to each core a number of sites, which it then goes through and integrates. Due to the complexities of interfacing between the GPU and normal CPU processing, along with necessarily very careful memory management, integrating the lattice using CUDA is more difficult than using CPU parallelisation, and only provides meaningful advantages with very large lattices. Up until every GPU core is in use (for our application, that corresponds to at least 5120 atoms in the lattice since we are using a GPU with 5120 cores), all computations take the same amount of time – with the lower clock speed, this means that we are unlikely to be doing better than in the CPU parallelisation regime. However, once this point is reached, there will be the expected linear relationship between compute time and lattice size.

As with the CPU performance, scaling results for an implementation of the dynamical model of graphene in CUDA are presented in Section 6.2.

**General Optimisation and Implementation Details**

Apart from the parallelisation of the integration code, here some of the other computational details are presented. As previously mentioned, numerical integration of equations of motion is an intensive task largely limited by the number of arithmetic operations required at each iteration. This means that a vital step in the optimisation of our code is to first ensure that the equations are written in a computationally efficient form – shortening equations as much as possible, checking that all cancellations are performed, and minimising the number of complex operations (e.g. trigonometric or exponential) that are performed at each step.

Beyond this, several optimisation steps were taken in the course of the development of the integration code for the DNA and graphene models. They are listed here, as they are in general very effective steps to take when optimising code of this sort.

1. Focussing on the integration steps, which are performed millions of times in the course of a run. Making sure that only absolutely necessary computations were repeated each time; anything that can be taken out of this step should be.

2. Choosing an efficient integrator for each model. This involved running comparative tests to identify the best option given the specifics of the model, the size of the time step required, the accuracy required, and so on.

3. Identifying quantities that were calculated repeatedly, and pulling them out at the beginning of each step as pre-calculated constants.

4. Once a basic version of the code is working, rework the equations to find any ways of saving time; perhaps certain elements of the integration can be computed together efficiently (for instance sines and cosines in several programming languages).

5. Code profiling to identify bottlenecks, and inform the decisions of where to focus attention. Tools such as gprof and NVprof are useful in this regard.

6. Becoming familiar with the compiler-specific optimisations available, in terms of vectorisation of operations, interprocedural optimisation, and optimisations for certain processors.

7. For parallelised code, running scaling tests to have a clear idea of an effective number of cores to use for a given system size. This is always a trade-off between the optimal efficiency, which is running in serial, and how long the code takes to run.

For these projects, a combination of codes written in Fortran90, C++, Python, Mathematica and Octave were used. The simulations were performed using Fortran90 (DNA models) and C++ (graphene model), compiled using IFort and ICC/ICPC respectively. Data analysis used primarily Python, with Fortran90, Octave and Mathematica where useful. In particular, the Python libraries `numpy`, `scipy` and `matplotlib` were used extensively.

Linux shell utilities such as bash, AWK, sed, and grep were invaluable in manipulating files, arranging data, and generally saving large amounts of time on repetitive tasks.

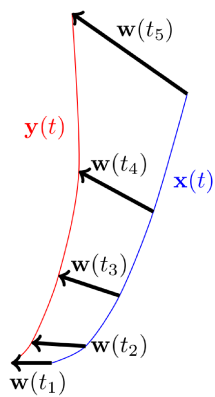

FIGURE 1.2: A schematic depiction of a reference orbit $x(t)$ in blue, and a perturbed orbit $y(t)$ in red, in a dynamical system's phase space. The evolution of the deviation vector $\boldsymbol{w}(t)$ is shown at a few points in time. This illustrates the idea behind evolving a deviation vector and the computation of the mLE as the mean rate of exponential growth of this vector.

The heavy computational requirements of producing statistically sound results means that the kind of optimisation discussed here is imperative. The efficiency brought by basic parallelisations and code optimisations enables the creation of orders of magnitude of more data than could otherwise be obtained, which in turn makes it possible to produce substantially more sound results.

## 1.3   The Maximum Lyapunov Exponent

If our main goal is to understand chaos in different systems, and to investigate the effects of various physical factors on the chaotic behaviour of real molecules, then we need to have a robust method of identifying and quantifying chaos. The fundamental principle of chaos that concerns us most here is the element of deterministic unpredictability. While all the systems we study have perfectly well defined equations of motion that determine exactly the trajectories of any chosen initial condition, what is important is that it is almost impossible to predict *a priori* what a given change in initial condition will produce in the exact details of the produced orbit. So what we will go into some detail about here is the identification of these diverging trajectories from initially nearby points. For a thorough review on this topic, detailing mathematical as well as practical points for Lyapunov exponents, see [29] and references therein.

It is fairly clear that if we can identify this behaviour of nearby initial conditions evolving into wildly different orbits, then we have found chaos. This is the fundamental idea behind Lyapunov exponents; measuring the exponential rate of divergence of two orbits. In Fig. 1.2, the evolution of two nearby orbits in the phase space of a dynamical system and the resultant deviation vector is illustrated schematically. The idea of studying the average growth rate of small deviations from a reference orbit was introduced more than a century ago by Lyapunov (in Russian, for the English translation, see [30]), where he formulated the Lyapunov characteristic exponents (LCEs) as precisely this measure of the asymptotic average rate of expansion of the distance between two orbits.

A major landmark in the history of chaos quantification was a breakthrough in 1968

by Oseledec [31], where his Multiplicative Ergodic Theorem opened the way for the computation of the LCEs, and in particular the largest of the LCEs, the maximal Lyapunov characteristic exponent (mLCE), or just the maximal Lyapunov exponent (mLE). This led to the development of techniques to estimate the mLE, based on the relationship between the LCEs and the exponentially fast divergence of perturbed chaotic orbits. The first estimation of the mLE, $\chi_1$, was presented as the asymptotic value when $t \to \infty$ of a *finite time mLE* (ftmLE), $X_1(t)$ [32], which can be computed directly based on the mean exponential rate of growth of the deviation, the details of which were outlined in the seminal work of Benettin et al [32, 33]. The "standard method" developed in this work has remained the most commonly used technique for computing mLEs, proving to be a very effective tool in a diverse array of dynamical contexts [29].

Before further discussion of the mLE, let us actually define it. Given some reference orbit, e.g. $x(t)$ in Fig. 1.2, and a deviation from this trajectory called $w(t)$ defining a nearby orbit $y(t)$ (Fig. 1.2), which evolves through time, then the mLE is defined to be the quantity

$$\chi_1 = \lim_{t \to \infty} \frac{1}{t} \ln \left( \frac{\|w(t)\|}{\|w(0)\|} \right), \qquad (1.13)$$

where $\|\cdot\|$ denotes the Euclidean norm (although any choice of norm produces the correct result [33]). Based on this formulation, we can draw some conclusions, many of which were discussed originally in [32]. Firstly, the fact that there is a prefactor of $1/t$ when we are taking the infinite time limit means that for any non-growing or slowly-growing deviation vector, the mLE value is going to be zero. This corresponds to the statement that for regular orbits the mLE is zero, since the implication of a slowly growing deviation is that the orbit is not chaotic. For a chaotic orbit we have an exponentially growing deviation vector [32], and then despite taking the logarithm of $\|w(t)\|$ this will continue to grow in time, and will counteract the $1/t$ to produce a constant value. So we have the second statement that chaotic orbits produce positive mLE values. Immediately we now have our desired chaos identifier; provided that we can somehow determine the mLE of a given orbit, we know whether or not it is chaotic. This alone is extremely useful, especially in "mixed" systems, where regions of the phase space exhibit chaotic motion and other regions are regular. By calculating mLE values in different regions, we can then identify islands of stability in a chaotic sea, or regions of chaos in a generally regular phase space.

The second powerful aspect of the mLE is that it not only identifies chaos, but it quantifies the chaoticity. We can see that if a deviation of orbit $x_A$, $w_A(t)$, grows faster than the deviation from an orbit $x_B$, $w_B(t)$, that the resultant mLE of orbit $x_A$ is going to be larger, simply by the definition. This means that by calculating the mLE for different orbits in a system, we can identify stable and chaotic regions, but we can also distinguish the more and less chaotic behaviours. The estimation of the mLE also facilitates the comparison of the chaoticity of different states of the same system (for physical systems such as studied in this work, these states could correspond to the different energy or temperature values), and even the comparison between the chaoticity of different systems.

As a consequence of these properties, the mLE is a tremendously useful tool for studying dynamical systems, when it can be calculated. The obvious catch with this statement is being able to calculate the mLE. In many cases it is completely impossible to theoretically calculate the rate of growth of the deviation vector, which in turn renders the mLE incalculable. In other cases even if some estimates can be made for the growth of the deviation vector, the exact calculation of the infinite limit creates problems.

As introduced in [33] and discussed in clear detail in [29], there have been very effective algorithms developed to numerically estimate the value of the mLE. If we are going to use a numerical approach, the first issue we have to clear away is the infinite limit, as infinite compute times are not possible to achieve. To this end, the ftmLE, an estimation of the mLE with the infinite limit truncated to a certain time, is used. This ftmLE is defined according to

$$X_1 = \frac{1}{t} \ln\left(\frac{\|\boldsymbol{w}(t)\|}{\|\boldsymbol{w}(0)\|}\right), \tag{1.14}$$

where from (1.13) we can see that $\mathcal{X}_1 = \lim_{t\to\infty} X_1$.

The evolution of the ftmLE itself is useful, even apart from the asymptotic behaviour which is its most important aspect. For clearly chaotic orbits, the ftmLE can generally be used fairly straightforwardly – if the mLE is a positive constant, then it follows that at some point the ftmLE must converge to that constant value. So then by pursuing the numerical integration of two nearby initial conditions, we can continue until reaching a point where it is clear that the ftmLE is practically no longer changing with time. For regular orbits, or effectively regular orbits, it becomes a little less clear. While one could in principle continue a computation until the ftmLE reaches some sort of "numerical zero", based on the machine precision of the computer, this could take an extraordinary amount of time. Fortunately though, it was shown early on [32] that for regular orbits, the ftmLE should tend to zero with a gradient of $1/t$. This can be seen from the definition of the ftmLE, (1.14), since a small growth in the deviation vector size will get completely lost in the overwhelming strength of the prefactor of $1/t$, leading to the evolution of the ftmLE being dominated by the slope of $1/t$. Additionally, we note that as long as the deviation vector grows linearly in time, the ftmLE will evolve according to a modified power law of $\ln(t)/t$, which will become closer to $1/t$ in the long time limit. This fact is particularly useful in light of the behaviour of so-called "weak chaos" regimes [34], where the ftmLE appears to be continuously decreasing, but this is not a reflection of a regular system. As a result of this phenomenon, it has proved useful both in studying one dimensional lattices [23] and two dimensional lattices [35] to be able to compare the slope of the ftmLE with the expected $1/t$ gradient. These works further confirm the usefulness of studying the behaviour of the ftmLE, as well as the final mLE.

A practical concern with the computation of the ftmLE is the fact that we are trying to investigate an exponentially growing variable (the norm of $\boldsymbol{w}$), over potentially very long times while we wait for the ftmLE to converge. The combination of exponential growth and long times is a recipe for numerical overflow disaster, and we need to find a way to circumvent this eventuality. As usual, the mathematical details can be found in [29], but the final outcome of practical importance is as follows. Instead of computing the ftmLE $X_1$ according to the norm of the deviation vector only at the final time, it is possible to instead iteratively update the ftmLE at each time step of the numerical integration, and renormalise the deviation vector (retaining its direction) at each step. With this, it is possible to write the ftmLE after $n$ integration steps of the system with a time step of $\tau$ as [29]

$$X_1(t = n\tau) = \frac{1}{n\tau} \sum_{i=1}^{n} \ln\left(\frac{\|\boldsymbol{w}(i\tau)\|}{\|\boldsymbol{w}(0)\|}\right). \tag{1.15}$$

So after each integration time step, the value of the deviation vector norm is used to update the ftmLE estimate, and then the deviation vector is renormalised, thus avoiding the problem of exponentially growing values.

As a final comment on the use of the mLE in dynamical systems, let us note that it is apparent from (1.13) that the units of the mLE are inverse time. By inverting the $\chi_1$ value, we find the so-called Lyapunov time $T_L$, which is an estimation of how long it takes for the system to become chaotic [29]. This is a useful quantity in several contexts, providing both an expected horizon for chaotic behaviour to appear in dynamical systems such as astronomical motions [36], and a characteristic time scale for chaos in physical models.

What has not actually been discussed yet is *how* exactly this deviation vector is computed through time. Here we present the two most common methods of evolving the deviation vector: The two particle method (2PM) simply using two nearby initial conditions and evolving them separately, and the tangent map (TM) method using the so-called variational equations.

### 1.3.1 Two Particle Method

The simplest, most direct approach to finding the evolution of the deviation vector is to directly integrate two nearby initial conditions. Since this approach involves effectively simulating two nearby particles, this is known as the two particle method (2PM). In this process, we choose an initial condition we would like to investigate, and create the system with these coordinates. Then we initialise the deviation vector, typically by randomly perturbing each coordinate, and then normalise the vector to the required norm. Finally, we create a second instance of the lattice with initial conditions described by the deviation from the initial lattice.

More clearly, say we have a phase space point $x(0) = x_0$, containing the initial position and momentum coordinates of the orbit we wish to characterise as regular or chaotic. We then choose the deviation vector we wish to evolve, $\delta x(t = 0)$. Using this deviation, we can define a second orbit $y(t)$ with initial condition $y(0) = x_0 + \delta x(0)$. These two initial conditions can be integrated simply according to the Hamiltonian equations of motion of our system, and in general at some time $t$ we can find the current state of the deviation vector by finding the difference between the two trajectories, $\delta x(t) = y(t) - x(t)$. In the evolution of these two trajectories, we are able to perform the renormalisation procedure discussed above, avoiding the problem of exponential growth of the deviation vector.

In an ideal world, this would be the end of it; we can choose our deviation however we wish and then simply evolve the two initial conditions. However, there are some considerations that need to be made. Foremost among these is that we have to have a *nearby* initial condition for the second trajectory. This means we need to somehow choose a sufficiently small deviation vector that our integration will correctly simulate two almost equivalent initial conditions. A first thought would be why not make this deviation as small as possible? After all, we would really like to simulate an infinitesimally small deviation. What we are able to do is to minimise the deviation size based on the available machine precision. Due to the renormalisation procedure, which needs to be performed accurately, a lower limit on the norm of the deviation vector has been suggested to be about $10^{-8}$ [37]. Any smaller deviation will start to be affected by machine precision errors since the square of the norm will be smaller than $10^{-16}$, which is beyond the accuracy possible with double precision floating point numbers. The trade-off for a small deviation vector is that the accuracy demanded of the integration is correspondingly higher – for a deviation vector norm of $10^{-R}$, the relative energy error of the integration should be on the order of $10^{-R-1}$ or smaller to maintain an accurate simulation of the two nearby orbits [37]. This accuracy is only possible through the decreasing of the integration time step (assuming that the integration scheme is optimal for this application), which means longer integration times. In some applications a larger deviation vector may produce accurate results, which allows for faster integration times. The exact choice of deviation vector thus has to be decided on a case by case basis.

When using this method, it is also advantageous to only update the ftmLE computation and renormalise the deviation vector after a gap of several time steps. By renormalising the extremely small deviation vector at every time step, there is a substantially increased chance for floating point roundoff errors to accrue, yielding an inaccurate estimation of the mLE. Managing floating point errors is particularly important since chaotic systems are inherently very sensitive to small changes in the coordinates. As such a small accrual of inaccuracies in the deviation vector size will very likely become apparent in a badly behaved ftmLE. Much like the size of the deviation vector, an optimal renormalisation time will depend on the system and the time step used. Approximately one time unit is recommended in [37], but with very long integration times or large time steps, longer renormalisation times may be necessary. Once again, the aim in choosing renormalisation times and time steps is to balance long-term accuracy with the desire for precision.

So implementing the 2PM, we have a direct way of estimating the mLE. While it is an old method [33], it is generally accurate and still relevant in many applications today (see e.g. [38]).

## 1.3.2   Tangent Map Method

A major advance in the road to fast and accurate computations of the mLE for dynamical systems, and Hamiltonian systems in particular, was brought about by the introduction of the so-called *variational equations* to integrate the evolution of the deviation vector through time [39]. The central idea here is rather than evolving two trajectories in the phase space and finding the difference between these two orbits, to directly simulate the evolution of a deviation vector in the tangent space. This is especially useful in the case of Hamiltonian systems, where the phase space has many useful properties, and we are able to study the tangent space with relative ease.

So if we wish to evolve a deviation vector $\boldsymbol{w}(t)$ from a reference orbit $\boldsymbol{x}(t)$, then we can do this using the variational equations. Following [29], we can define the Hamiltonian equations of motion for our solution $\boldsymbol{x}(t)$ of an $N$-dimensional system (with a correspondingly $2N$-dimensional phase space) as

$$\boldsymbol{x}(t) = \boldsymbol{g}(\boldsymbol{x}) = \mathbf{J}_{2N} \cdot \frac{\partial H}{\partial \boldsymbol{x}} \tag{1.16}$$

where $\partial H / \partial \boldsymbol{x}$ is the $2N$ dimensional vector given by

$$\frac{\partial H}{\partial \boldsymbol{x}} = \left( \frac{\partial H}{\partial q_1}, \frac{\partial H}{\partial q_2}, \ldots, \frac{\partial H}{\partial q_N}, \frac{\partial H}{\partial p_1}, \frac{\partial H}{\partial p_2}, \ldots, \frac{\partial H}{\partial p_N} \right)^T , \tag{1.17}$$

with the $q_i$ and $p_i$ the position and momenta coordinates of the vector $\boldsymbol{x}$. The matrix $\mathbf{J}_{2N}$ is the antisymmetric matrix

$$\mathbf{J}_{2N} = \begin{pmatrix} \mathbf{0}_N & \mathbf{I}_N \\ -\mathbf{I}_N & \mathbf{0}_N \end{pmatrix}, \tag{1.18}$$

where $\mathbf{0}_N$ is the $N \times N$ matrix of zeros and $\mathbf{I}_N$ is the $N$-dimensional identity matrix. We can see that (1.16) is of course just the vectorised form of the well-known Hamiltonian equations of motion

$$\dot{q}_i = \frac{\partial H}{\partial p_i}, \qquad \dot{p}_i = -\frac{\partial H}{\partial q_i}. \tag{1.19}$$

Now to study the tangent dyanamics of this orbit $\boldsymbol{x}(t)$ we need to take another derivative to find the tangent vector, and use this to evolve the deviation $\boldsymbol{w}(t)$ forward in time. To this end, we define the variational equations (the equations of motion for the deviation vector) as

$$\dot{\boldsymbol{w}}(t) = \frac{\partial \boldsymbol{g}}{\partial \boldsymbol{x}}(\boldsymbol{x}(t)) \cdot \boldsymbol{w} = \mathbf{J}_{2N} \cdot \mathbf{D}^2\mathbf{H}(\boldsymbol{x}(t)), \tag{1.20}$$

where $\mathbf{D}^2\mathbf{H}(\boldsymbol{x}(t))$ denotes the Hessian matrix. We note that this is simply the linearisation of the evolution of the nearby orbit, $\boldsymbol{x} + \boldsymbol{w}$, arising from Taylor expanding the equations of motion for $\boldsymbol{x} + \boldsymbol{w}$ and taking only linear terms in $\boldsymbol{w}$. This leaves us with $2N$ coupled linear differential equations for the motion of $\boldsymbol{w}(t)$, albeit with coefficients that have to be computed after each time step of the integration, as in general they depend on the current state of the orbit $\boldsymbol{x}(t)$.

There are several advantages to this approach, the most significant of which are related to the main aims of any numerical integration procedure: Speed and accuracy. Since we have more or less exact equations for the evolution of the deviation vector, we are freed from a number of the constraints of the 2PM. Firstly, we no longer have to worry about the size of the deviation. Any deviation norm will be correctly evolved by the variational equations, and since we are not simulating a second integration in the phase space, we are at liberty to choose any reasonable deviation vector norm. It is generally convenient (both mathematically and computationally) to set the norm of the deviation vector to unity. The fact that the size of the components of the deviation vector are no longer on the same order as the machine precision of our computers also means that we can renormalise the deviation vector after every time step at no cost to the precision of the integration. The second main advantage is that the variational equations provide a much more precise estimation of the deviation vector's evolution, and hence of the mLE itself [29]. We are consequently able to use results found through this method with a great deal of confidence (having of course ascertained that the equations are correct!), rather than having to take much care as with the 2PM that no numerical errors are creeping in to the process.

The major drawback to this method is that it requires the computation of the Hessian matrix of the Hamiltonian function. In many cases (especially for low dimensional systems and 1D lattices) this is not a problem, and explicitly finding the Hessian matrix and consequently the variational equations is fairly straightforward. In other cases though, finding these second derivatives may involve a significant amount of work. This then results in the very pragmatic trade off between mathematical difficulty in the derivation of the variational equations, and the numerical challenges of the 2PM. For a given system, it makes sense to consider both options and choose the method that will pose the fewest practical problems in the course of computing the mLE, based on the complexity of the Hamiltonian and the computational power available.

We note that there are other methods for computing the mLE from the deviation vector evolution, such as singular value and QR decomposition procedures (see for example [29] and references therein) which are nevertheless closely related to the "standard method", but in all the investigations carried out in our work we have used either the 2PM or the variational equations.

### 1.3.3 The Deviation Vector Distribution

Before we abandon the deviation vector as nothing but a tool for computing the mLE, it is worth taking a closer look at what else we can learn about the chaoticity of the system from it. It is well established that in a chaotic system, the deviation vector eventually aligns with a direction defined by the mLE, corresponding to the most chaotic behaviour of the system [29]. From this it is natural to look more closely at this direction of the deviation vector, as by considering it we can find useful information about the instability and the regions of active chaos [34]. This concept also fits the intuitive understanding of the deviation vector as a measure of chaos – consider for example a one dimensional lattice. If at a particular site the component of the deviation vector remains very small, it implies that for that site the two nearby orbits yield very similar results. Thus the system is in some sense "locally" relatively stable at that site. Conversely, if at a given site the deviation vector grows extremely fast, then we know that there the nearby orbits are producing very different results, commensurate with chaotic behaviour. So we can track these components of the deviation vector, and use them to identify chaotic hotspots in our systems [23, 34, 35, 40].

We define the deviation vector distribution (DVD) as the normalised distribution of the deviation at each site,

$$\xi_i = \frac{w_i^2 + w_{i+N}^2}{\sum_{j=1}^{N} w_j^2 + w_{j+N}^2}, \tag{1.21}$$

where the deviation vector $w$ has $N$ components of displacement deviations $\delta q$ followed by $N$ momentum deviations $\delta p$. Thus $\xi_i$ measures the (squared) relative total "deviation size" in the $i^{\text{th}}$ direction, or corresponding to site $i$ in a lattice. By computing $\xi$ through the course of the deviation vector evolution, we can build up a picture of how the instability spreads and moves through the lattice. Wherever the DVD is largest corresponds to the highest concentration of chaoticity, and the regions where the DVD is small are regions of relative stability.

The DVD has been used to study the phenomenon of breaking Anderson localisation in disordered one dimensional lattices [34], as well as to probe the underlying mechanisms of chaotic wave packet spreading in both one dimensional [23] and two dimensional lattices using various models, such as the Klein-Gordon and DNLS models [35], as well as Hamiltonian mean field [40] and Hertzian and Fermi-Pasta-Ulam-Tsingou models [41]. These studies have provided good evidence that the DVD can be a useful measure in understanding chaotic dynamics in a variety of systems.

# Chapter 2

# Modelling DNA

Now we turn to discussing the material of the research, through a brief introduction to the basics of DNA and a summary of the progress of the most relevant DNA models. This chapter will outline the structure of DNA, as far as the dynamics are concerned, and discuss the DNA model. This will include an overview of the ways that DNA has been modelled in the past, with an eye to how the models used in this investigation fare in reproducing experimental results.

## 2.1   Structure of DNA

The rich biochemical structure of DNA is deserving of several textbooks' worth of discussion (e.g. [1, 42, 43]), but fortunately for this specific application many of the biological details can be omitted. Indeed, the general model used for our studies, the Peyrard-Bishop-Dauxois (PBD) model, makes some drastic simplifications for the purposes of accurately capturing the dynamical behaviour [44–46]. The model itself will be discussed in the next section, but here we will first look at the basic structure of DNA. For a thorough review of DNA in the context of nonlinear dynamics, Peyrard's 2004 review paper is highly recommended [47].

Double-stranded DNA exists in the fairly well-known double helix structure. Figure 2.1 illustrates this basic formation. The fundamental building blocks of the strand are tripartite nucleotides – comprised of a phosphate group, a sugar ring, and a nitrogenous base. These monomers form the overall DNA polymer, the double-stranded DNA helix. The strands (made up of the phosphate and sugar backbones) are bound together by hydrogen bonds that form between the bases of the nucleotides. The discovery of this base-pair structure of DNA in 1953, and the relative simplicity of their nature, was groundbreaking in the field of biochemistry [48]. There are four bases found in these nucleotides: adynine, guanine, thymine, and cytosine. These bases fall into two groups; the purines (adenine and guanine) and the pyramidines (thymine and cytosine). Purines have a double-ringed structure, while the pyrimidines have a single-ringed structure. One of the more surprising aspects of Watson and Crick's discovery was that these bases form pairs very strictly with one complementary base. Rather than either purine bonding with either of the pyrimidines, they discovered that the only two possible naturally occurring base pairs are adenine-thymine (AT) and guanine-cytosine (GC). These base pairs turn out to be the essence of the genetic code that governs characteristics of living organisms, with the sequence of AT and GC pairs encoding essential information. They are also of obvious dynamical significance – while the remainder of the nucleotide is the same for every site along the strand, these base pairs provide the bond force connecting the two polymers by forming hydrogen bonds. The next notable property is that between AT base pairs two hydrogen bonds form, while between GC pairs there are three hydrogen bonds. This means that apart from the genetic significance of the sequence of base pairs, due to the inhomogeneity of the bond strengths, the sequence of base pairs also affects the dynamics of the DNA strand. Here as far as the dynamics go we are primarily concerned with studying the displacements of these base pairs, how far apart the bases move from each other in the natural course of their time evolution, which is of course significantly influenced by the strength of the base pair bonding.

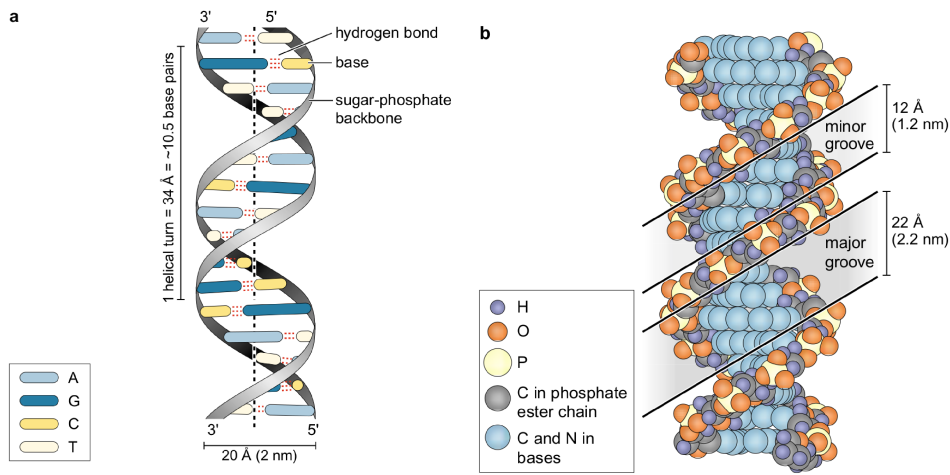

FIGURE 2.1: Double helix structure of DNA. The coiling allows for very efficient compactification. Each 'rung' in this twisted ladder is a base pair, connected by two or three hydrogen bonds. Figures from [42].

One of the most significant forces in the DNA double strand is the stacking interaction between base pairs. This is essentially a force between neighbouring base pairs arising from the interaction of the bases' electron clouds. As a consequence of the electron charge distribution across the base pairs, and also the fact that the base pairs' surfaces are hydrophobic, in order to optimise the energy arrangement the base pairs twist relative to each other, so that there will be an angle of around 30° between neighbouring bases. This is in large part what causes the double helix structure, as the bases twist against each other, forcing the backbone to form a helix, caused by the base pairs in some sense trying to minimise surface contact with the surrounding solvent. The energetically favourable arrangement aligns the planar aromatic rings of the nucleotide bases, and consequently causes the electronic stacking interaction, which depends on the charge distribution of each base pair. This interaction accounts for much of the coupling between neighbouring base pairs, and is the primary mechanism which needs to be modelled in DNA dynamics [1, 42, 43].

### 2.1.1 Significance of Open Base Pairs and Denaturation

In discussing dynamics of DNA, it is naturally important to identify where the biological significance of DNA (as a genetic code) meets the dynamical aspects that can be studied. This meeting point comes in with DNA transcription, the process whereby a particular DNA sequence "instructs" proteins in their development, defining the sequence of amino acids in the protein by the formation of a particular polypeptide chain, which then undergoes protein folding. In this process of communication between DNA and protein, the DNA strand opens up, under the influence of the RNA polymerase enzyme, to copy the base sequence of the relevant gene to a strand of RNA [1]. In particular, when transcription occurs, the RNA polymerase moves along the DNA strand, opening up the double helix to expose the base sequence to the RNA molecule. Specific base pair sequences are known to denote the transcriptions start and end sites, rather than amino acids, and the RNA polymerase moves between these two set sequences, allowing a single gene to be transcribed at a time. This RNA molecule will then provide the genetic information for the polypeptide chain, the sequence of amino acids, and thus the final protein product.

What all this means is that the opening of base pairs, indeed base pairs opening wide enough for the RNA copying to occur, is an essential part of the everyday activity of DNA molecules (even though these openings are created by an enzyme). This activity of base pairs is precisely the opportunity for the physicists, and particularly the dynamicists, to provide input. The opening and closing of base pairs is a fundamentally dynamical process – while in the case of DNA transcription it requires an external agent to perform the zipping and unzipping, similar openings occur entirely spontaneously in the form of thermal fluctuations. These thermal fluctuations are termed "bubbles", and can take various forms; breathers in a lattice, long-lived stable openings, travelling waves similar to transcription bubbles, erratic openings, or any number of ways [49–53]. Wherever there is energy in the strand, these openings can occur. The tendency of these openings to even exist with very large separations, sometimes larger than 7Å, or three times the natural length of a hydrogen bond [47, 54], means that nonlinear aspects of the dynamics will certainly be explored. While the dynamics of the stacking force and hydrogen bonds may be near linear at short distances, when probing the large displacement regimes, nonlinearity becomes important, as will be seen in the following section on the different models used to describe DNA dynamics.

A particularly important property of DNA, from a biological as well as dynamical standpoint, is the phenomenon of denaturation, the complete separating of the double helix [55]. In some sense this is an extreme case of bubbles forming, and encompassing the entire strand. The experimental approach to measuring denaturation is perhaps surprisingly simple for such microscale work. Due to the electronic structure of the nucleotide bases, there is a dramatic increase in the absorption of ultra-violet light with wavelength of 260nm when the base pairs separate, in the region of a 30–40% increase [56]. This allows the fraction of open base pairs to be quite accurately measured experimentally – as the solution is slowly heated, the amount of light absorbed can be tracked, and this corresponds precisely to the fraction of open base pairs. Experiments have been carried out on a variety of synthesised DNA strands, initially on homopolymers, strands with a single base pair type, but soon expanding to heterogeneous strands under various conditions (see [47, 56, 57] and references therein). This thermal denaturation is a remarkable process, since as the temperature of DNA in a solvent is increased, there is an abrupt increase in the fraction of base pairs that are separated [56, 57]. The type of solvent, salinity and various other factors can slow the process somewhat, but the overall picture is very clear, of a sudden change. Figure 2.2 shows an illustration of the ultra-violet absorption curve as the temperature increases.

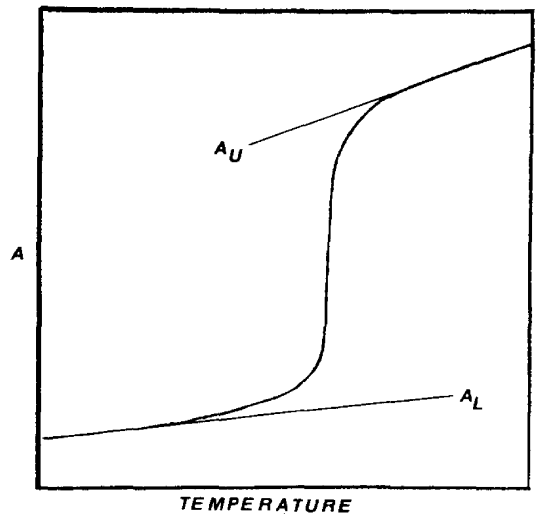

FIGURE 2.2: A representative plot of the amount of ultra-violet radiation absorbed in a particular DNA strand, as a function of the temperature. There is at first a slow linear increase in absorption, but a clear sudden increase around the critical temperature $T_c$. This corresponds exactly to the shape for the fraction of open base pairs. From [56].

Due to this sharp increase, it is convenient to define the critical denaturation temperature $T_c$ as the temperature at which precisely half the base pairs in the strand have separated. The fact that the region of denaturation is so small means that this single temperature is sufficiently precise. There has been a great body of experimental and theoretical work devoted to the study of melting points of different DNA sequences, the effects of alternations, solvent, base pair types, and strand length among a host of others. See for instance [56–60] and references therein. Biologically, denaturation occurs in the process of DNA replication, where the DNA strand separates completely to allow the entire sequence to be read and copied [1].

The denaturation of DNA has all the characteristics of a thermodynamic phase transition. As such melting is of interest as it is a genuine phase transition, from one phase (the connected double helix) to another phase (where the strands are completely separated), that can be modelled as a one dimensional system [61]. This phase transition will be discussed further in Chapter 4, in the context of nonlinear dynamics and chaos in DNA.

## 2.2 The Peyrard-Bishop-Dauxois Model of DNA and its Precursors

Over the last fifty years, there have been a variety of approaches to modelling DNA mathematically. Recently models have been proposed using *ab initio* calculations, attempting to capture the complex dynamics by accurately modelling each contribution to the overall forces [62], but by and large the most effective models have approached the problem from a more coarse-grained, base pair centred standpoint [47, 63].

As with any effort in mathematically modelling a physical system, the primary concern is in being able to reproduce physical results. In this case of DNA, one main result checked against is the denaturation temperature – as discussed above, there are experiments finding fairly precise melting temperatures for different base pair compositions. It is thus possible to model DNA by declaring that the most important aspect is whether or not a given base pair is open or closed, and then to reproduce some physical measurements of critical melting temperatures.

There are of course several ways of approaching this task, including a significant body of work focussing on solitons as explanations for open base pairs [64, 65], as well as the self-consistent phonon method [66]. But since here we are primarily concerned with discrete models, and of course particularly the PBD model of DNA, the discussion will be focussed on the precursors to this model and the model itself.

## 2.2.1 The Poland-Scheraga Model

The starting point for this type of model was a relatively simple Ising-like model, the Poland-Scheraga model developed in 1966 [67]. The Poland-Scheraga model treats each base pair as a lattice site that can be either open or closed, in the same way as the ferromagnetic Ising model has two-state magnetic dipole moments of $\pm 1$. This type of model has also been called a "helix-coil" model, as the two states (closed and open) correspond to sections of the DNA molecule that are in a helix form (closed base pairs) and a coil (open base pairs, the helix is separated) [63]. The essential idea of this model is to provide a rate ratio for the helix to grow or to shrink. If the helix grows, then more base pairs close, and this is often called zipping. Correspondingly, should the helix shrink, more base pairs are opening and the process is known, unsurprisingly, as unzipping. The main idea of this model is to accurately reproduce melting curves, essentially reporting the correct numbers of open base pairs. It has been moderately successful in this [56, 68, 69], and is still relevant for testing more recent results against [60]. The model was also improved by considering the problem from a random walk point of view, and particularly self-avoiding random walks, to describe the melting transition more accurately by Fisher [70]. However, despite this success it does not consider the dynamics at all, beyond considering the movement of bubbles in the double strand, and so it does not have the same scope as a more detailed model would have.

## 2.2.2 The First Peyrard-Bishop Model

The Peyrard-Bishop (PB) model arises from developing the Ising-like PS approach, and has been successfully used to reproduce experimental results of the thermodynamic properties of DNA [44, 45]. Essentially, this model takes the binary, discrete, variable at each site of a base pair being open or closed, and now allows this quantity to vary continuously. This moves the model from purely describing the presence of bubbles and denaturation to actually modelling the nonlinear dynamics of the base pair fluctuations. This of course can be used to study the open/closed nature of base pairs, through choosing thresholds for open base pairs (this will be discussed in detail in Chapter 5), but inherently allows a more thorough investigation of the nonlinearity of the base pair interaction as pointed out by Prohofsky [71]. A significant advantage of the model is that it presents a Hamiltonian dynamical system, which allows for efficient and effective analysis of the dynamics in general, and the chaoticity in particular.

In the basic version of this PB model, the effects of the left-right ordering of base pairs on the dynamics is not considered. So an AT base pair is functionally identical to a TA base pair, and similarly GC and CG. The reduction from the complex molecular DNA structure to a straightforward lattice is illustrated in Fig. 2.3. The overarching idea is to reduce the biological complexity, which is tremendously hard to model in faithful detail, to a simpler structure that is amenable to modelling through common molecular dynamics techniques. In this regime, the biological complexities will be expressed through carefully chosen and fitted parameters, optimising an approximation rather than attempting to exactly reproduce details.

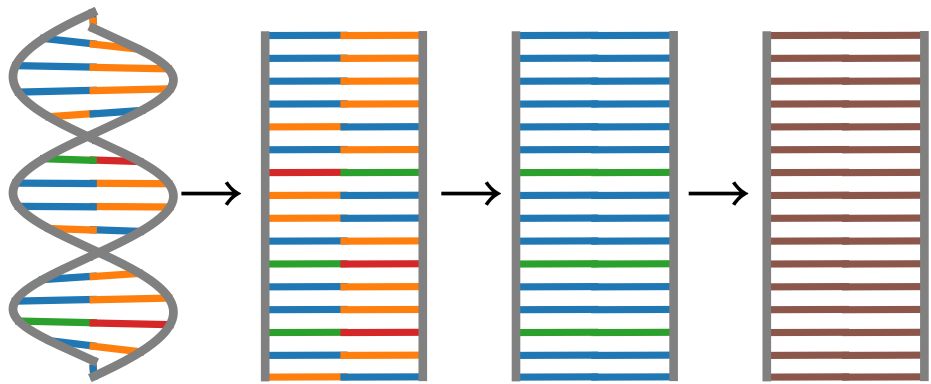

FIGURE 2.3: The simplification process for the PB model of DNA. The initial double helix is "flattened" into a straight line, with the twisting forces accounted for in the potential. Next, the precise details of AT/TA and GC/CG base pair orientation are ignored, resulting in a simple one dimensional lattice of two possible element types – AT and GC base pairs. The final simplification is to ignore even this distinction, and create a homogeneous lattice of base pairs, where the type is neglected. We will see how more developed models step these simplifications back, and take more detail into account.

Now that the structure is simplified to a comprehensible lattice, we can move on to describe the potential used to model its dynamics. The first, fundamental, Hamiltonian deals with the bases themselves as the elementary entities, and so the dynamical variables $u_n$ and $v_n$ are the displacements from equilibrium of the $n$th base in each strand. Here, we have not yet moved to the base pair as the fundamental object – so these $u_n$ and $v_n$ displacements describe the positions of the actual bases themselves, in each backbone strand. In these variables, for an $N$-base pair double strand the Hamiltonian reads [44]

$$H = \sum_{1}^{N-1} \frac{m}{2} \left( \dot{u}_n^2 + \dot{v}_n^2 \right) + D \left( e^{-a(u_n - v_n)} - 1 \right)^2 + \frac{k}{2} \left[ \left( u_n - u_{n-1} \right)^2 + \left( v_n - v_{n-1} \right)^2 \right], \quad (2.1)$$

where we have assumed that the masses of each nucleotide have the same value $m$. The two potential terms model the on-site potential for each base pair and the stacking interaction respectively. The on-site potential is modelled by a Morse potential [72]. This has been used extensively in modelling of chemical potentials [49, 73], and accurately imitates the combination of hydrogen bond attraction and phosphate interactions, providing a simple but effective way of approximating the base pair forces. Crucially, it has a sharp increase for negative displacements [see Fig. 2.6(a)], which is physically desirable – bases should very rarely come into close contact. The potential has a minimum at zero displacement, corresponding to the desired behaviour that the base pairs should return to precisely zero displacement. It also has a plateau at large displacements. This means that as the hydrogen bonds stretch significantly, the force pulling the base pair back to equilibrium diminishes, exactly as occurs when the bonds stretch too far and break. In the case of a simple harmonic potential, which has many of the desired characteristics – the minimum at zero, and prohibiting small displacements – the large stretching of base pairs would not be possible, as there is no plateau for large displacements. This makes the Morse potential a qualitatively appealing option for describing the base pair interaction. Here, the parameters of the Morse potential (the depth $D$ and characteristic width $a$) do not depend on the site at all, meaning that the Morse potential represents an average potential for both base pair types. This is an obvious simplification, but it facilitates analytical treatment of the system [44, 47].

The other term of the potential simply approximates the stacking interaction between adjacent bases as a straightforward harmonic coupling. As we will see, this is a definite oversimplification, but as with the average Morse potential this allows for relatively simple mathematical analysis.

Returning to the discussion of the strand dynamics, a change of variable simplifies the analysis significantly, as what we are primarily interested in is the displacement between base pairs – thus, we are concerned with *out-of-phase* motion of the two strands. If we denote the in-phase motion by the variable $x_n$, and out-of-phase motion by $y_n$, i.e.

$$x_n = \frac{(u_n + v_n)}{\sqrt{2}}, \qquad y_n = \frac{(u_n - v_n)}{\sqrt{2}},$$

then we can focus only on the out-of-phase variables $y_n$ to provide the dynamical information about the displacements between the base pairs. This then gives the Hamiltonian for the out-of-phase displacements after rescaling the parameter $a$ as

$$H = \sum_{n=1}^{N-1} \frac{p_n^2}{2m} + D(e^{-ay_n} - 1)^2 + \frac{k}{2}(y_n - y_{n-1})^2. \tag{2.2}$$

It is important to note that after this change of variables, the variable $y_n$ does not exactly represent the physical displacement between base pairs – in order to recover the actual displacement, the factor of $\sqrt{2}$ needs to be reapplied. This then defines a dynamical system which can be studied numerically and theoretically. In [44], where the model was introduced, dynamical quantities were studied using statistical physics techniques, and particularly the transfer integral method (see [47, 63] and references therein). One main result produced was to estimate the mean stretching of the bonds between base pairs, depending on the chosen coupling constant. Figure 2.4 shows this result, and most significantly shows the sharp increase in the mean stretching near the phase transition point. While this result does not explicitly relate to the fraction of open base pairs and thus denaturation, it certainly gives an indication that the model approximately exhibits the desired behaviour at the transition.

## 2.2.3 Peyrard-Bishop-Dauxois: Nonlinear Coupling

While this first model of the nonlinear dynamics of DNA is useful for studying various properties, and admitted extensive theoretical treatment as a straightforward model with harmonic coupling, it does have limitations [45]. There was consequently a need to modify the model, with the aim of accurately reproducing more physical results – while it roughly predicted the melting transition, through the fraction of open base pairs as well as a peak in the specific heat [45], the melting temperatures suggested by this analysis is much higher than the temperatures observed in reality.

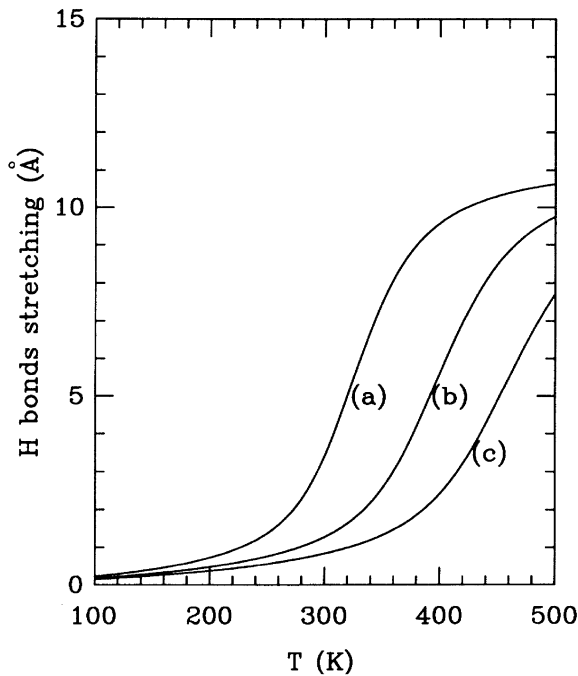

FIGURE 2.4: The mean stretching $\langle y_n \rangle$ of base pairs in DNA strands, according to the initial PB model. We see that there is a sharp increase near some critical temperature, for all values of the coupling constant $k$ considered [44].

The solution to this discrepancy between theory and experiment was to update the model, based on the observation that the force of the stacking interaction is dependent not only on the individual bases and their electronic arrangement, but actually on the base pairs themselves. Apart from supporting the philosophy that the fundamental unit structure in DNA dynamics is the base pair, this arises from the fact that when base pairs separate and break their connecting hydrogen bonds, the distribution of the electrons on the bases changes [46]. This change in electron distribution results in a decreased stacking force.

The idea here is that when there is a significant stretching in a base pair, it will weaken the coupling force between it and its neighbours. To implement this in the model would require some kind of adjustment to account for this new behaviour. The particular solution proposed by Dauxois, Peyrard and Bishop [46] is to add a nonlinear prefactor to the harmonic coupling. The new coupling potential is

$$W = \frac{k}{2}\left(1 + \rho e^{-\alpha(y_n + y_{n-1})}\right)(y_n - y_{n-1})^2. \tag{2.3}$$

This coupling displays the desired behaviours: If the displacements of both interacting base pairs are small, then the prefactor gives an increase (compared to the original

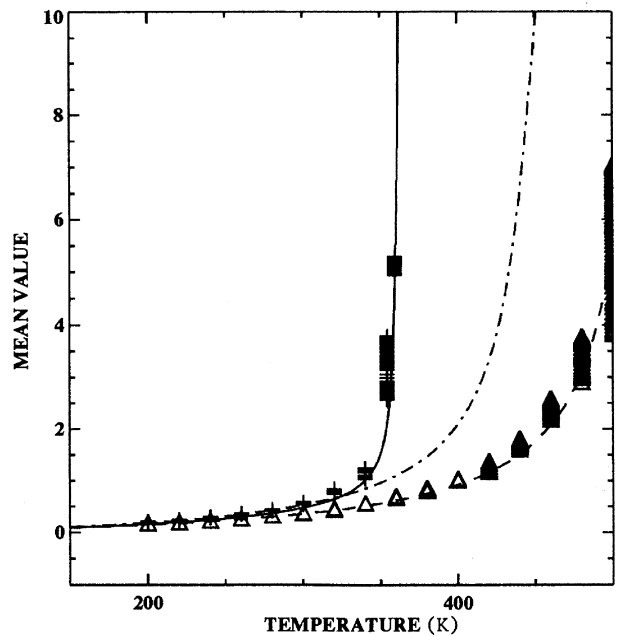

FIGURE 2.5: Mean stretching of base pairs in the PBD model. The solid line and plus signs correspond to the nonlinear coupling, exhibiting the desired sharp increase below 400K. The dash-dot corresponds to the harmonic coupling, and the dash-dot line and triangles to the harmonic coupling with a constant prefactor of 1.5 (equivalent to the nonlinear potential with $\alpha = 0$ and $\rho = 0.5$). The lines come from transfer integral calculations, and the symbols from molecular dynamics simulations [46].

PB model) of $\frac{k}{2}(1 + \rho)$, while when either base pair stretches significantly, the prefactor reduces to $\frac{k}{2}$, giving a much smaller stacking force.

On one hand, this is a minor adjustment. After all, the only change is that there is now a prefactor in front of the coupling term. On the other hand however, this does add the complexity of a nonlinear coupling, which turns out to be significant. One of the major issues with the previous PB model was that it did not reproduce melting temperatures at all closely, and exhibited a slower transition than expected. Now however, the improved PBD model actually performs much more reasonably. Figure 2.5 shows the increase in mean stretching of base pairs with temperature, for the nonlinear coupling, as well as the linear coupling with two different constant prefactors. We see that in the two harmonic cases, even worse with an increased constant factor, the transition is at very high temperatures (substantially above experimental melting temperatures), and is still slow, with the mean stretching increasing slowly. The nonlinear case gives us exactly what we would like though – a sharp transition, slightly above physiological temperatures. While the model as it currently stands does not account for the different types of base pair, the predicted melting temperature is at least now in the correct region.

There are however further steps to take before this model can start reproducing multiple specific experimental melting points. The first limitation is that there is no accounting for heterogeneity in the strand, despite knowing that AT and GC base pairs have fundamentally different dynamical contributions. The second requirement is that the parameters need to be chosen carefully, in order to produce accurate results for a variety of DNA sequences. Through allowing for heterogeneity in the Morse potential, allowing the parameters $D$ and $a$ to depend on the base pair type, and fitting to experimentally determined denaturation curves, parameters have been found that produce generally accurate results for a diverse array of base pair sequences [74].

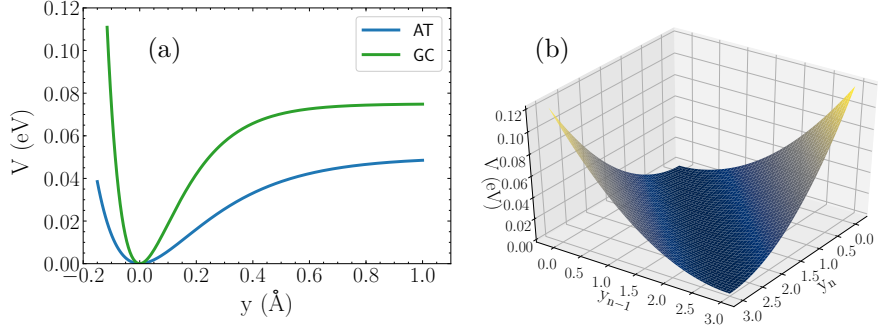

FIGURE 2.6: (a) The Morse potential for an AT base pair (blue) and a GC base pair (green). We see the effect of the coefficient $D_n$, dictating the height of the plateau. The higher plateau of the GC base pair is indicative of the stronger attractive on-site force. The constant $a_n$ affects the narrowness of the well. (b) The coupling potential between two base pairs. If the displacements are equal (i.e. no relative displacement), then the coupling potential contributes nothing, as would make sense. As displacement grows, the coupling potential increases exponentially.

The parameters chosen by Campa and Giansanti [74] are not significantly different to the original proposed PBD parameters, but the addition of heterogeneity allows for a much more precise reproduction of experimental results. The PBD potential, now considering all these factors, is

$$V = \sum_{n=1}^{N} D_n (e^{-a_n y_n} - 1)^2 + \frac{k}{2}(1 + \rho e^{-\alpha(y_n + y_{n-1})})(y_n - y_{n-1})^2. \quad (2.4)$$

Here we see the combination of the heterogeneous Morse potential (note the subscripts on the parameters $D_n$ and $a_n$), and the nonlinear coupling. The parameters of Campa and Giansanti have become accepted as near optimal [54, 60, 75], and are the parameters used throughout this investigation. The parameters are given in Table 2.1.

| Parameter | Value |
|-----------|-------|
| $k$ | $0.025\text{ev}/\text{Å}^2$ |
| $\rho$ | 2 |
| $\alpha$ | $0.35\text{Å}^{-1}$ |
| $D_{AT}$ | $0.05\text{eV}$ |
| $D_{GC}$ | $0.075\text{eV}$ |
| $a_{AT}$ | $4.2\text{Å}^{-1}$ |
| $a_{GC}$ | $6.9\text{Å}^{-1}$ |

TABLE 2.1: Optimised parameters for PBD simulations.

Figure 2.6 shows the potential energy contributed by each term. Figure 2.6(a) illustrated the Morse potential, for the two sets of parameters. We see that the larger well depth $D$ for the GC base pairs naturally results in a higher plateau, corresponding to the greater amount of energy required to break the triple hydrogen bonds. The characteristic width of the GC potential is also narrower, with the well consequently being much sharper than the AT case. So while it requires more energy to break the hydrogen bonds, the physical displacement required to reach the flattened region is in fact lower than it is for an AT base pair. The coupling potential is inevitably more difficult to visualise, as it depends on both neighbouring base pairs. In Fig. 2.6(b) we see the coupling energy as a function of the displacement of two base pairs. This plot shows basically the expected behaviour – a valley of zero energy when the displacements are equal, and so in relative equilibrium, and nonlinear increases as soon as the displacements differ.

### 2.2.4 The Extended Peyrard-Bishop-Dauxois Model

While the heterogeneous PBD model with a nonlinear coupling is an effective model for DNA, being used in studies of various properties of DNA (e.g. [50–52, 60, 76]), there are still dynamical aspects of DNA that are not explicitly addressed in this potential. One such aspect is the dependence of the stacking interaction on the base pair sequence. It is known that the stacking interaction changes depending on the actual bases involved, and the stacking energy of neighbouring base pairs can vary significantly [77, 78]. This stacking interaction depends not only on the base pair type (AT or GC), but also on the exact ordering of the bases. So now whether a base pair is AT or TA becomes significant, as this affects the stacking force. This is actually a large part of the motivation behind introducing a sequence-dependent term to the stacking force, as the difference between CC/GG and CG/GC stacking energies is significant [75]. Under the PBD model's framework, these two sequences are considered precisely equivalent – the details of CC/GG or CG/GC are completely lost in the final simplification step. However, if we wish to accurately model these different behaviours, we need to account for this sequence dependence. It is noteworthy that the difference in melting temperatures between these two sequences – one where one strand consists of purely guanine bases and the other of cytosine, and one where both strands alternate – is around 22K, which is significant (see Fig. 2.7 for an illustration of these two sequences).

In 2009, an extended PBD model (ePBD) was proposed, that took the sequence-dependence of the stacking interaction into account [75]. Here the stacking potential is adapted very slightly to account for this dependence by allowing the constant $k$ to vary according to the sequence. The updated stacking potential is then

$$W = \frac{k_n}{2} \left( 1 + \rho e^{-\alpha(y_n + y_{n-1})} \right)(y_n - y_{n-1})^2, \tag{2.5}$$

where the index $n$ on the constant $k$ allows for the coupling between every stacked base pair to be different. This means that $k$ of course takes on ten possible values, one for each possible step from one base pair to another. To find values of $k_n$ that accurately model both homogeneous and heterogeneous sequences, the authors of [75] fitted the melting temperatures obtained by the simulations to experimentally observed melting temperatures for specific sequences. Use was also made of previously available stacking constant ratios, and the known stacking energies. The optimised values of $k_n$ are presented in

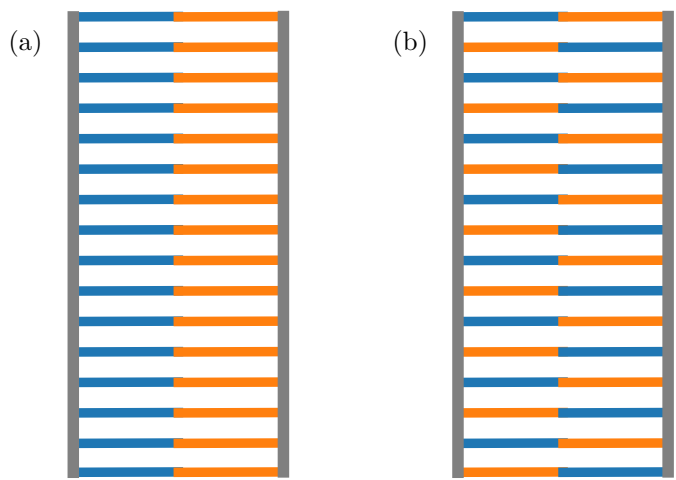

FIGURE 2.7: (a) A sequence with guanine bases on the left strand (blue), and cytosine on the right (orange). This corresponds to the CC/GG case, where the stacking energy and melting temperature is much lower. (b) A sequence with alternating guanine and cytosine bases. This (CG/GC) is a much more robust arrangement, as the stacking energies are higher, resulting in a significantly higher melting temperature.

|   | C | G | A | T |
|---|---|---|---|---|
| C | 0.0192 | 0.028 | 0.025 | 0.0229 |
| G | 0.0249 | 0.0192 | 0.019 | 0.0226 |
| A | 0.0226 | 0.0229 | 0.0228 | 0.023 |
| T | 0.019 | 0.025 | 0.0193 | 0.0228 |

TABLE 2.2: Numerical values of the sequence-dependent stacking constants $k_n$ in eV/Å$^2$, following the base sequence of the top strand. The row denotes the first base, and the column denotes the second base in the step. So $k_{AG}$, the stacking constant for the step from A to G, is found in the third row (A), and second column (G).

Fig. 2.8 as they are given in [75]. The numerical values are given in Table 2.2. We see that there is a noticeable difference between the extremes, and also that the CC/GG – CG/GC discrepancy is illustrated in the vastly different stacking constants. The other notable fact is that the stacking constants are almost all lower than the value used in the PBD model, suggesting that overall the base pairs are slightly more weakly coupled in the extended model.

The purpose of all this effort in extending the model is to more accurately reproduce physical results, particularly melting temperatures. The ePBD model has been successfully shown to match experimental melting temperatures for several sequences, as well as to produce other results that are biologically consistent and useful [54, 75]. In further chapters, the behaviour of this extended model will be compared to the simpler PBD model, both in chaotic dynamics and in physical properties such as melting temperatures and bubble formation.

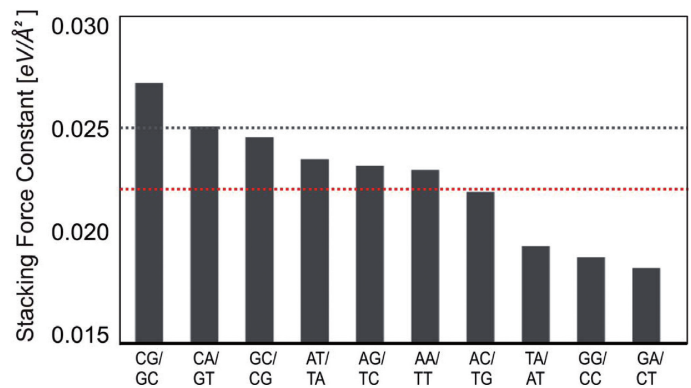

FIGURE 2.8: The fitted values of $k_n$ for each dinucleotide step. We see that the coupling between CG/GC base pairs is unusually strong, with the GG/CC stacking unusually weak, resulting in the dichotomy noticed in their melting temperatures. The black dashed line denotes the stacking constant of the PBD model (with no sequence-dependence), and the red line the average of all the stacking constants. Figure from [75].

# Chapter 3

# The Alternation Index and its Probability Distribution

Before actually getting into the dynamical study of DNA, we will take a rather mathematical side path in order to prepare an important tool for the dynamical study. In the course of studying the effects of heterogeneity of DNA, we need to have some understanding of what it is precisely that makes one DNA sequence "heterogeneous", and another sequence "homogeneous". As discussed in some depth in the preceeding chapter, we know that there are two kinds of base pair. It follows quite naturally that the heterogeneity, and disorder, in this DNA model comes from the combination of the two types of base pair.

It is intuitively quite clear that a sequence of only AT base pairs should be considered to be homogeneous. There is no disorder whatsoever, and no distinction between given sites in the lattice (here the one dimensional lattice is of course the DNA double strand). Similarly for a sequence of purely GC pairs, while it would be dynamically different to the AT sequence, it is nonetheless very homogeneous. The extension of this natural step is to introduce heterogeneity based on the mixing of these base pairs in a single sequence. So a sequence comprised of say fifty GC base pairs and thirty-five AT base pairs would be considered heterogeneous and disordered – it is now significant which lattice site we consider, as it will have different dynamical properties depending on what base pair type it is hosting.

However, this still leaves us a little short of a complete description of heterogeneity. We can separate homogeneous and heterogeneous sequences, but we would very much like to be able to compare the heterogeneity of two sequences. In this chapter, we will introduce the idea of the alternation index as a measure of this heterogeneity, and present the mathematical machinery used to derive a completely rigorous probability distribution for it. This chapter is based on the work in [79].

## 3.1   The Alternation Index

We will consider some particular DNA sequences to get some idea of what we are looking for. But first, in order to make things clearer for this section, in all representations AT and GC base pairs will be represented by black and white circles. This is motivated by the fact that for now we are forgetting all the DNA details – all we want to understand is the mixing of base pairs. So a simple and clear representation serves our purposes neatly. Filled black circles or beads will represent GC base pairs, and empty circles or white beads will represent AT base pairs. Figure 3.1 shows one particular arrangement of base pairs. In this configuration there are ten base pairs, four AT and six GC. Throughout this investigation we will quantify the ratio of AT to GC by referring to the percentage of AT base pairs in the chain. So in this case, where four of the ten base pairs are AT, the AT percentage $P_{AT} = 40\%$.

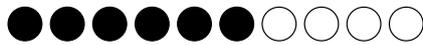

FIGURE 3.1: A schematic representation of a DNA sequence. Here we have ten base pairs, four of which are AT (white beads), and six of which are GC (black beads). Thus the AT percentage $P_{AT} = 40\%$. The base pairs are grouped together, so there is in some sense minimal mixing of base pairs.

We can take the same base pairs and mix them around a bit, and get a large number of different arrangements (such as in Fig. 3.2). These arrangements are referred to as disorder realisations, when linking back to the idea of DNA as a one dimensional lattice. So for a given number of base pairs in the sequence, and a given AT percentage, there are multiple disorder realisations possible.

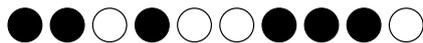

FIGURE 3.2: Here we still have ten base pairs, four of which are AT, and six of which are GC, but now there is some mixing – there are three distinct clusters of GC base pairs, and three distinct clusters of AT base pairs. So with the same AT percentage and number of base pairs, we have a sequence that we would expect to be meaningfully different to the previous arrangement in Fig. 3.1. There is a more thorough mixing in this arrangement, which we consider to be more homogeneous.

The question now becomes what is different between these realisations? We know that the sequence of base pairs is genetically significant, which is encoded in the fact that the PBD model accounts for base pair types in the Morse potential. It is also natural than when considering heterogeneity the AT percentage is significant – of course sequences with different compositions will yield differing dynamical behaviours. But even within the same AT percentage, as these two realisations have illustrated, quite different arrangements are possible.

In order to quantify these differences in arrangements we will introduce the *alternation index*, $\alpha$. This alternation index simply counts the number of times the DNA sequence changes base pair type, as it is traversed. So for instance, consider Fig. 3.1. Starting from the left, we find that from the third to fourth base pairs, the base pair type switches from GC to AT. So this is the first alternation. Then from the seventh to the eighth base pair, once again we have an alternation back from AT to GC. This is the second and final alternation. We now also realise that boundary conditions play a significant role in the alternation index – if we were to impose periodic boundary conditions on the DNA molecule then we would end up where we started. This will be discussed in more detail later where we do make use of periodic boundary conditions.

Applying the same counting process, we see that the sequence in Fig. 3.2 has six alternations. This tells us that the more "well-mixed" a sequence is, the higher the alternation index. We can also see that for this case, of four AT base pairs mixed between six GC base pairs, there is a maximum possible value of $\alpha$ of eight – this corresponds to every AT base pair being isolated from each other and dispersed among the GC mass. Of course, whether the majority of base pairs are GC or AT is irrelevant – if we had four GC base pairs and six AT base pairs, the maximum value of $\alpha$ would still be eight. Similarly, we can see that the minimum number of alternations in a heterogeneous sequence (so at least some of both AT and GC base pairs) is one, or two if we impose periodic boundary conditions. So we now have a way of quantifying how heterogeneous a sequence is – the smaller the alternation index, the more "chunky", and hence heterogeneous, it is.

What we would like to do then, is to study the dynamics of DNA based on this measure of the heterogeneity. This is the motivation for introducing the alternation index. So having defined it, we are in principle in a position to study the effects of heterogeneity – we can vary both the AT percentage and the alternation index, and see what results we find. However, we would like to perform a more informed investigation. In particular, we would like to know which values of $\alpha$ are more probable, and which values are extremely improbable. Ideally this would be in the form of a probability distribution function (PDF), where the exact most likely value can be found, and the probability of extreme cases can be quantified. As it turns out, this is a surprisingly difficult probability distribution to find. Especially when imposing periodic boundary conditions, finding a clean way of calculating the number of possible combinations is extremely tricky. There have been calculations making use of Markov chains to approximate the probabilities of different arrangements of base pairs in non-periodic sequences [80–82]. However, in our periodic case there turns out to be a mathematically beautiful, and most importantly exact, way of calculating the number of different arrangements, and hence finding the probability of each.

### 3.1.1 Formulation of the PDF Problem

We are about to dive in to a mass of mathematics, which we will ultimately use to describe the probability of various values of $\alpha$, in a sequence of $N_{AT}$ AT base pairs and $N_{GC}$ GC base pairs. But before being subjected to pages of group theory, let us see the motivation for using all this machinery. Our aim is essentially simple: Take a number of base pairs, $N$, of which $N_{AT}$ are AT pairs and $N_{GC}$ are GC. If we were to pick a completely random arrangement of these base pairs, we would like to know what the probability is for some particular number of alternations $\alpha$. Here we are imposing periodic boundary conditions, so we additionally require that the first and last base pairs are neighbours.

This distribution depends on two variables, $N_{AT}$ and $N_{GC}$ (note $N = N_{AT} + N_{GC}$). It would be natural to consider a binomial distribution, or some kind of general statistical distribution. However, as it turns out trying to select only sequences with the same alternation index is a rather difficult task in this approach. We instead consider the problem from a more first principles standpoint: What we actually need is the number of possible arrangements of base pairs with a given alternation index, for every possible alternation index. From this we could simply normalise the distribution and we would have our desired PDF.

A crucial fact to recognise in this problem is that our difficulty (restricting sequences to a given alternation index) in fact allows for a simplification which leads to a new type of solution. Our periodic DNA sequence is, for the purposes of this study, equivalent to a necklace of black and white beads. So associating black beads with GC base pairs and white beads with AT base pairs, we can address this as a classical combinatorical question, rather than a purely statistical one. With this in mind, we can recognise that actually a necklace of black and white beads can be written as a necklace of *containers* of beads, with each container holding a certain number of base pairs. Let us consider some examples to illustrate this idea.

Figure 3.3 shows the periodic representation of the sequence shown in Fig. 3.2. This step is fairly straightforward – we are simply connecting the boundary beads to form a loop.

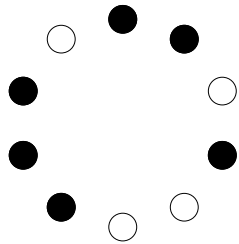

FIGURE 3.3: The necklace of black and white beads corresponding to the arrangement in Fig. 3.2. The sequence has just been looped around so that the first and last beads are neighbouring.

So having this necklace, we can represent it through a sequence of containers rather than beads (Fig. 3.4). This necklace of containers takes each homogeneous chunk of the sequence and compresses it to a single element, which gives the colour of the beads contained in that chunk as well as the number of beads. So the two white beads compress to a single white container labelled with "2". Note that even the single beads are now represented through containers, even though the number of beads contained is simply one.

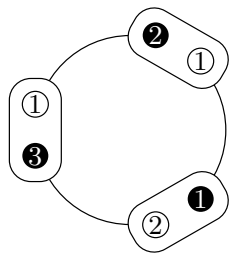

FIGURE 3.4: The representation of the necklace in Fig. 3.3 as a necklace of containers. The number on each container corresponds to the number of beads "contained" within it. So each homogeneous sequence of beads collapses to a single element, which then represents that sequence as the number of beads in it. The containers are grouped in pairs to show the alternations more clearly – each pair of containers corresponds to two alternations. Note also that we could expand this necklace of containers unambiguously into the original necklace of black and white beads.

The important thing to note here is that the number of containers is equal to the alternation index. This is where we are taking advantage of the fact that the actual number of beads in each container does not affect the alternation index. So as long as the correct number of beads are somehow distributed among these containers (of course with at least one bead in each), then we are satisfying the criteria of set numbers of black and white beads and a set number of alternations. This reformulation of the problem is actually significantly easier to approach.

The approach we use to count the possible number of bead distributions for each value of the alternation index is based in Pólya counting theory [83]. In the following sections we will introduce the necessary tools to solve the problem – group theory and especially cyclic groups, finite power series, and the crux of the theory, Burnside's Lemma and Pólya's Enumeration theorem with an extension for bipartite sets.

## 3.2    Some Mathematical Tools

Let us dive into some mathematics, primarily from the field of group theory, to establish the requisite tools to address the rather complex problem of counting necklaces subject to constraints. These topics are covered in most algebra texts; see for instance [84, 85]. The discussion here is based largely on the contents of those works.

### 3.2.1    Groups: Permutations, Cycles and Properties

Mathematical groups are well-known abstract entities, consisting of a set with an operation. So formally, given a set $A$ with a binary operation $\cdot$, they are considered in combination to be a group if the following hold [84]:

1. $\exists i \in A \mid \forall a \in A \; i \cdot a = a \cdot i = a$ (Existence of an Identity).

2. $a \cdot (b \cdot c) = (a \cdot b) \cdot c \; \forall a, b, c \in A$ (Associativity).

3. $\forall a \in A \; \exists b \in A \mid a \cdot b = b \cdot a = i$ (Existence of Inverses).

**Definition 3.2.1** (Group). *Any set with a binary operation satisfying properties 1-3 is a group.*

A particularly useful, and hence well-studied, branch of group theory is that of permutation groups, where the operation on the set of items is a permutation of the group items. These permutations are rearrangements of the elements of the set. So the elements of the set $C = \{1, 2, 3\}$ could be written in six different ways: 123, 132, 321, 312, 231, 213. Each of these ways of writing the elements is a permutation of the set $C$.

The *symmetric group* on a set $A$ is the set of permutations of $A$.

**Definition 3.2.2** (Symmetric Group). *The symmetric group on a set A, $S_A$ is defined as*

$$S_A = \{\sigma : A \to A \mid \sigma \text{ a permutation}\}$$

This group is important for a number of reasons; not only is it useful in itself, but also every group whose elements interchange set elements in any way is a subgroup of the symmetric group, as all possible interchanges are contained in the set of permutations.

Within these permutations, it is useful to introduce the notion of *cycles*. A cycle, as the name suggests, is a permutation that is cyclically periodic under repeated application. We say that a given permutation is an *n-cycle* if after $n$ applications of the permutation the set elements have returned to their original positions.

**Example 1.** *Consider the set $A = \{1, 2, 3, 4\}$, and a permutation $\sigma$ that swaps the first and third elements and swaps the second and fourth elements, i.e. $\sigma(1) = 3$, $\sigma(2) = 4$, $\sigma(3) = 1$, $\sigma(4) = 2$.*

In this example, we see that clearly after two applications of $\sigma$, the set returns to its original state: $\sigma(\sigma(A)) = A$. We also see that there are two distinct cycles occurring within $\sigma$. The first cycle sends $1 \to 3$ and $3 \to 1$, while the second cycle sends $2 \to 4$ and $4 \to 2$. We denote these in *cycle notation* as (13) and (24) respectively. This is equivalent to calling them (31) and (42), as these mean the same thing – interchanging $1, 3$ and $2, 4$ respectively. What is also clear is that each of these cycles are 2-cycles. After two iterations, each element is back where it started. Thus, we can write $\sigma$ as the composition of two 2-cycles, $\sigma = (13)(24)$.

**Definition 3.2.3** (Cycle). *A permutation $\sigma \in S_A$ is called a cycle if there are distinct elements in the set $A = \{a_1, a_2, \ldots, a_m\}$, such that*

1. *$\sigma(a_i) = a_{i+1}$ for $i \in \{1, 2, \ldots, m-1\}$.*

2. *$\sigma(a_m) = a_1$.*

3. *$\sigma(a) = a$ for any element of $A \notin \{a_1, a_2, \ldots, a_m\}$.*

So any $n$–cycle can be written as $(a_1 \, a_2 \, \ldots \, a_n)$. In a natural way, we say two cycles are disjoint if they do not interact at all. So if we have two disjoint sets $\{a_1, a_2, \ldots, a_m\}$ and $\{b_1, b_2, \ldots, b_k\}$, then the cycles $\sigma_1 = (a_1 \, a_2 \, \ldots \, a_m)$ and $\sigma_2 = (b_1 \, b_2 \, \ldots \, b_k)$ are also disjoint.

With these definitions, and using the fact that for this problem we are only concerned with finite groups, we can state a theorem that will be very useful in further considerations.

**Theorem 1** (Cycle Decomposition Theorem). *For a finite set $A$, any permutation $\sigma \in S_A$ can be written as a composition of a finite number of pairwise disjoint cycles. So*

$$\sigma = (a_{11} \, a_{12} \, \ldots \, a_{1m_1})\ldots(a_{k1} \, a_{k2}\ldots \, a_{km_k}). \tag{3.1}$$

*This decomposition is unique, up to ordering of the cycles.*

Being able to write permutations as a product of distinct cycles turns out to be crucial for the machinery in the coming sections, as we will soon see.

## 3.2.2 Group Action and Cycle Index

In order to understand the next tool in the necklace counting process, we need a basic notion of a *group action*. A group action of some group $G$ on a set $B$ is a mapping of the elements of $G$ to the symmetry group $S_B$. This mapping is in fact a homomorphism that identifies the group elements of $G$ with permutations in $S_B$, in some sense allowing communication between the group and the set. We label the group action of $G$ on $B$ as $A_G$ on $B$.

For instance, let us take $G$ as the cyclic group of order four $C_4$, which is the group of rotations for sets of four elements. Then the action of $G$ on a set of four points $P = \{p_0, p_1, p_2, p_3\}$ would be to rotate the points through their various positions. So we have the group action $A_G$ on $P$, where we abbreviate notation by saying that the action of a group element $g$ on an element $p$ is $g\,p$, rather than $A_G(g)(p)$. If $g_1$ is the element of $G$ that performs a single rotation, then $g_1 p_i = p_{(i+1)\%4}$. Similarly, $g_2 p_i = p_{(i+2)\%4}$ and so on. This is the idea of a group action, where the group elements act on the set to produce permutations of the set.

Having this group action, we can define some important concepts. The first we will address is the idea of *fixed points*.

**Definition 3.2.4** (Fixed Points). *Given a group $G$ acting with action $A_G$ on a set $B$, the set of fixed points of an element $g \in G$ is denoted by*

$$\text{Fixed}_g = \{b \in B \,|\, g\,b = b\}.$$

This simply defines the set of all elements that are left invariant by the action of the given element. For this reason this set is also referred to as the *invariant set* of $g$.

In the context of counting distinct necklaces, there is an intuitive link here between the ideas of "elements that are unchanged by a permutation" and "necklaces that are the same when permuted". This will of course become clearer in due course.

The next subset to consider is the *orbit* of some element of the set.

**Definition 3.2.5** (Orbit). *The orbit of an element $b \in B$ under the action of a group $G$ is the set of all elements that $G$ maps $b$ to*

$$\text{Orbit}_b = \{g\,b \,|\, g \in G\}.$$

While this may be less clear at first thought than the idea of fixed points, it is also not an intuitively strange idea. It is somewhat analogous to the range of a conventional mathematical function.

Thirdly, we will consider the subgroup called the *stabiliser* of an element $b$.

**Definition 3.2.6** (Stabiliser). *The stabiliser of an element $b \in B$ under the action of a group $G$ is the group of all elements of $G$ that leave $b$ unchanged.*

$$\text{Stab}_b = \{g \in G \,|\, g\,b = b\}$$

This is of course similar to the set of fixed points – where the fixed points are set elements unchanged by a given group element, the stabiliser is comprised of group elements that are idempotent on a given set element.

The final concept to introduce here is the cycle index [86]. The cycle index encodes combinatorial information in the form of a polynomial, and in particular the coefficients of each term in the polynomial provide information about the action of groups on sets. When we make use of the cycle index for counting distinctly coloured necklaces, its utility will become more readily apparent. For now though, we settle for a definition and some illustrative examples.

**Definition 3.2.7** (Cycle Index). *Given a finite group $G$, acting on a finite set $B$ with action $A_G$. If the cardinality of $B$ is $n$, then the cycle index of the group action is given by*

$$Z_G(y_1, y_2, ..., y_n) = \frac{1}{|G|} \sum_{g \in G} y_1^{c_1(g)} y_2^{c_2(g)} ... y_n^{c_n(g)}.$$

In this definition, the $c_i$ refer to the number of $i$—cycles in the (unique) disjoint cycle decomposition of the permutation $A_G(g)$. So for example if there are twelve cycles of length seven, $c_7 = 12$.

Some examples of cycle indices for groups that are relevant to our counting problem are given below [87].

- The cyclic group $C_n$, which is used to account for rotational symmetries:

$$Z_{C_n}(y_1, y_2, ..., y_n) = \frac{1}{n} \sum_{p|n} \varphi_t(p) y_p^{n/p}. \tag{3.2}$$

  Here $\varphi_t(p)$ represent Euler's totient function, counting the number of integers less than $p$, that are also relatively prime to $p$.

- The dihedral group $D_n$, which adds reflectional symmetries to the rotations of the cyclic group:

$$Z_{D_n}(y_1, y_2, ..., y_n) = \begin{cases} \frac{1}{2n} \sum_{p|n} \varphi_t(p) y_p^{n/p} + \frac{1}{2} y_1 y_2^{(n-1)/2}, & \text{for } n \text{ odd}, \\ \frac{1}{2n} \sum_{p|n} \varphi_t(p) y_p^{n/p} + \frac{1}{4} \left( y_1^2 y_2^{(n-2)/2} + y_2^{n/2} \right), & \text{for } n \text{ even}. \end{cases} \tag{3.3}$$

  In our bead-counting context, the two cases for odd and even $n$ correspond to the slightly different meaning of a reflection for necklaces with odd or even numbers of beads. This is discussed in more detail later, when the dihedral group is used for the counting of distinct necklaces.

We have now introduced, albeit briefly, the fundamental building blocks needed in the process of counting distinct necklaces. There are some other concepts that will be defined as they come up in the natural flow of outlining the counting theory, rather than front-loading this chapter with an enormous mass of abstract ideas without context.

## 3.3 Burnside's Lemma and Pólya Counting

With these tools, we turn back to the discussion of different ways of arranging beads in a necklace. To complete the change from DNA sequences to pure combinatorics, we will refer to these arrangements of beads as *colourings*. This is a very natural change of terminology – after all, we can consider the base necklace to be an empty template of $N$ blank beads. Then to create some particular necklace we colour $N_{AT}$ beads with white, and $N_{GC}$ with black. So by colouring the "blank template", we have formed our necklace of beads.

Let us recapitulate the aim in this new language: We wish to count the number of *distinct* colourings of the necklace subject to some restrictions, with colours of black or white. We do now need to spend some time thinking about these restrictions. Firstly there are the natural restrictions mentioned previously, that we have $N_{AT}$ white beads and $N_{GC}$ black beads, and a total of $\alpha$ alternations in the chain. But there are further, more subtle restrictions we need to impose. For instance, in principle the two necklaces in Fig. 3.5 are different, although they are in practice the same necklace. We need to account for this symmetry somehow. All necklaces that are equivalent under some rotation should be subtracted from the count of total possible arrangements.

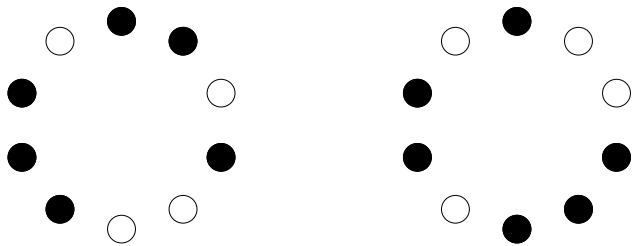

FIGURE 3.5: Two apparently distinct necklaces. However, it is clear that they are simply rotated versions of the same necklace. Harking back to the original problem, periodic DNA molecules are clearly unaffected by a rotation like this. So for our purposes we would like to consider these two necklaces to be equivalent.

Furthermore, for the PBD model of DNA, there is a reflectional symmetry. Due to the fact that it does not matter which way we traverse the sequence in this model, the two necklaces in Fig. 3.6 are functionally equivalent. This finally completes our set of restrictions on the necklace. The problem now is to count all possible colourings, or permutations of beads in the necklace, and then to discount all duplicate colourings.

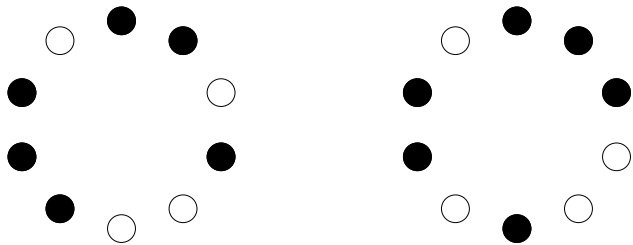

FIGURE 3.6: Two arrangements which are again in principle different, but in actual fact are considered equivalent as they are simply reflections of each other.

### 3.3.1 Counting Distinct Colourings: Burnside's Lemma

The basic lemma underlying the theory of counting distinct colourings is often referred to as Burnside's Lemma [88]. But in order to make use of this, we need first to translate the colouring problem to a problem using sets and groups. So we shall consider a set $B$, containing all the beads we wish to colour. The cardinality of this set is of course $N$. Then we wish to colour each of these beads with an element of the set of possible colours $\Omega$. In our case this set $\Omega$ contains just two colours, black and white, but all that follows applies to more general sets of colours. Our colouring is then a mapping from $B$ to $\Omega$, giving each bead in $B$ a colour from $\Omega$. We will denote this map as $\varphi : B \to \Omega$. With this description of the colourings, we can label the set of all possible colourings $\varphi$ from $B$ to $\Omega$ as $\Omega^B$. Explicitly, $\Omega^B := \{\varphi : B \to \Omega\}$

Next, we will use a group action to account for the symmetries in the necklaces. If we have some symmetry group $G$ with a group action $A_G$, then this group action on the set $B$ describes the symmetry we need to account for. Intuitively, the chosen group $G$ will be associated with certain symmetries. The group action $A_G$ on the set $B$ will then, in the course of the algorithm, allow us to eliminate these repeats. What is important to note here is that if we have this action $A_G$ on $B$, which maps to permutations of $B$, then these permutations can apply in exactly the same way to the colours associated with the permuted elements. So if some group element $g$ takes an element $b \in B$ to $c \in B$, then the same element $g$ would map $\varphi(b) \in \Omega$ to $\varphi(c) \in \Omega^B$. Concisely, the group action $A_G$ on $B$ gives rise to an associated group action $\tilde{A}_G$ on $\Omega^B$ such that $g \varphi : b \mapsto \varphi(g\,b)$.

Now we have the descriptions we need – we have a set of elements, and a set of possible colourings, as well as a group which will account for the relevant symmetries. Note that the number of distinct possible colourings is equal to the number of distinct orbits of the group action $\tilde{A}_G$. Intuitively, the key point is that all colourings that are equivalent under symmetry are included in a single orbit. This is the first step to reducing the number of excess colourings. The second step is of course to eliminate orbits that themselves are duplicates. So finally counting only these distinct orbits, we have actually counted all the truly independent colourings of the necklace.

Burnside's lemma gives us a very efficient way to calculate the number of distinct orbits of a group action on a set [88]:

**Lemma 1** (Burnside's Lemma). *In the group action of a finite group $G$ on a set $S$, the number of distinct orbits is given by the average number of fixed points of the elements in $G$.*

$$N_{DO} = \frac{1}{|G|} \sum_{g \in G} \left| \text{Fixed}_g \right| \tag{3.4}$$

On the face of it, this may not appear particularly helpful, but in many cases it is actually deceptively simple. To illustrate the process more clearly, consider a simple example.

**Example 2.** *Compute the number of cyclically distinct arrangements of black and white beads in a five-bead necklace, i.e. colourings that are equivalent up to rotations of the necklace.*

In this simple five bead case, we can actually explicitly construct all the colourings that are not rotations of another. All the possible arrangements, without any restrictions, are shown in Fig. 3.7.

Figure 3.8 shows the arrangements of necklaces that are not rotationally equivalent. We can see that there are eight such distinct colourings, and one can convince oneself of this by comparing every possible colouring in Fig. 3.7 and seeing that the remaining necklaces are in fact not unique under rotation.

Now, let us see what Burnside's lemma gives us. The formula requires the cardinality of the symmetry group, and the number of necklaces fixed or left invariant by *each* element of the group. Since the symmetry in question is rotation, we will use the cyclic group of order five, $C_5$. There are of course five elements in this group, rotations of one through four places as well as the identity permutation which leaves each bead as it is. So now we have

$$N_{DO} = \frac{1}{5} \sum_{g \in C_5} |\text{Fixed}_g|. \tag{3.5}$$

For each element in $C_5$, i.e. each rotation, we need to count how many necklaces (out of all the possible necklaces) remain unchanged by a rotation. Since there are five beads of two colours, the total number of possible necklaces is $2^5 = 32$. If we now consider each rotation case by case:

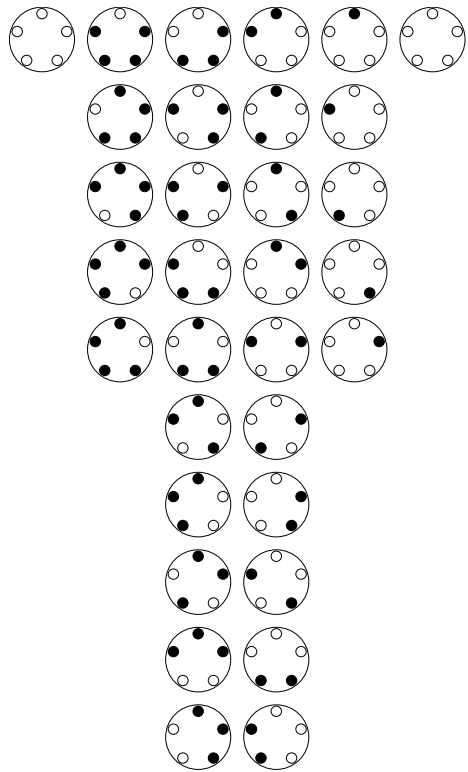

FIGURE 3.7: All possible arrangements of the five bead necklace. It is immediately clear for instance that all necklaces with four black beads and one white bead are in fact rotationally equivalent.

- Rotation by 0 places (Identity). As this leaves every necklace invariant, $|\text{Fixed}_{R_0}| = 32$.

- Rotation by 1 place. Now the only necklaces left invariant are the pure black and pure white necklaces. Rotating every other necklace gives a slightly different necklace each time. So then $|\text{Fixed}_{R_1}| = 2$.

- Rotation by 2 places. Once again, because of the odd number of beads in the necklace, there are only the same two invariant necklaces, giving $|\text{Fixed}_{R_2}| = 2$.

- Rotation by 3 places. As above. So $|\text{Fixed}_{R_3}| = 2$.

- Rotation by 4 places. Again, as above. $|\text{Fixed}_{R_4}| = 2$.

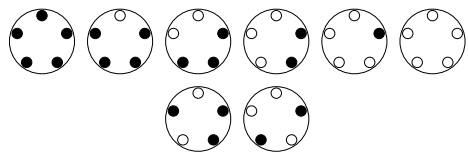

FIGURE 3.8: Rotationally distinct arrangements of black and white beads in a five bead necklace. Clearly none of these necklaces can be constructed merely by rotating another. However, if we consider any other arrangement of beads, we will find that it is simply a rotation of one of these eight fundamental necklaces.

Now, we can compute the number of distinct orbits using (3.5). So we get

$$N_{DO} = \frac{1}{5}(32 + 2 + 2 + 2 + 2) = \frac{40}{5} = 8,$$

which is precisely the number we found by manual comparisons. □

This example illustrates the general idea of using Burnside's lemma, and a very mechanical approach to calculating the number of distinct orbits. Fortunately though, in general there are useful mathematical tools that allow us to calculate these numbers of invariant elements without having to consider which necklaces remain invariant under each permutation. Here we can make use of the cycle index defined in Section 3.2 and restate Burnside's lemma as

$$N_{DO} = Z_G(k, k, ..., k), \tag{3.6}$$

where the symmetry is given by a group $G$, and there are $n$ beads coloured with $k$ colours. Note that the cycle index takes $n$ arguments, so in (3.6) $k$ enters $n$ times into the bracket.

If we now apply this process to our example, we have a much easier route to the number of distinct colourings. Recalling from earlier (3.2) the cycle index of the cyclic group, we can calculate the cycle index of the order five group as

$$Z_{C_5}(x_1, x_2, x_3, x_4, x_5) = \frac{1}{5}x_1^5 + \frac{4}{5}x_5, \tag{3.7}$$

since the only integers in $\{1, 2, 3, 4, 5\}$ that divide 5 are 1 and 5, and $\varphi_t(5) = 4$ since $1, 2, 3$ and 4 are all relatively prime to 5.

So then to find the number of distinct colourings of a five bead necklace coloured with two colours from Burnside's lemma is very straightforward:

$$\begin{aligned} N_{DO} = Z_{C_5}(2, 2, 2, 2, 2) &= \frac{1}{5} \cdot 2^5 + \frac{4}{5} \cdot 2 \\ &= \frac{32 + 8}{5} \\ &= 8, \end{aligned} \tag{3.8}$$

just as we had before.

Among the obvious advantages of this approach in general is that we can trivially compute the possibilities for different numbers of colours by changing $k$. While we do need to compute the cycle index for each value of $n$ we are interested in, since we have a general formula for most symmetry groups this is not a major hardship, and certainly far less work than manually finding the fixed points of every permutation in $G$.

So we now have a robust algorithm to calculate the number of distinct necklaces of $n$ beads with $k$ colours, with various symmetries. This is great progress. If all we wanted was to find the number of possible DNA sequences with given length, this would suffice. However it addresses neither of our chief concerns: Nowhere does this method account for specific numbers of black and white beads, nor does it consider the number of alternations of these beads. Some more work is required in order to include these restrictions.

### 3.3.2 Weighted Colourings and Pólya's Enumeration Theorem

The first topic we will address is that of specifying set numbers of black and white beads. In order to account for numbers of different kinds of beads, we need a mathematical way of identifying the bead types. The way we do this is by introducing the concept of a *weight* for each colour. So for each colour $c$ in our set of colours $\Omega$, we associate with it a weight $w_c \in \mathbb{N}$. The function that assigns these weights is simply $w : \Omega \to \mathbb{N}$, where $w(c) = w_c$.

The weight of a particular colouring $\varphi : B \to \Omega$, we can find the total weight of this colouring by taking the sum of the weights of all the colours:

$$\left| W_\varphi \right| = \sum_{1 \leq i \leq |B|} w\left(\varphi(b_i)\right). \tag{3.9}$$

These weights then in some way take note of the colours of each bead in a given colouring.

What is significant about the total weight is that it allows us to identify colourings with certain numbers of colours. More concretely, take the pertinent case of two colours, black and white, for a necklace of $N$ beads in a set $B$. Assign $w_{\text{white}} = 1$ and $w_{\text{black}} = 2$, with $N_{\text{white}} + N_{\text{black}} = N$ in some particular colouring $k$. Then the total weight will be given by

$$|W_k| = \sum_{i=1,N} w(k(b_i)), \; b_i \in B$$
$$= 1 N_{\text{white}} + 2 N_{\text{black}},$$

and with the restriction that $N_{\text{white}} + N_{\text{black}} = N$, there is of course only one total weight that possible for a specific choice of $N_{\text{white}}$ and $N_{\text{black}}$. This means that we can rephrase the problem of counting arrangements with $N_{\text{white}}$ white beads and $N_{\text{black}}$ black beads as a problem of counting the number of distinct colourings with a given weight. For instance, in the above case, if we were to set $N = 10$ and $N_{\text{white}} = 3$, $N_{\text{black}} = 7$, then we are looking for colourings of 10 beads satisfying

$$|W_k| = 1 \cdot 3 + 2 \cdot 7 = 17.$$

So now we can move onwards in this direction, of looking at colourings and weights, aiming to find some way of counting these distinct colourings.

What has not appeared in the above discussion of course is this notion of distinct colourings. We still want to account for equivalence classes of colourings, and only count non-equivalent colourings. Of course, counting non-equivalent colourings is the same as counting the equivalence classes of the colourings, up to some desired symmetry. So now we are back in the territory of Burnside's lemma, where we wish to account for these symmetries with some group action. If we have some symmetry group $G$ acting on our set of beads $B$, then what is important to note is that the orbits of $A_G$ on $B$ all have the same total weight. This is the crux of the idea – because in a given orbit, we have the same set of coloured beads being shuffled around (rotated, reflected, etc.), the total weight of course remains the same. The same beads, however permuted, have the same total weight.

So in order to count the number of distinct colourings, we need to count the number of distinct orbits under the desired group action. Correspondingly, to count distinct colourings with given numbers of beads, we need to count the distinct orbits with the requisite total weight.

In order to generalise Burnside's lemma to this case, we will use Pólya's powerful theorem for counting distinct colourings with chosen weights [83]. Before moving to the actual theorem however, we do need to introduce one more piece of mathematics: *Generating functions*. These are essentially simple objects. The idea of a generating function is to encode information about a certain problem or process in the coefficients of a power series [87]. Frequently they are used in counting problems, where these coefficients can represent solutions for different cases. This is the application used here – we will see that the computation reduces to finding the desired coefficient in the generating function series. An ordinary generating function takes the form [87]

$$f(x) = \sum_{n=0} c_n x^n, \tag{3.10}$$

which is a power series in $x$ with coefficients $c_n$.

We are however interested in a more specific class of generating functions. What we want to investigate are colourings with a particular total weight. So at this point, we need a generating function for a given assignment of weights. Take an assignment of weights $w$, which assigns each colour in $\Omega$ a natural number in $\mathbb{N}$. If there are a finite number of colours with each weight, i.e. the inverse function $w^{-1}(n)$ maps to a finite number of colours, then we can define a generating function for this weight assignment as the following:

$$f_w(x) = \sum_{n=0} |w^{-1}(n)| x^n, \tag{3.11}$$

where $|w^{-1}(n)|$ is the number of colours with a particular weight $n$.

Now we are ready to understand Pólya's enumeration theorem. While there are further mathematical tricks that will be introduced in due course to facilitate rapid calculations, we can present the framework of the theorem.

**Theorem 2** (Pólya's Enumeration Theorem). *Take a set of objects $B$, a set of colours $\Omega$, a function $w : \Omega \to \mathbb{N}$ assigning weights to each colour with an associated generating function $f_w$, with a group $G$ acting on the set $B$ through the action $A_G$. This action passes to a group action $\tilde{A}_G$ on the set $\Omega^B$, and we have that the generating function for the number of distinct orbits of $\tilde{A}_G$ by total weight is*

$$f_{DO}^{\tilde{A}_G}(x) = Z_G(f_w(x), f_w(x^2), \ldots, f_w(x^n)), \tag{3.12}$$

*where $Z_G$ is the cycle index of the group $G$.*

This theorem allows to obtain information about the number of colourings with a desired total weight from the cycle index of the symmetry group. As a generating function, the coefficients of this series give the number of colourings with a certain *total weight*. So the number of colourings with a total weight of $k$ is given by the coefficient of the $k^{th}$ term in the polynomial.

To illustrate this, let us return to the earlier example (Example 2), where we computed the total number of possible colourings of cyclically distinct necklaces with two colours. Now let us restrict further, and say we want only necklaces with three white beads and two black beads. Note that this restriction was not available to us in Burnside's lemma, so we do in fact need to use Pólya's enumeration theorem.

We will once again assign weights very simply: $w(\text{white}) = 1$, $w(\text{black}) = 2$. So we have that the total weight $W = 1N_{\text{white}} + 2N_{\text{black}}$, and $N = N_{\text{white}} + N_{\text{black}}$. Recalling our reformulated aim, which is to count colourings with a given total weight, we need to identify the desired total weight in this case. Since we want $N_{\text{white}} = 3$ and $N_{\text{black}} = 2$, we have $W_{32} = 1 \cdot 3 + 2 \cdot 2 = 7$. The next step is then to produce the generating function polynomial, and find the coefficient of the term in $x^7$, as this coefficient will tell us the number of colouring with total weight of seven.

The ingredients we need are 1) the relevant cycle index, and 2) the generating function for these weights. We of course already have the cycle index (3.7), so let us look at the generating function. The generating function will be as given in (3.11). In our case, we have two colours, so the only weights that have colours assigned to them are the numbers 1 and 2. So expressing our generating function, we will have

$$f_w(x) = \sum_{n=0}^{\infty} |w^{-1}(n)| x^n = |w^{-1}(1)| x^1 + |w^{-1}(2)| x^2. \tag{3.13}$$

Now we can use the fact that each weight corresponds to only one colour – the weight 1 corresponds to the colour white, and the weight 2 to the colour black. This means that $|w^{-1}(1)| = 1$ and $|w^{-1}(2)| = 1|$, and our generating function becomes simply

$$f_w(x) = x + x^2. \tag{3.14}$$

So we now have all the pieces to construct the final polynomial. Pólya's enumeration theorem tells us that

$$f_{DO}^{\tilde{A}_G}(x) = Z_G(f_w(x), f_w(x^2), \ldots, f_w(x^n))$$
$$f_{DO}^{\tilde{A}_{C_5}}(x) = Z_{C_5}(f_w(x), f_w(x^2), f_w(x^3), f_w(x^4), f_w(x^5))$$
$$= Z_{C_5}(x + x^2, x^2 + x^4, x^3 + x^6, x^4 + x^8, x^5 + x^{10})$$
$$= \frac{1}{5}(x + x^2)^5 + \frac{4}{5}(x^5 + x^{10})$$
$$= x^5 + x^6 + 2x^7 + 2x^8 + x^9 + x^{10}. \tag{3.15}$$

This is now our generating function polynomial for distinct colourings under rotational symmetry, by total weight. So we can read off the coefficients and recover the number of distinct colourings with any distribution of black and white beads. Our original aim was $N_{\text{white}} = 3$, $N_{\text{black}} = 2$, with total weight of seven. Clearly there are two possible distinct colourings in this case, as the coefficient of the $x^7$ term is equal to two. Pleasingly, we also get all the information about other arrangements at the same time – so we can find that the number of arrangements with two white and three black beads is also two for instance. We can in fact gain all the information we gained by painstakingly constructing Fig. 3.8, just by reading each coefficient.

So this illustrates the power of the method. We can, in general, use this to find the number of distinct colourings with particular numbers of black and white beads. But of course, this is not the end – we have a further restriction, as well as a difficulty on the horizon. The further restriction is that we want to enforce a certain number of alternations between differently coloured beads. Thus we need to move from the notion of coloured beads to the containers of beads introduced in Section 3.1.1, where the number of alternations maps directly to the number of containers. The challenge which quietly raised its head in the course of the example above is the expansion of the polynomial. While in this case we had a very small necklace, and could expand the polynomial without undue difficulty, for large polynomials this becomes computationally very expensive.

We will address this, and some other computational details, in due course. But for now, we need to turn to this idea of the containers, and how we can use Pólya's enumeration theorem to help us find what we want.

### 3.3.3 Pólya's Enumeration Theorem for Bipartite Sets

So let us recall the container construction described in Fig. 3.4. The obvious advantage of this description is that if we are working with these containers, then the problems about the alternation index are immediately solved – if we have $2M$ alternations, then we simply have $2M$ containers, $M$ containers of each colour. So far so good, there is in principle no problem with using these containers as the fundamental elements of our necklace. But what else do we require in order to use Pólya's counting theorem? The symmetry group is fine, we can still choose the same group to account for relevant symmetries. So if we have this, and of course the corresponding cycle index, all that remains is to figure out the weights. Conveniently, there is also a very natural way to treat the weights – instead of weighting white and black beads with 1 and 2 respectively, we will now colour each container with *the number of beads it contains*. So if a container holds seven beads, its weight is seven. Notice a critical point from this assignment of weights: the total weight (the crucial element we are looking at) of the black containers will give the total number of black beads, and the same for the white containers.

What is not clear in this situation is how to deal with the different coloured containers; how does the symmetry group act on the containers? And how does Pólya's theorem apply if we have to treat the different colours separately? To this end, we need to consider the special case of *bipartite* sets and group actions.

So let us take a set $B$, which is separated into two partitions, so $B = Y \sqcup Z$, where $\sqcup$ denotes a disjoint union. In our case, we can see that our set will be the set of containers, and the partitions the sets of black and white containers. What we then need to consider is a colouring that colours elements of the set $Y$ from a set $\Omega_Y$, and colours elements of $Z$ from $\Omega_Z$. Thus a valid colouring $\varphi$ according to this bipartite viewpoint is any colouring $\varphi : B \to \Omega_Y \sqcup \Omega_Z$ if it takes only elements of $Y$ to $\Omega_Y$ and elements of $Z$ to $\Omega_Z$, with no crossover. So concisely, $\varphi(a) \in \Omega_Y \iff a \in Y$ and $\varphi(a) \in \Omega_Z \iff z \in Z$.

Now that we have made sure that our colouring preserves the bipartite nature of the set, all that remains is to discuss the group action, which should do something similar. Our only additional requirement on the group action, in order for us to be able to state a version of Pólya's theorem for bipartite sets, is that the group action needs to maintain the separation of the sets. So we will call this kind of group action *partition-preserving*.

**Definition 1** (Partition-Preserving Group Action). *Given a group $G$, and a partitioned set $B = Y \sqcup Z$, we say that the group action $A_G$ is partition preserving if for each $g \in G$, $ga \in Y \iff a \in Y$ and $g \in G, ga \in Z \iff a \in Z$.*

With this requirement, the group action in some way splits into two effectively identical actions acting simultaneously on the two partitions. In particular, it means that the cycles of the cycle decomposition of elements of the group action are themselves partition-preserving. So if we write the decomposition of $A_G(g)$, for some element $g \in G$, we will have $c_1 \cdot c_2 \cdot c_3 \cdots c_m$, and each $c_i$ permutes elements of the same set partition. Not all cycles have to be in the same partition – $c_j$ and $c_k$ can permute elements in partition $Y$ and $Z$ respectively, but no cycle can itself move elements between the partitions.

We will then denote by $c_m^Y(g)$ the number of $m$—cycles in the cycle decomposition of $A_G(g)$, that are confined to the partition $Y$. Similarly, $c_m^Z$ gives the number of $m$—cycles in partition $Z$. From this we have a fairly complete bookkeeping of the group action on the partitioned set. Particularly, we are able now to define the *bipartite cycle index*, which will allow us to move on to the final statement of a Pólya counting theorem for these bipartite sets.

**Definition 2** (Bipartite Cycle Index). *Given a set $B = Y \sqcup Z$ of $n$ elements, and a group $G$, the* bipartite cycle index *of a partition-preserving group action $A_G$ on $B$ is defined as*

$$\tilde{Z}_G(y_1, \ldots, y_n, z_1, \ldots, z_n) = \frac{1}{|G|} \sum_{g \in G} y_1^{c_1^Y(g)} \ldots y_n^{c_n^Y(g)} z_1^{c_1^Z(g)} \ldots z_n^{c_n^Z(g)} \tag{3.16}$$

Now, we have all the tools to state our bipartite counting theorem [79].

**Theorem 3** (Bipartite Pólya Enumeration Theorem). *Take a finite group $G$, with a partition-preserving group action $A_G$ on a finite set $B = Y \sqcup Z$. Let the set of colours be $\Omega = \Omega_Y \sqcup \Omega_Z$, with weights $w_Y : \Omega_Y \to \mathbb{N}^+$ and $w_Z : \Omega_Z \to \mathbb{N}^+$ and corresponding generating functions given by $f_Y$ and $f_Z$. If the set of valid colourings of $B$ is $\varphi$, then the group action $A_G$ passes naturally to a group action $\tilde{A}_G$ on $\varphi$, and the generating function by total weight for the number of distinct orbits of $\tilde{A}_G$ is*

$$f_{DO}^{\tilde{A}_G}(a) = \tilde{Z}_G\left(f_Y(a), \ldots, f_Y(a^m), f_Z(a), \ldots, f_Z(a^m)\right). \tag{3.17}$$

So now we are able to use this theorem to solve our actual problem: We can identify terms with the desired total weight in each variable, which corresponds to the number of black and white beads, and the number of alternations is given by the total number of elements (so the cardinality of the set $B$ will be equal to $\alpha$). All that remains is to construct the bipartite cycle index for the dihedral group, since that is the symmetry we are concerned with, and then to find efficient ways of computing the coefficients of relevant terms of the polynomial.

## 3.4 A Probability Distribution for the Alternation Index

Now, in order to solve our original problem, we need to adapt the relevant symmetry group to the bipartite structure of the necklace of containers. The symmetry group under consideration is still the dihedral group, as this accounts for the rotational and reflective symmetries in our necklace. It is worth noting here that the PBD model of DNA does not have a prescribed direction of traversal – whether one moves clockwise or counter-clockwise along the sequence, the dynamics are identical. This means that reflectionally equivalent sequences are dynamically equivalent. In DNA models with a set direction of traversal (such as the ePBD model) there is no longer this reflective symmetry, as a reflection would mean essentially moving along the sequence in the opposite direction. In these models the symmetry group simplifies to the cyclic group. However, in this study we are sticking to the PBD model, and so we use the dihedral group.

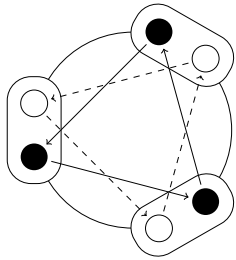

FIGURE 3.9: An illustration of a necklace of six containers, showing the action of a rotation on the containers. The black and white containers are rotated independently.

In order to construct the bipartite cycle index of the dihedral group $D_{2M}$, we look at the two components separately: The contribution from the cyclic component, and the contribution of reflective symmetry. The action of rotations on the containers is very straightforward. Figure 3.9 shows the partition-preserving rotations, simply moving each container around, while maintaining the separation of black and white beads.

So the group generated by these rotations, the cyclic group, has the usual cycle index form, but adapted to the bipartite nature by the inclusion of two sets of variables. Recall (3.2), for the cycle index of the cyclic group in one set of variables. We adapt it to the bipartite case by simply adding in the second set of variables in a natural way:

$$\tilde{Z}_{C_M}(y_1, y_2, ..., y_M, z_1, z_2, ..., z_M) = \frac{1}{M} \sum_{p|M} \varphi_t(p) \, y_p^{M/p} z_p^{M/p}. \tag{3.18}$$

This accounts for half of the elements of $D_{2M}$, as the dihedral group consists of half rotations and half reflections.

Accounting for the reflections requires a little more thought however. Consider the two cases given in Figs. 3.10 and 3.11, corresponding to reflections with $M$ odd or even. Recall that there are a total of $2M$ containers in the necklace. We can see that in the odd case, the reflections are fairly straightforward affairs – one white container stays where it is, one black container stays where it is, and every other container switches places with the container opposite (forming 2-cycles). Thus the total cycle index for the case of $M$ being odd is

$$\tilde{Z}_{D_{2M}}(y_1, ..., y_M, z_1, ..., z_M) = \frac{1}{2M} \sum_{p|M} \varphi_t(p) \, y_p^{M/p} z_p^{M/p} + \frac{1}{2} y_1 z_1 y_2^{(M-1)/2} z_2^{(M-1)/2}. \tag{3.19}$$

Now we need to consider the slightly more complex case of $M$ being even. For this, we see that in half the reflections we fix two white containers and in half the reflections we fix two black containers. In all these cases the rest of the containers split into 2-cycles. So to account for this we will have two terms in the cycle index – one to account for the half fixing two white containers, and one to account for the half fixing two black containers. The final form for even $M$ is then

$$\tilde{Z}_{D_{2M}}(y_1,...,y_M,z_1,...,z_M) = \frac{1}{2M}\sum_{p|M}\varphi_t(p)\,y_p^{M/p}z_p^{M/p} + \frac{1}{4}\left(y_1^2 y_2^{(M-2)/2}z_2^{M/2} + z_1^2 z_2^{(M-2)/2}y_2^{M/2}\right). \tag{3.20}$$

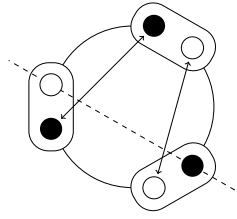

FIGURE 3.10: A reflection with $M$ odd. Note that this fixes one black and one white container.

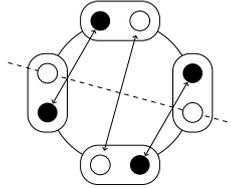

FIGURE 3.11: A reflection with $M$ even. This now fixes one pair of either black or white containers.

Now armed with our bipartite cycle index, we are in principle set to create polynomials, and find numbers of possible arrangements from the coefficients. Particularly, we also have as a corollary from our bipartite Pólya counting theorem the bivariate generating function [79]

$$f_{DO}^{\tilde{A}_G}(a,b) = \tilde{Z}_G\left(f_Y(a),...,f_Y(a^m),f_Z(b),...,f_Z(b^m)\right), \tag{3.21}$$

which gives a generating function for the number of distinct colourings of the set $B$ by total weight in $\Omega_Y$ and $\Omega_Z$.

We still have not addressed the issue of actually computing the coefficients of the terms in this polynomial, but let us first look at an example to illustrate why we very quickly reach a point where brute force techniques do not work.

**Example 3.** *Find the number of distinct necklaces with seven white beads, six black beads, and ten total alternations.*

So, we have $\alpha = 2M = 10$, $N_W = 7$, $N_B = 6$. Thus we have the case where $M$ is odd, and we can immediately identify our generating function using the bipartite Pólya's theorem with the cycle index given by (3.19).

$$
\begin{aligned}
f_{DO}^{\tilde{A}_G}(a,b) &= \tilde{Z}_G\big(f_Y(a),\ldots,f_Y(a^m),f_Z(b),\ldots,f_Z(b^m)\big) \\
&= \tilde{Z}_{D_{10}}\big(f_W(a),\ldots,f_W(a^5),f_B(b),\ldots,f_B(b^5)\big) \\
&= \frac{1}{10}\sum_{p|5}\varphi_t(p)f_W^{5/p}(a^p)f_B^{5/p}(b^p) + \frac{1}{2}f_W(a)f_B(b)f_W^2(a^2)f_B^2(b^2) \\
&= \frac{1}{10}\Big[(a+a^2+\ldots)^5(b+b^2+\ldots)^5 + 4(a^5+a^{10}+\ldots)(b^5+b^{10}+\ldots)\Big] \\
&\quad + \frac{1}{2}(a+a^2+\ldots)(a^2+a^4+\ldots)^2(b+b^2+\ldots)(b^2+b^4+\ldots)^2. \quad (3.22)
\end{aligned}
$$

Here we the generating functions $f_W$ and $f_B$ are used as introduced earlier, simply power series in their argument. Note here that while in previous examples we truncated the power series to only two terms, since in principle we can have any number of colours here (recall that the colours are the number of beads in each container, which for long necklaces could become arbitrarily large), we will work with the full power series. This leaves us with some difficulties. The first is that we have infinite series to multiply – this is obviously impossible. Since we are only interested in the coefficient of terms in $a^7 b^6$ (recall this is the total weight required for the given numbers of beads), we can truncate the series though, so that we only expand terms that can contribute this term. In this case, we can truncate the first polynomials (which are raised to the power of five) at fourth order since higher powers have no effect on the end polynomial below terms of ninth order. We are not actually concerned with the second pair of polynomials – since the powers only appear in multiples of five, none of these will have an impact on our targetted term. The final four polynomials can also be truncated. Since the highest power concerned is eight, the two squared polynomials can be truncated to two terms, since all other terms will ultimately enter in order nine or above. Similarly, the non-squared polynomials can be truncated to four terms, as after this the added terms do not have an impact. So for our purpose, we just need to expand the following polynomial, and extract the coefficient of $a^7 b^6$:

$$
\begin{aligned}
&\frac{1}{10}\Big[(a+a^2+a^3+a^4)^5(b+b^2+b^3+b^4)^5\Big] \\
&\quad + \frac{1}{2}(a+a^2+a^3+a^4)(a^2+a^4)^2(b+b^2+b^3+b^4)(b^2+b^4)^2. \quad (3.23)
\end{aligned}
$$

Needless to say, this is best performed using a computer algebra system. Expanding the two multiplications using Sage [89] we find the relevant parts of the expansion to be

$$
\frac{1}{10}\big(\ldots 75a^7 b^6\ldots\big) + \frac{1}{2}\big(\ldots 3a^7 b^6\ldots\big). \quad (3.24)
$$

But we now have the final result – the coefficient of $a^7 b^6$, the number of necklaces with seven white and six black beads up to dihedral symmetry, is $75/10 + 3/2 = 9$.

That was a tiresome calculation; we had to decide how to truncate polynomials, then algebraically expand nontrivial expressions. If this is the process for a small necklace of thirteen beads, it is clear that a radically new approach is needed for necklaces of hundreds or thousands of beads (let alone the fact that actual DNA sequences can be on the order of millions of base pairs). Particularly, the computation time for expanding $f_w(x)^n$ grows exponentially with $n$, so it is imperative that we find a better method. To save us this trouble, we will introduce some more mathematical short cuts. The first of these is to consider our generating functions as formal power series. The theory of formal power series (see e.g. [90]) gives a large number of properties that we can potentially take advantage of, but in particular we will use the fact that there is a form of the binomial theorem that holds.

**Lemma 2.** *If $(1-x)^{-n}$ denotes the formal inverse of $(1-x)^n$, then we have that*

$$(1-x)^{-n} = \sum_{i=0}^{\infty} \binom{n+i-1}{n-1} x^i, \tag{3.25}$$

Just as the binomial theorem allows us to calculate binomial coefficients, so we are able to use this theorem, with some massaging, to find the coefficients in the power series expansions. The key here is that the computation of these binomial coefficients is performed in linear time (computation time increasing as a linear function of $n$), so this is a major boost in performance from the naive expansion of enormous polynomials. Particularly, we have another lemma we can state about the powers of the generating functions $f_w(x)$.

**Lemma 3.** *The generating function $f_w(x)^n$ can be written as a formal power series as*

$$f_w(x)^n = \sum_{i=0}^{\infty} \binom{n+i-1}{n-1} x^{n+i}, \tag{3.26}$$

This lemma is straightforward to prove from Lemma 2. Since $x f_w(x) = x^2 + x^3 + \ldots = f_w(x) - x$, we can see that clearly $f_w(x) = x(1-x)^{-1}$, which gives us that $f_w(x)^n = x^n(1-x)^{-n}$. Thus, we now have the result from Lemma 2. $\square$

Putting these results into the final form, for finding relevant coefficients, we have that the coefficient of $x^p$ in the polynomial $f_w(x^a)^b$ is given by

$$\left[ f_w(x^c)^d \right]_p = \begin{cases} 1, & \text{if } d = 0 \text{ and } c = 0, \\ 0, & \text{if } d = 0 \text{ and } c > 0, \\ 0, & \text{if } d > 0 \text{ and } c \nmid p \text{ or } p < cd, \\ \binom{p/c-1}{d-1}, & \text{otherwise.} \end{cases} \tag{3.27}$$

and that the coefficient of $x^p$ in the polynomial $f_w(x^{a_1})^{b_1} \cdot f_w(x^{a_2})^{b_2}$ is given by

$$\left[ f_w(x^{c_1})^{d_1} \cdot f_w(x^{c_2})^{d_2} \right]_p = \sum_{i=0}^{p} \left[ f_w(x^{c_1})^{d_1} \right]_i \left[ f_w(x^{c_2})^{d_2} \right]_{p-i} \tag{3.28}$$

where the coefficients on the right hand side are computed according to (3.27).

Now, despite the daunting appearance of these formulas, we have a computationally efficient and fundamentally straightforward way of computing the coefficients of the polynomial terms. The complexity of the general formulas diminishes rapidly when confronted with concrete cases, so let us now illustrate their use by redoing example 3.

The cycle index is given in (3.22). Once again, let us look at this one polynomial set at a time. Recalling that we want terms in $a^7 b^6$, for the first pair, we have

$$(a + a^2 + \ldots)^5 (b + b^2 + \ldots)^5.$$

So the based on the procedures above, we want to find $\left[ f(a)^5 \right]_7$, the coefficient of the term in $a^7$, for the generating function $f(a)$ raised to the fifth power. This means that in (3.27), we have $c = 1$, $d = 5$, $p = 7$, and hence

$$\left[ f(a)^5 \right]_7 = \binom{7-1}{5-1} = \binom{6}{4} = 15.$$

Similarly for $b$, we find

$$\left[ f(b)^5 \right]_6 = \binom{6-1}{5-1} = \binom{5}{4} = 5,$$

and thus the total coefficient of $a^7 b^6$ in the first polynomial pair is $15 \cdot 5 = 75$. As before, the second polynomial pair contributes nothing, as according to (3.27) all the coefficients are 0.

The final polynomial, from the reflective component, requires us to make use of (3.28), as we have a product of generating functions with different powers. Considering first the series in $b$. What we want is

$$\left[ f(b) \cdot f(b^2)^2 \right]_6 = \sum_{i=0}^{6} [f(b)]_i \left[ f(b^2)^2 \right]_{6-i},$$

and after checking each term in the sum, we realise that the only nonzero term comes from the case where $i = 2$, and this gives us

$$\left[ f(b) \cdot f(b^2)^2 \right]_6 = \binom{2-1}{1-1} \binom{4/2-1}{2-1} = \binom{1}{0} \binom{1}{1} = 1 \tag{3.29}$$

after applying (3.27). Now to this we need to add the contribution from the series in $a$. Here we now have

$$\left[ f(a) \cdot f(a^2)^2 \right]_7 = \sum_{i=0}^{7} [f(a)]_i \left[ f(a^2)^2 \right]_{7-i},$$

and we need to work through each term in the series to find the coefficient. Here we have nonzero contributions from the terms where $i = 1$ and $i = 3$, corresponding to terms from $a \cdot a^6$ and $a^3 \cdot a^4$. So the overall contribution is

$$\left[ f(a) \cdot f(a^2)^2 \right]_7 = \binom{0}{0} \binom{2}{1} + \binom{2}{0} \binom{1}{1} = 2. \tag{3.30}$$

Now adding together all the contributions, we find that in the end we have

$$\frac{1}{10}(75) + \frac{1}{2}(1 + 2) = 9$$

total necklaces, just as before with the full expansion.

So this gives us an efficient way of calculating the number of possible distinct necklaces with a given number of white beads, a given number of black beads, and a set number of alternations from one colour to another. In order to produce the distribution then, all one needs to do is to select the numbers of black and white beads, and then step through every allowed value of $\alpha$ for those beads. Then normalising the number of possible necklaces for each $\alpha$ by dividing through by the total number of necklaces across all values of $\alpha$.

This was a fairly indirect route to finding what looked like being a relatively simple probability distribution. Especially since, as we will see in the next subsection, Monte Carlo simulations suggested that the distribution should be more or less normal. But this approach, for all its apparent abstract complexity, does illustrate the beauty and power of applying branches of pure mathematics to very practical, applied problems. Ultimately, we have successfully created an efficient probability distribution for DNA sequences, as desired.

## 3.5   Numerical results

Here we present some plots displaying properties of this distribution, as numbers of AT and GC base pairs are varied, as the ratios are varied, and so forth.

We begin by comparing the theoretical results to Monte Carlo simulations. This has a twofold purpose – firstly it serves as a check for the theory, that we have not got lost somewhere in the mathematics. Secondly, it gives an idea of how many random realisations are necessary in order to more or less cover the spectrum of possible cases. Figure 3.12 shows the comparison between the theoretical calculation and Monte Carlo simulations. We see that the results are very close, which is a useful validation of the theory. It is also worth noting that since the possible number of DNA sequences with $N_{AT} = N_{GC} = 50$ and $\alpha = 50$ is on the order of $10^{25}$, which makes it unsurprising that the Monte Carlo approximation is imperfect.

However, in order to check the convergence of the Monte Carlo method and understand how the number of simulations affects the accuracy, we can compute the total absolute difference between the true distribution and the approximate distribution. We define this difference $d$ as

$$d(N_{MC}) = \sum_{\alpha} |P_{MC}(N_{MC}, \alpha) - P(\alpha)|. \tag{3.31}$$

Here $P_{MC}(N_{MC}, \alpha)$ is the probability of finding a sequence with $\alpha$ alternations, using Monte Carlo simulations with $N_{MC}$ realisations. We see in Fig. 3.12(c) that as $N_{MC}$ is increased, we see the expected improvement in accuracy. There is however a significant slowing down in the improvement around $N_{MC} = 10000$, and after 20000 there is little change. This suggests that using only 10000 random runs, or even only 5000, will give generally good enough statistics to cover the spectrum of $\alpha$ values fairly well.

The other point that arises immediately after looking at Fig. 3.12(a) and (b) is that the shape of the distribution is more or less normal. This suggests that we should be able to fit the distribution with a Gaussian function, which would give us some useful insights into the properties of the distribution as the parameters $(N_{AT}, N_{GC})$ change. Figure 3.13

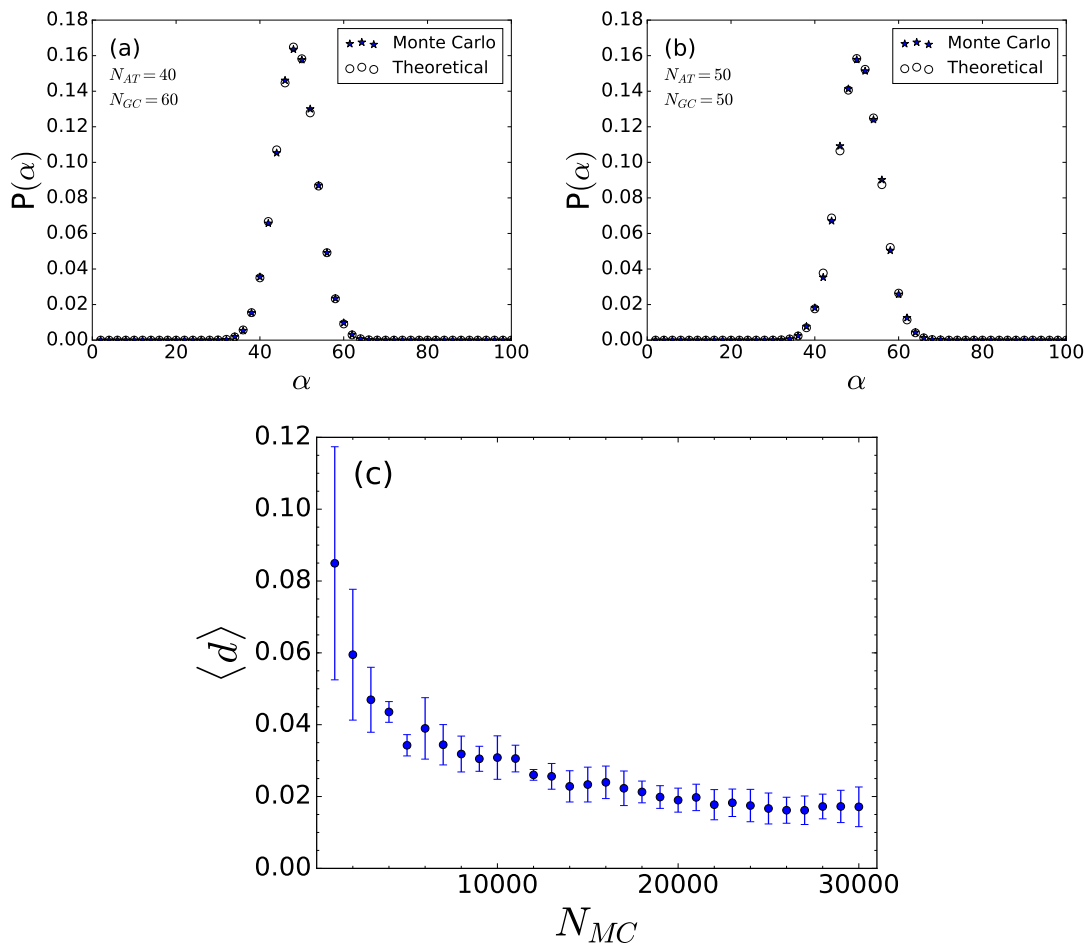

FIGURE 3.12: Comparison of the theoretically determined probability distribution (empty circles) and Monte Carlo generated results using 20 000 realisations (filled stars) for two cases. (a) $N_{AT} = 60$, $N_{GC} = 40$ and (b) $N_{AT} = N_{GC} = 50$. We see that the two methods agree fairly closely, as desired. The Monte Carlo results will never agree exactly, due to finite statistics, but these results do validate the theoretical method and show that with a finite number of Monte Carlo simulations a good approximation to the true distribution can be made. Panel (c) Shows the total absolute difference $d$ for increasing numbers of statistics. The points are averaged over five Monte Carlo runs with each number of realisations, and the errorbar shows the standard deviation from this averaging.

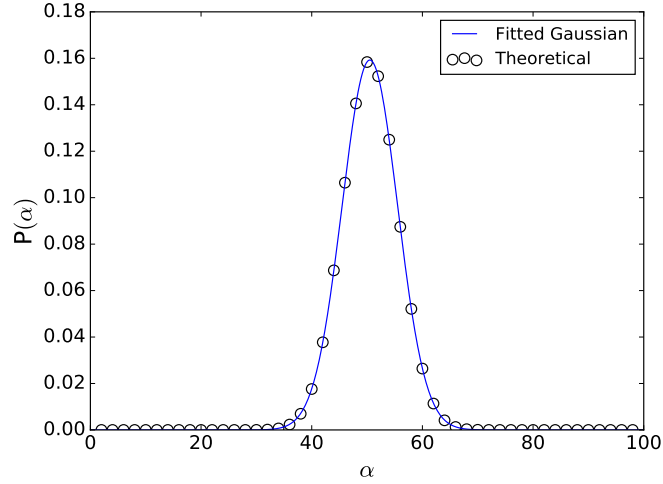

FIGURE 3.13: A fitting of the probability distribution in the $N_{AT} = N_{GC} = 50$ case with (3.32). We see that a normal, Gaussian distribution does fit well indeed. For this case, we find that the mean value is at $\alpha_0 = 50.5$ and the standard deviation $\sigma = 5.1$.

shows the fitting of the data with a Gaussian function of the form

$$P(\alpha) = \frac{1}{\sigma\sqrt{2\pi}} e^{-\left(\frac{\alpha - \alpha_0}{2\sigma}\right)^2}. \tag{3.32}$$

We can explore some features of the distribution by fixing the number of AT base pairs, and then varying the number of GC base pairs. Of course the same results would be found if the number of GC base pairs was held constant. We would expect that in cases where there are an overwhelming number of one type of base pair (say there were five AT base pairs scattered amongst two thousand GC base pairs) that the alternation index would almost always be maximal (in the example, ten). Similarly, it would be natural for the distribution to be more spread out when the number of base pairs are similar. Figure 3.14 shows some cases of these distributions. Our expectations are indeed borne out, as we see the extreme cases being substantially narrower than the more balanced cases. This behaviour motivates a slight further investigation; what is the general trend of the parameters of the Gaussian fitted function? We are expecting that as $N_{GC}$ is varied, at the extreme ends the standard deviation would be small and the maximum probability relatively large. The mean value should increase, until for very large $N_{GC}$ it approaches near saturation, and should be close to 200.

In Fig. 3.15 we see more or less the expected results. The mean value increases rapidly at first, but then exhibits a very slow asymptotic tendency to the maximum allowed value of $\alpha_0 = 200$. The rapid increase continues past the point of balanced base pairs, only slowing down dramatically in the region of $N_{GC} \approx 500$. The standard deviation shows the expected behaviour at the extremes; much smaller at either end of the spectrum, and largest in the region where $N_{AT} \approx N_{GC}$. There is a slowing decrease towards large $N_{GC}$, suggesting that it will only extremely slowly start to resemble a very sharp distribution. The maximal value follows a similar but inverted trend, giving high probabilities for the unbalanced cases, with a minimum where $N_{AT} = N_{GC}$.

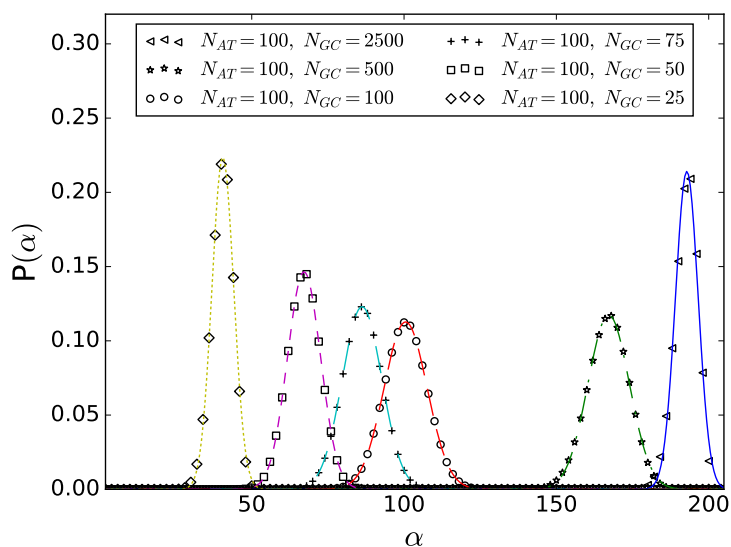

FIGURE 3.14: Probability distributions for DNA sequences with different numbers of GC base pairs, while $N_{AT} = 100$ is kept constant. As anticipated, the two extreme cases (very few and very many GC base pairs) correspond to the narrowest, tallest distributions, while the more balanced middle region exhibits broader, shorter distributions. Plotted points are the exact distribution, with the curves denoting Gaussian fits.

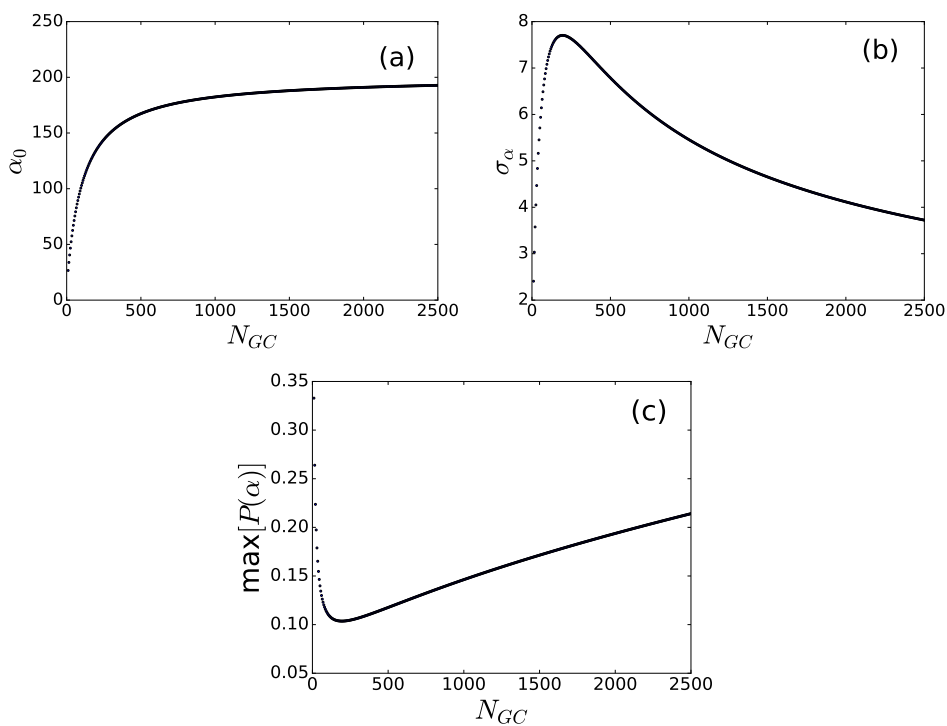

FIGURE 3.15: The variation with the number of GC base pairs $N_{GC}$ of (a) the mean value of the distribution, (b) the standard deviation of the distribution, and (c) the maximum probability of the distribution.

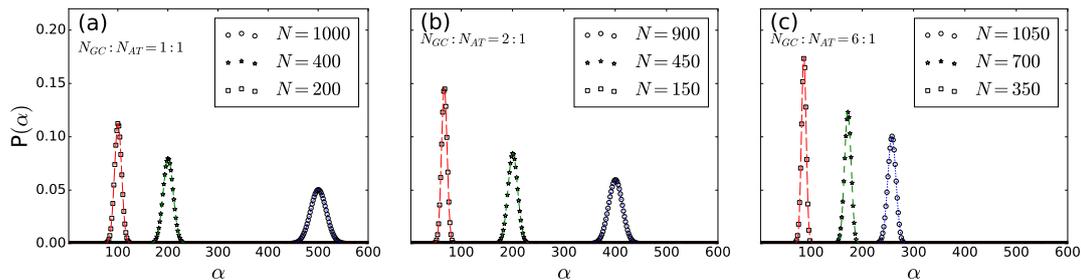

FIGURE 3.16: Probability distributions for three cases of total number of base pairs, for the ratio of $N_{GC} : N_{AT}$ equal to (a) 1 : 1, (b) 2 : 1, (c) 6 : 1. Plotted points are the exact distribution, with the curves denoting Gaussian fits.

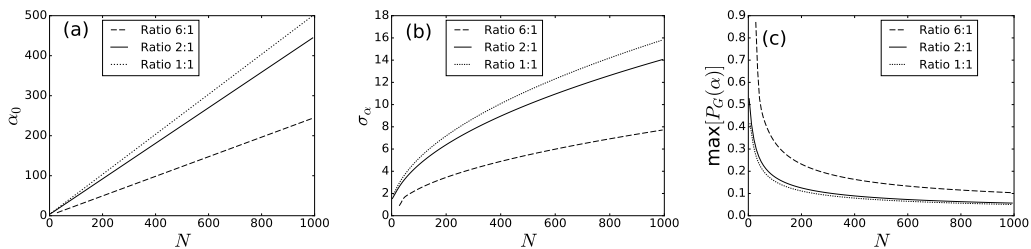

FIGURE 3.17: The variation with total number of base pairs $N = N_{AT} + N_{GC}$ of different parameters of the Gaussian fit (a) The mean value, (b) the standard deviation, and (c) the maximum probability.

The final property we investigate is keeping the ratio of $N_{GC} : N_{AT}$ constant, and increasing the total number of base pairs. Here the expectation is a little less obvious; the mean value should increase as $N = N_{AT} + N_{GC}$ increases, but the other parameters are less clear. We see some of the distributions plotted in Fig. 3.16. It is immediately clear that as we saw previously, the more unbalanced ratios give narrower distributions. The larger ratios also appear to show a slower increase in the mean value.

Once again the changes in parameters $\alpha_0, \sigma_\alpha$ and maximum probability have been plotted, in Fig. 3.17. Here we see that in fact the mean value of the fitted Gaussian increases linearly in all cases. The slope does however depend on the ratio; for the ratio of 1 : 1, the slope $m = 0.5$. For 2 : 1, $m = 0.45$, and for 6 : 1, $m = 0.25$. For the standard deviation, we see a nonlinear increase, more rapid for small $N$, but then slowing down for longer sequences. As before the maximum probability more or less displays an inverse behaviour, with a sharp decrease for small $N$ and then decreasing very slowly for larger values of $N$.

## 3.6 Conclusion

Using methods of Pólya counting theory, and extending Pólya's Enumeration Theorem to bipartite sets using adapted cycle indices along with generating functions as formal power series, we have constructed an extremely efficient algorithm for computing the probability distribution of the alternation index $\alpha$ as the numbers of AT and GC base pairs in a DNA sequence are varied. Monte Carlo simulations have validated our theoretical results, and show that around 5000 random realisations are sufficient to approximate the distribution of the alternation index. This distribution has various properties which can be studied, and significantly can be fitted with a Gaussian function which enables a large amount of investigation. Most importantly for the primary task of this thesis however, we now have a valid way of answering the question "what are the most and least likely values of $\alpha$ to occur in a random DNA sequence", and we are able to quantifiably measure the impact of the alternation index and hence heterogeneity on the dynamics of DNA molecules.

# Chapter 4

# Chaotic Dynamics of DNA Models

In this chapter, we will finally move on to some nonlinear dynamics, and study the effects of heterogeneity on the chaoticity of DNA molecules in the PBD model. The mLE is computed to quantify the chaoticity, and the variation of this mLE with AT content is studied across a spectrum of energy densities. The influence of heterogeneity is probed more deeply through studying the effect of the alternation index $\alpha$ that we have introduced in Chapter 3 on the chaoticity. Further, we study the behaviour of the mLE as the temperature increases past the melting temperature, comparing this regime to very low temperatures. This is followed by a discussion of the distribution of deviation vector (DVD), and what can be gleaned about the chaotic hotspots in the strands, especially focussing on biologically significant sequences. Finally, some similar results (mLE as well as DVD) are produced for the sequence-dependent ePBD model, and compared to the original model.

The work in this chapter is based around the results presented in [91].

## 4.1   Background

As has been outlined in Chapter 2, the study of DNA through mathematical models is a well-established endeavour. Here we now turn particular focus to the studies of the *nonlinear* dynamics of DNA. As noted earlier, this is a crucial aspect of the modelling DNA behaviour, responsible for reproducing most of the significant properties of DNA. These include the biologically important formation of long-lived bubbles at transcriptionally active sites [54] (more about this in Chapter 5). The nonlinearity of the model also introduces the localisation of energy, and particularly the formation of nonlinear localised modes [92]. The oscillations near the minimum of the Morse potential well lead to periodic breather solutions, related to the low-frequency vibrational modes present in DNA molecules. The existence of these localisations in the absence of disorder can in some sense be attributed to "dynamical disorder", due to the nonlinear potential. This nonlinearity causes certain sites to become self-trapping high amplitude, low frequency breathers, which are unable to disseminate their energy due to the dynamical restrictions on the phonon spectrum [92]. The presence of these nonlinear fluctuations impacts charge transport in DNA molecules where "vibrational hotspots", regions with a high propensity for bubbles or breathers to form (i.e. localised nonlinear modes), can cause anomalous subdiffusion of the charge [93, 94]. It is again interesting that the nonlinearity can fill the role typically associated with disorder, as in these studies of charge transport only homogeneous DNA polymers have thus far been considered [94]. Most critically of course, the nonlinearity leads to what we noted previously, the abrupt denaturation curve which is so important to meaningfully modelling the biodynamics of DNA [46].

Here we should also discuss the phase transition mentioned in Chapter 2. The existence of phase transitions in one dimensional systems with short-range interactions has been the topic of much discussion in the past. In a large number of cases, these transitions are forbidden due either to the findings of van Hove [95], or by the argumentation of Landau [96]. As such, the emergence of the DNA melting transition in the PBD model as an apparent first order phase transition is a slight surprise – even in the homogeneous cases studied early on [44, 45], the abrupt denaturation process demonstrated properties of a true thermodynamic transition. However, due to a number of factors, the arguments against phase transitions in one dimensional systems to not actually apply to the PBD model in question. In particular, van Hove's theory demands that the potentials all be functions of $(y_n - y_{n-1})$, so the presence of an on-site potential which is only a function of $y_n$ means that we cannot necessarily apply this theory [97, 98]. The second requirement for the absence of a phase transition in a one-dimensional system is that the energy of the domain walls between different regions in the lattice should be finite [96]. In the DNA case, since the energy of domain walls between regions of open and closed base pairs is infinite, the phase transition is not in fact forbidden by this theory [47, 97]. Thus we have a true phase transition [97], and further study of the dynamics of the PBD model of DNA to understand this phase transition has provided useful results [47].

## 4.1.1 Chaotic Dynamics of the PBD Model

Focusing in even more closely from general nonlinear dynamics to the chaotic behaviour of the PBD model, there has been fairly little research into identifying and quantifying the chaotic behaviour of the model. However, an initial investigation of the chaotic dynamics has been performed [99]. In that investigation, the chaoticity is quantified through the mLE, which is estimated using a combination of the transfer integral method and Riemannian geometry. Here the harmonic coupling is used to expedite the analytical calculations, but numerical results are presented for both the harmonic and anharmonic couplings. The result of that work was the finding that the mLE could serve as a dynamical order parameter, showing all the requisite characteristics. The mLE increases with energy density (or equivalently temperature), and demonstrates a change in behaviour near a critical value corresponding to the phase transition. This result is shown in Fig. 4.1, reproduced from [99]. The harmonic coupling was found to lead to a second-order phase transition, while the anharmonic coupling shows a more first-order-like transition. This additional thermodynamical significance associated with the mLE provides another point of importance to the study of the chaotic dynamics of DNA molecules [99].

It is with this in mind that the work presented here was undertaken. These early results indicate that there is value in a thorough understanding of the chaotic dynamics at play in DNA dynamics, and the work of Barré and Dauxois [99] with the homogeneous model suggests that there may be more depth to be gained by studying more complete versions of the model with the advantage of modern computational resources.

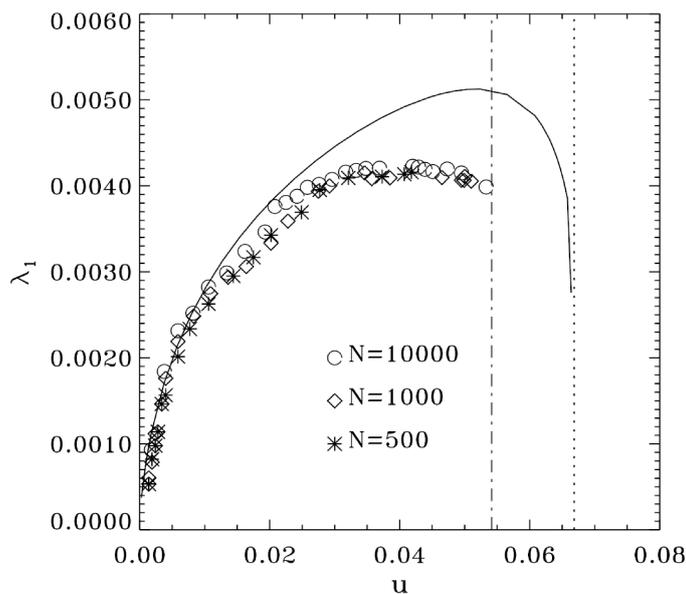

FIGURE 4.1: The computation of the mLE for the homogeneous PBD model of DNA, as a function of energy density. The solid line depicts the change in the mLE as calculated for the harmonic coupling. The symbols show the numerical computation of the mLE for the anharmonic coupling, with different numbers of base pairs. The stars correspond to 500 base pairs, the diamonds to 1000 base pairs, and the circles to 10000 base pairs. It is noteworthy that the number of base pairs apparently does not affect the value of the mLE at all. From [99].

## 4.2 Effect of AT/GC composition

Our aim now is to understand the overall effect of heterogeneity on the chaotic dynamics of DNA. The two key ingredients here are of course *heterogeneity* and *chaoticity*. We will explore some different ways of understanding the heterogeneity, but with regard to chaoticity the primary measure we will use is the mLE, as introduced in Section 1.3. The quantification of heterogeneity is slightly less clear, however. Firstly, let us note the origin of the heterogeneity (or, similarly, disorder) in the DNA molecule. While we treat each base pair as having identical masses, the different on-site constants in the PBD potential (2.4) result in disorder arising from the AT and GC base pairs. Thus we have two homogeneous cases (those of pure AT and pure GC sequences), and then everything else must be considered as a heterogeneous lattice. Since this composition in terms of AT and GC base pairs is the starting point of the heterogeneity, it is logical to begin our consideration of the chaotic dynamics from this point. As before, we denote this composition by using the percentage of the DNA sequence that is made up of AT base pairs, $P_{AT}$. So we now have a fairly clear first aim: To vary this $P_{AT}$, and calculate the mLE for various percentages. Since the interest in the mLE is partly based on its properties as an order parameter, it is also useful to compute the mLE across a spectrum of energy densities, allowing us to understand the behaviour of the mLE as both AT content and energy density varies.

### 4.2.1 Computing an average mLE

In order to build a statistically robust picture of the mLE's behaviour, we need to perform sufficiently many simulations and combine them effectively. The procedure here was to perform, at each energy density, and for each AT percentage, one hundred total simulations. These simulations are further split up into ten realisations (base pair configurations) which each then have ten runs with different initial conditions. This allows the separate investigation into whether the individual realisation give significantly differing mLE values, which led to the investigation of the alternation index, and the results presented in Section 4.3. For each of these simulations, the finite time mLE (ftmLE) $\chi(t)$ was computed and its evolution tracked. Figure 4.2 shows the evolution of one such run. In this plot we see that $\chi(t)$ displays the behaviour expected for a chaotic trajectory, initially decreasing before saturating to a positive value. This final value is the estimate for the mLE of this trajectory.

The ultimate article of interest in this is of course essentially only this final value of the ftmLE. It is however instructive to observe these individual curves, as the choice of the final integration time needs to be made with the complete saturation of the mLE in mind. In Fig. 4.3(a) we can see that after $10^4$ ps, there is still a noticeable deviation between the various curves. We would like, as far as possible, to eliminate any kind of finite time inaccuracy from our calculations, and be sure that the only source of spreading of values is actually inherent to the system. To this end, after looking at the curves and checking their variation and saturation, the final time of $10^5$ ps was chosen as the point at which the ftmLE has practically ceased to change in time, as here there are very few noticeable changes in the curves. Figure 4.3(b) shows the ftmLE averaged over multiple initial conditions and realisations $\langle \chi(t) \rangle$ (shown by the solid dark line), along with the individual curves that make up the average (shown by lighter grey lines). The error bars in the average ftMLE are simply the standard deviation of the 100 runs making up the average. This shows quite clearly that the final values of $\chi(t)$ are very similar in all cases, giving reason to believe that it is reasonable to assign a single mLE value to each point with given $E_n$ and $P_{AT}$.

In all cases, the initial conditions are given by the displacements $y_n = 0$ for all base pairs, starting in static equilibrium, and the initial momenta are chosen randomly from a normal distribution at all sites. Other initial momentum conditions yield the same results; for instance using a uniform distribution instead of a normal distribution, or even single site excitations. Single site excitations require a longer equilibration time (both thermally and with regard to the mLE), and can be problematic at very high energies, where this requires a single site to have an enormous amount of momentum. Periodic boundary conditions were imposed, and one hundred base pairs were used for these computations. We note that increasing the number of base pairs does not change the mLE values.

Finally, in order to produce a final measurement of the mLE at a given energy density, for a given AT percentage, we average the last points of the ftmLE for each run, giving an average value $\chi_1$. The standard deviation of this average gives the uncertainty, completing the measurement.

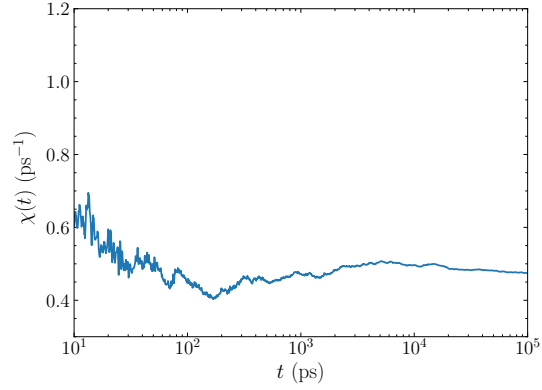

FIGURE 4.2: The time evolution of the finite time mLE $\chi(t)$ at $E_n = 0.0475\text{eV}$ with $P_{AT} = 30\%$, for a single initial condition.

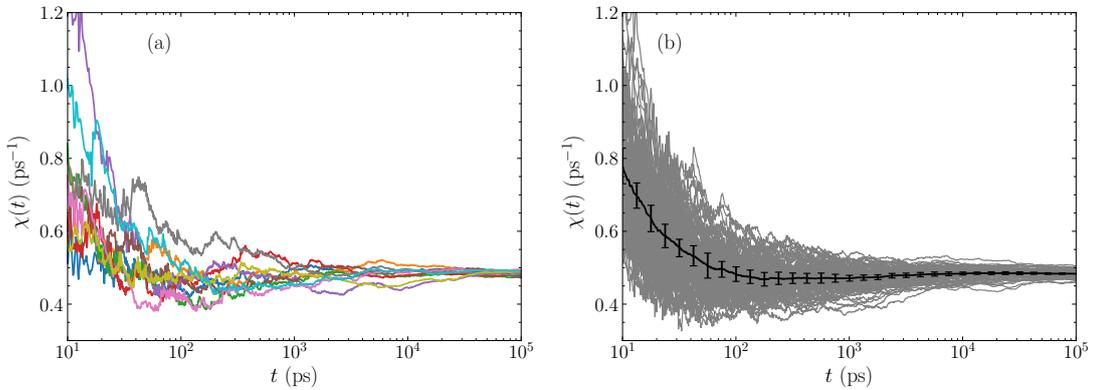

FIGURE 4.3: The time evolution of the finite time mLE at $E_n = 0.0475\text{eV}$ with $P_{AT} = 30\%$, for (a) ten initial conditions, and (b) ten initial conditions each for ten realisations (the light grey lines), averaged into one curve (the solid black line). We see that for every initial condition, and every realisation, the mLE is quite close together, resulting in an average value that represents this energy density and $P_{AT}$ fairly well. The error bars of the average value are the standard deviation of the 100 runs making up the averaged curve.

## 4.2.2 Average mLE Results

Having this machinery to estimate the mLE $\chi_1$ for any chosen parameters, we can now look at some of the results. The first thing to consider is just a simple computation of $\chi_1$ for a large number of values of $E_n$, for different AT percentages. Given the dynamical differences between the base pair types encoded into the model, we would expect different behaviours between percentages.

Figure 4.4 shows this result. We see that there is a different behaviour depending on $P_{AT}$, as well as an increase in the chaoticity with increased energy density. There are however two distinct regimes at play here: The first where the mLE increases with energy density, and then the second where the mLE flattens out before rapidly dropping to zero. We will split the discussion to analyse these sections. This is expedited by the fact that there is a physical justification for separating the regions – the first is the physically significant region before the melting transition, the complete separation of the DNA

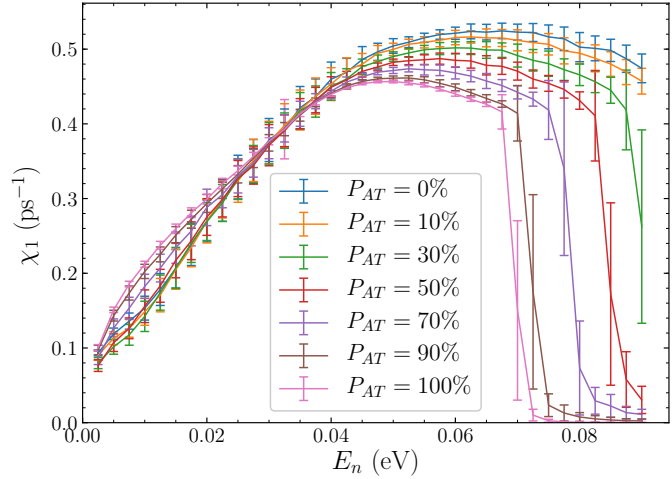

FIGURE 4.4: The average mLE $\chi_1$ for several AT percentages, across a spectrum of energy densities $E_n$. Line connections guide the eye only. The error bars correspond to the standard deviation in the average calculation of $\chi_1$.

double helix, and the second is what happens after this complete denaturation.

**Pre-Melting mLE Behaviour**

The first point here is to establish where in fact the melting transition takes place. The melting temperature (in Kelvin) of a general DNA molecule in the PBD model is given by [60]

$$T_m = 365 - 0.4 P_{AT}, \qquad (4.1)$$

following the experimentally verified linear variation, and a melting temperature of 365K for a pure GC sequence, and 325K for a pure AT sequence, again in agreement with experiments [60].

So now we know that we should cut the results at the melting temperature appropriate to each $P_{AT}$. But we do need to estimate the temperature of the system, since we are performing simulations in the microcanonical ensemble where energy rather than temperature is kept fixed. Thus we use the usual representation of temperature as the average kinetic energy in a one dimensional system scaled according to the Boltzmann constant

$$T = \frac{2\langle K_n \rangle}{k_B}, \qquad (4.2)$$

where the average kinetic energy $\langle K_n \rangle$ is calculated as

$$\langle K_n \rangle = \frac{1}{2mn} \sum_{j=1}^{n} p_j^2, \qquad (4.3)$$

with $n$ particles of mass $m$.

Thus for each simulation, we calculate the temperature, and then take an ensemble average for all (100) runs at that energy density and arrive an estimation of the average temperature corresponding to that energy density. With that, it is possible to identify the point where the temperature crosses the melting point, and then to stop the curve in the plot of the mLE before that point. The plot of $\chi_1$ as a function of energy density before the melting point is given in Fig. 4.5.

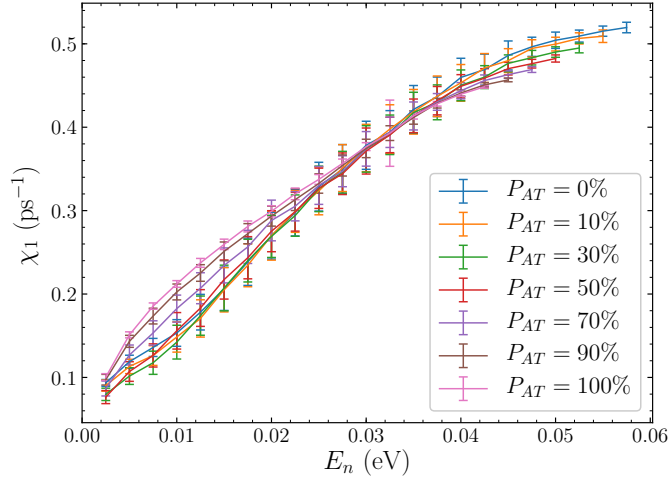

FIGURE 4.5: $\chi_1$ for each AT percentage, as a function of $E_n$, as in Fig. 4.4, but stopping when melting occurs.

Here we see that indeed the whole section of the plot where $\chi_1$ drops to zero, as well as some of the preceding decrease, is omitted. This now leaves us with the physically relevant region of interest. In this region there are apparently three distinct behaviours visible in the mLE. While there is still an overall trend of increasing chaoticity with energy density, we see that at lower energies, generally sequences with higher AT content are more chaotic. Below $E_n \approx 0.015\text{eV}$ we see some sort of crossing, where the strongly GC sequences actually become slightly more chaotic than the mixed sequences. Between $0.015 \lesssim E_n \lesssim 0.025\text{eV}$, there is a trend that higher $P_{AT}$ corresponds to a larger $\chi_1$ value. This is followed by a small region where the effect of the AT/GC composition is minimal; all cases more or less give similar values of $\chi_1$ between $0.025 \lesssim E_n \lesssim 0.035\text{eV}$ regardless of the $P_{AT}$ value. Thereafter comes a region where the rate of increase in $\chi_1$ slows down, as the curves flatten out towards the melting point. There is also a fairly clear trend here that the low energy behaviour has switched around, and now higher GC content corresponds to higher $\chi_1$.

The effect of the base pair composition on the chaoticity is consistent, with the more weakly bonded AT base pairs tending to exhibit more chaotic behaviour at low energies, and the GC base pairs taking longer to break apart and thus showing more chaoticity at higher energies.

Figure 4.6 shows the variation of the mLE as a function of the temperature, where the temperature is found from (4.2). The shape of the curves is slightly different, as the relationship between energy density and temperature is nonlinear (see Section 5.2.2). Due also to the fact that the AT percentages has an effect on the energy-temperature relationship, points in Fig. 4.6 at the same energy are not necessarily at the same temperature, especially as energy increases.

It is also good to note that the overall behaviour exhibited here is in agreement with the results of [99], showing the same shape of curves, but of course different values due to the heterogeneity and use of different parameters. In the work of Barré and Dauxois, only a homogeneous version of the PBD model was considered, thus omitting the different behaviours evident in the more general model. The other major difference comes from the use of different units, and particularly the resulting time units, which means that our results differ from those in Fig. 4.1 by two orders of magnitude.

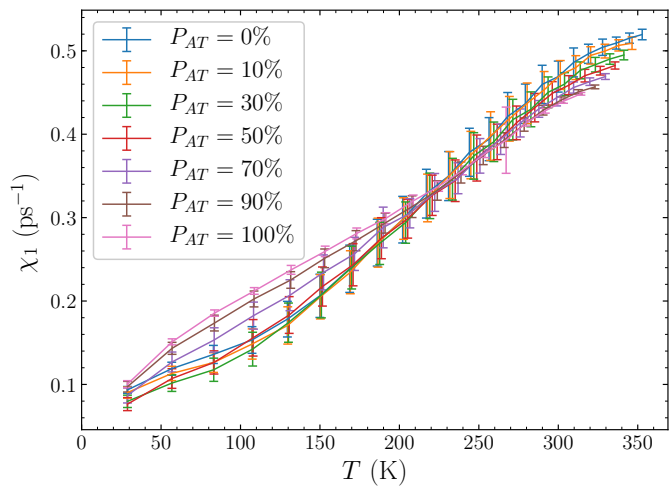

FIGURE 4.6: $\chi_1$ for each AT percentage, as a function of the temperature $T$, stopping at the melting temperature. The temperature is calculated according to (4.2) for each energy density.

**Post-Melting mLE Behaviour**

Some time after the melting point we see that the average mLE decreases suddenly to zero in Fig. 4.4. This drop to zero of course signifies a transition to a completely regular system. We see that there are also some intermediate points where the average mLE drops, but not all the way to zero. These can be explained by seeing that the system is at this point mixed – some initial conditions lead to chaotic trajectories, while others correspond to regular trajectories.

We do first have to note that this post-melting regime is physically irrelevant – the model makes no claim to accurately model the dynamics past the melting point. In fact, given that the two strands can completely separate and decouple, this situation is in some sense impossible to model. We can briefly consider the model from a purely dynamical perspective though, and look at what happens at this transition.

The regularisation of the motion can be explained from the potential. Recalling that the nonlinearity of both the coupling and the on-site Morse potential arises from exponential terms of the form $e^{-y_n}$ and $e^{-(y_n+y_{n-1})}$, we can see that as these displacements increase beyond a certain point, the exponential terms rapidly decrease to near zero, and the remaining terms are simply a constant and a harmonic coupling. This system is now completely linear, and we would expect completely regular behaviour. From this observation we can also understand the reason for the $\chi_1$ to increase more slowly at higher energy densities. As more of the base pairs remain in widely separated states (see Chapter 5 for a detailed discussion of the fraction of open base pairs), these base pairs are in some sense contributing linearly to the overall dynamics. So while the base pairs that remain dynamically bound are increasingly active, and thus more unstable and chaotic, the base pairs that reach the breaking point and separate form a stable dynamical behaviour in wide-open bubbles. They can of course fluctuate between open and closed, and different base pairs will contribute differently to the chaoticity at different times, but the average picture is more or less the same.

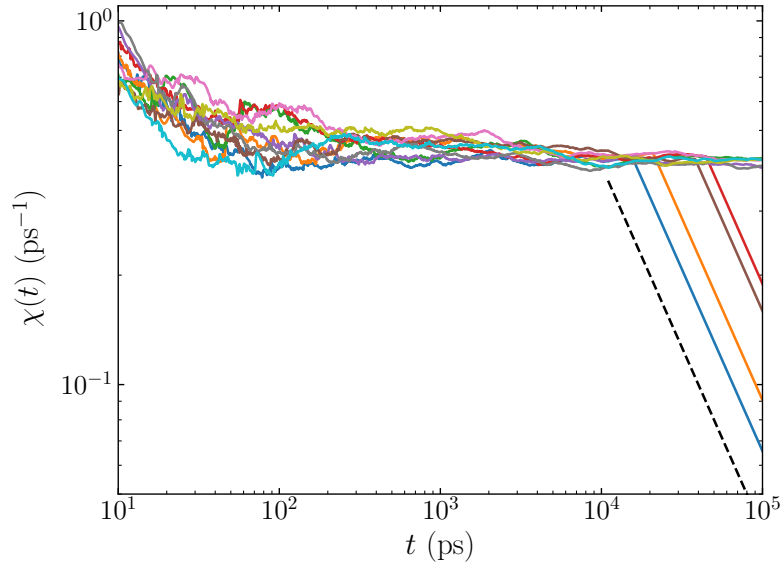

FIGURE 4.7: The evolution of ten finite time mLEs $\chi(t)$ for $P_{AT} = 70\%$ at $E_n = 0.0775$eV. Some initial conditions lead to positive mLE values, while others lead to a decrease towards 0 at the expected rate for regular trajectories.

Figure 4.7 shows the evolution of a few $\chi(t)$ values for different initial conditions at $E_n = 0.0775$eV for the $P_{AT} = 70\%$ case, which is above the melting point. Here we can see the hybrid behaviour of the system. Some of the ftmLE estimations remain positive, indicating that chaotic behaviour remains prevalent and the double strand is not completely denatured. Others however show the steady decrease towards zero, at a slope of $-1$ in the log-log scale, that implies regular behaviour. So here we have evidence of a mixed phase space; some initial conditions produce chaotic behaviour, and others regular. Increasing the energy density has introduced regions (however small) of stability to the phase space.

It is interesting to note that the system approaches linearity at both ends of the energy spectrum. While we see here that increasing the energy past the melting point results in a breakdown of the DNA structure and a linearisation of the potential, we can also look at the low temperature behaviour in Fig. 4.5. At very low energy densities we see a very small mLE, as the overall dynamics become less active, resulting in more stable behaviour. As the energy decreases to zero the mLE also decreases, but does not actually become 0 until the energy is 0, where there is no motion at all.

## 4.3   Heterogeneity: The Effect of Alternations

Having seen that the base pair makeup of the sequence is important for the dynamics of the system, we can now turn to a more detailed investigation of the heterogeneity.

One question that arises from the previous section is "where does the spread of mLE values come from?" Is it just a fact that slightly different arrangements of base pairs and slightly different initial conditions give different values for $\chi_1$? Is this variation inherent in any molecules, even very similar ones? Or, as we will investigate here, can we break down this spread of values into different classes of physically similar molecules, by considering the "clumping" of base pairs? To study this "clumping", we will make use of the alternation index $\alpha$ introduced in Chapter 3.

So now at last we can reap benefit of ploughing through the mathematics required to produce a probability distribution for $\alpha$. What we really want to know is how the heterogeneity affects the chaotic dynamics. For instance, if we had a DNA double strand consisting of alternating AT and GC base pairs, we might expect the dynamics to be rather distinct from that of a molecule with two solid chunks of AT and GC pairs. The alternation index allows us to quantify this heterogeneity, and so we can compare the different extreme cases of very well mixed, homogeneous sequences, and extremely chunky, heterogeneous, sequences. The probability distribution enables the identification of these extremes, as well as the most likely value of $\alpha$. In principle, one could produce a plot showing the variation of $\chi_1$ with $\alpha$ for all possible values of $\alpha$, but it is much more interesting to look at the extremes and see if there is any visible difference there before continuing with a series of long computations. So here results are presented for a collection of AT percentages, comparing the cases of

1. The most likely value of $\alpha$.

2. Extremely small values of $\alpha$ (more heterogeneous sequences).

3. Extremely large values of $\alpha$ (more homogeneous sequences).

Figure 4.8 show the value of $\chi_1$ for five different AT percentages. In each plot, the variation of $\chi_1$ with energy is shown for five values of $\alpha$. In order to cover the spectrum of possibilities we use the most likely value, the lowest possible value, the largest possible value, and two extra points near the extremes: the largest possible value less four, and the lowest possible value plus four. This means we will have a set of five cases, $\{\alpha_{min}, \alpha_{min} + 4, \tilde{\alpha}, \alpha_{max} - 4, \alpha_{max}\}$. These give an idea of the behaviour in the three regimes (unmixed, most probable, and well-mixed). We expect that the most probable case will reflect similar behaviours to the overall picture presented in Section 4.2, since randomly chosen sequences will of course be drawn more often than not from the centre of the distribution.

It is apparent from Figs. 4.8(a) and 4.8(e) that where there is a significant proportion of a single base pair type, the alternation index does not significantly affect the chaoticity. Thus we can see that if there are only ten AT (GC) base pairs in a sequence of ninety GC (AT) pairs, it does not make a difference to the chaoticity where the ten remaining base pairs are scattered. In the intermediate cases however [Figs. 4.8(b),4.8(c) and 4.8(d)], we do see some clear differences between the regimes. It is of course most notable in the 50% case, where the balanced number of base pairs lends extra significance to each of their position, we will first consider Fig. 4.8(c). Here the two well-mixed, high $\alpha$ cases (red and purple lines) clearly exhibit a smaller mLE than the other, more heterogeneous, cases. This is most pronounced at low energies, where below $E_n = 0.01\text{eV}$, the mLE of the $\alpha = 50$ case is more than double that of the $\alpha = 96$ and $\alpha = 100$ cases. This continues at higher energies, although less extremely. The green $\alpha = 50$ threads its way between the

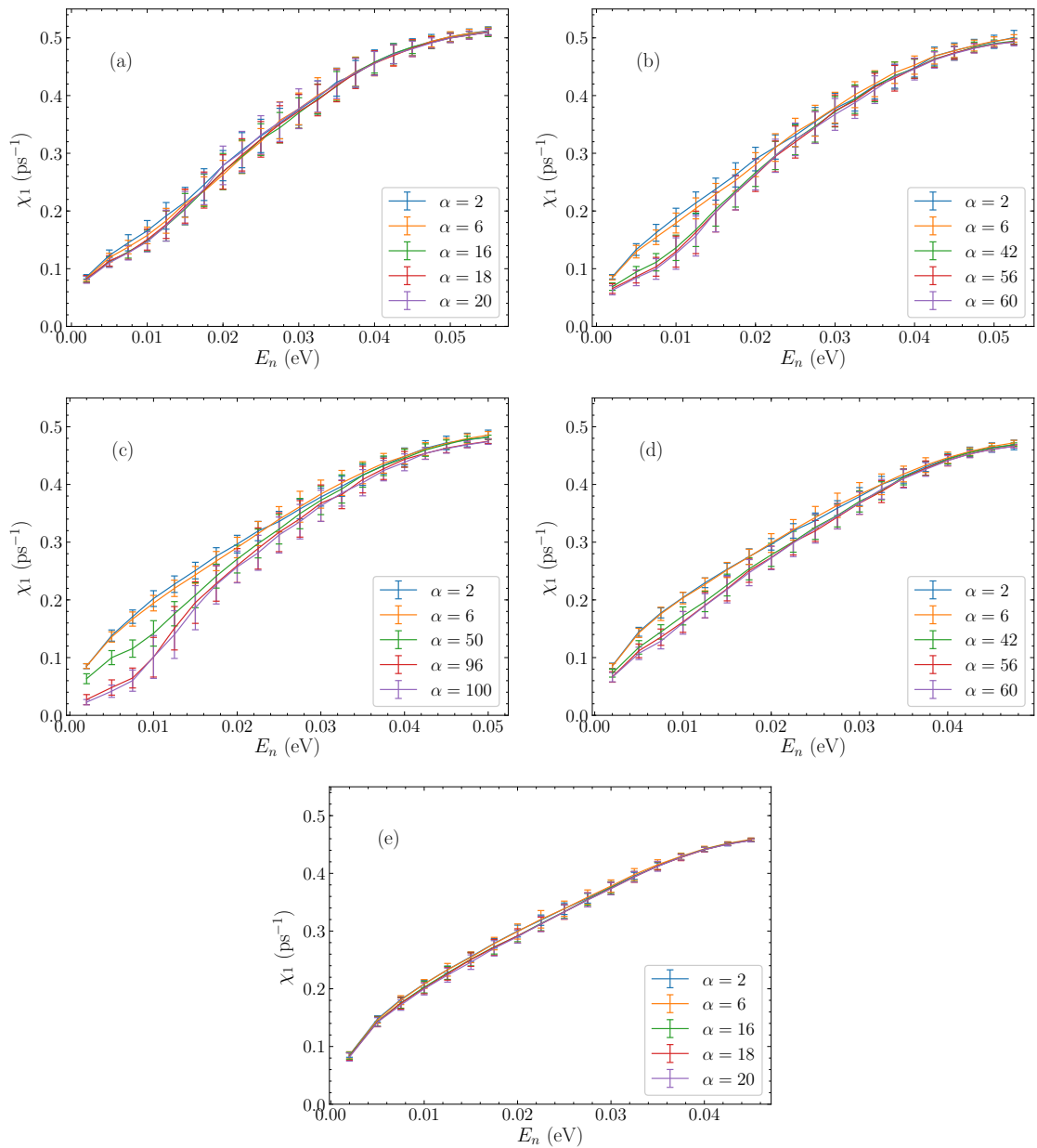

FIGURE 4.8: $\chi_1$ for five different values of $\alpha$, for five AT percentages. (a) $P_{AT} = 10\%$, (b) $P_{AT} = 30\%$, (c) $P_{AT} = 50\%$, (d) $P_{AT} = 70\%$, (e) $P_{AT} = 90\%$. In each case the $\alpha$ values are shown in the legend, corresponding to two extremely low values, two extremely high values, and the most probable value.

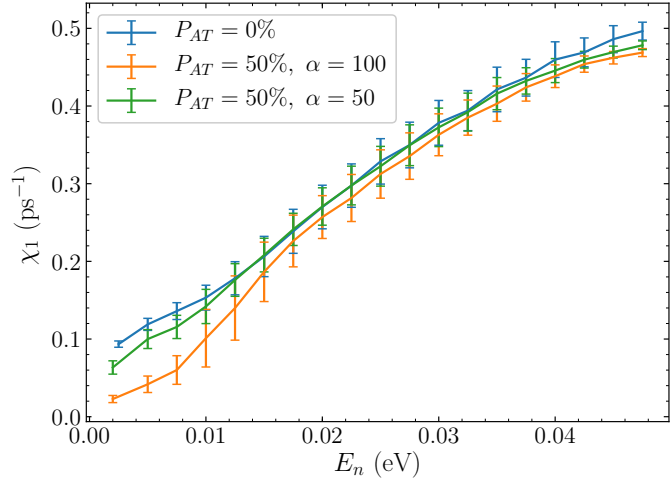

FIGURE 4.9: $\chi_1$ across the energy density spectrum, comparing the extremely
well-mixed 50% AT sequences with $\alpha = 100$ (orange) with the pure GC
sequence (blue). The $\alpha = 50$ curve for the 50% AT case is also shown
(green) to show the most common behaviour of this percentage.

high $\alpha$ lines and the low $\alpha$ lines, which are in turn substantially more chaotic. The blue
$\alpha = 2$ case and orange $\alpha = 6$ case show very similar characteristics, and are noticeably
more chaotic than the $\alpha = 50$ case, again with an emphasis on the lower energies. Here
the difference lasts all the way up to close to $E_n = 0.03$eV, before the lines all more or
less converge.

In general, we see that the more heterogeneous cases (small $\alpha$) lead to distinctly more
chaotic behaviour. Figs. 4.8(b) and 4.8(d) show for the $P_{AT} = 30\%$ and $P_{AT} = 70\%$ cases
respectively that the same trend holds true that the green $\alpha = 42$ case (the most probable)
stays between the two extremes, while the $\alpha = 2$ and $\alpha = 6$ cases remain more chaotic.
In these plots, the fact that the most likely value of $\alpha$ is relatively large is also apparent.
For instance, where in Fig. 4.8(c) the green line is, in the beginning at least, similarly far
from both extremes, in Figs. 4.8(b) and 4.8(d) the most probable case shows a definite
affinity for the higher $\alpha$ cases, with a much larger gap between the orange $\alpha = 6$ line
and the green $\alpha = 42$ line. This is most significant for lower energies, as we see a region
above $E_n \approx 0.035$eV where no matter what is changed, the mLE is more or less the same.

It is worth noting that even in this higher energy region, the trend remains that
the more heterogeneous cases are more chaotic. While the difference becomes small,
these green lines stay between the other cases throughout the energy spectrum. Moving
through this region of similar $\chi_1$, the lines do not really separate again, and we see more
or less the same behaviour for all values of $\alpha$ up to the melting point.

Overall, these results are fairly consistent – the more well-mixed a sequence is, the less
chaotic it is. This sits well with the intuition that a lattice with the disorder evenly spread
should be more stable than a lattice with extended regions of weaker on-site potentials,
where larger displacements are prone to occur.

What is also notable is to look at the particular behaviours of the extreme cases. Let
us focus in on the 50% AT case, as the effects of $\alpha$ on the mLE are strongest here. What
we would like to see is how the extreme cases of $\alpha$ compare to other AT percentages.
Figure 4.9 shows the most stable case for a sequence of 50% AT base pairs, $\alpha = 100$

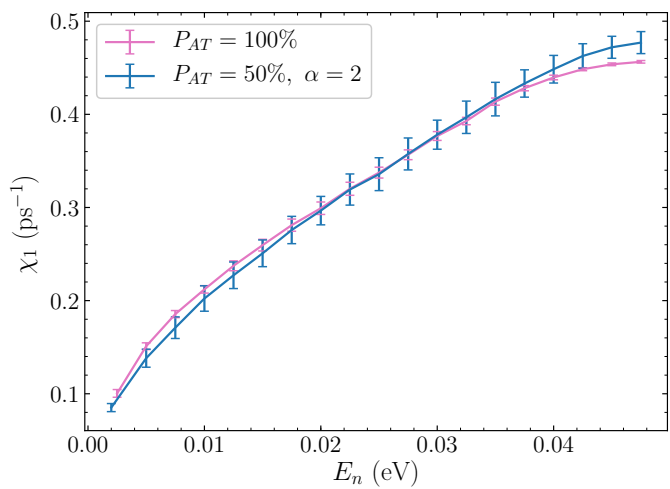

FIGURE 4.10: $\chi_1$ across the energy density spectrum, comparing the completely unmixed 50% AT case with $\alpha = 2$ (blue) with the pure AT case (pink).

(orange curve), compared to the pure GC sequences (blue curve). This emphasises how extremely stable the very homogeneous molecules are at low energies – here the $P_{AT} = 50\%$, $\alpha = 100$ curve is well below the pure GC curve up to large energy densities. We did see however, in Fig. 4.5 that at very low energy densities, the pure GC case is not the most stable. In that regime we should be comparing to the 50% case itself, where the most probable case is shown in Fig. 4.9 as the green curve. So we can combine this comparison with the result from Fig. 4.8(c) showing the large $\alpha$ cases to be noticeably more stable than the most probable case, and say that the most stable DNA configurations at low energy densities (or equivalently, low energies) are relatively homogeneous sequences with extremely well-mixed base pairs.

Additionally, we can study the opposite extreme – what happens when we have a complete separation of AT and GC base pairs ($\alpha = 2$)? In Fig. 4.10 we see the comparison of $\chi_1$ values between the $P_{AT} = 50\%$, $\alpha = 2$ case (blue curve) and the pure AT sequences, $P_{AT} = 100\%$ (pink curve), across the energy density spectrum. Here we see something quite interesting – the pure AT and unmixed 50% AT curves are actually very similar. The GC cluster still has some effect on the mLE, since the 50% curve does lie slightly below the pure AT curve, but overall the AT cluster is contributing most significantly to the chaoticity. As the energy increases, and the melting point of the pure AT sequence is being reached, the GC cluster reluctantly contributes to the dynamics, and we see the mLE increases past the value for the pure AT case, as the GC base pairs are excited into nonlinear behaviour.

This ties into the point that the length of the DNA sequence does not affect the molecule's chaoticity – so here the dynamics are approximately representative of a fifty base pair pure AT sequence, which is more or less the same as the one hundred base pair case. Intuitively this is justifiable, as we are saying that the chaotic dynamics are being dominated by the fifty AT base pairs, which are more easily excited than the GC cluster. So here we see that in just the single heterogeneous case of DNA sequences made up of half AT and half GC base pairs we can more or less see both the most and least stable behaviour exhibited by all DNA sequences, solely by varying the alternation index.

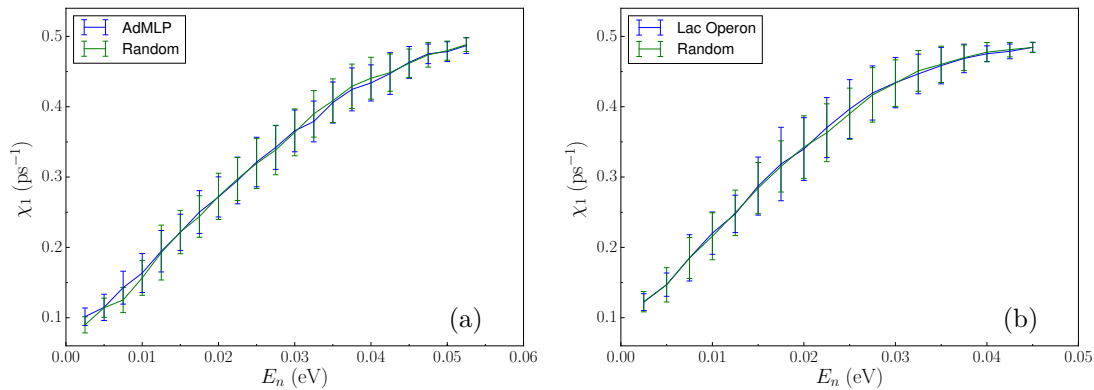

FIGURE 4.11: (a) $\chi_1$ values across the energy density spectrum for the AdMLP (blue), and random sequences with the same AT/GC composition, $P_{AT} = 33.7\%$. (b) $\chi_1$ values for the Lac Operon (blue) and random sequences with $P_{AT} = 51.9\%$ (green).

This confirms the early impression that the heterogeneous nature of DNA molecules has a strong influence on its dynamics and stability. Apart from the general effects of the AT/GC composition noted earlier in Section 4.2, we see here that the particular ordering of these base pairs is important in determining the chaoticity of the sequence.

The region in the energy/temperature spectrum where all AT/GC compositions yield the same mLE is noteworthy, especially as this trend survives into the more detailed study through the alternation index, with all cases there also showing this same behaviour. It is also very interesting to note the effect of the alternation index on the lower temperature region, where there is more scope for variation in the mLE, suggesting that the chaoticity of DNA sequences is more sensitive to the base pair ordering at low temperatures. By varying $\alpha$ in sequences of equal parts AT and GC content we can see more or less the entire spectrum of chaotic behaviour, from behaviour similar to pure AT sequences to behaviour showing properties of pure GC sequences. This does suggest that there is relatively significant variation possible in the chaotic dynamics due purely to the particular arrangement of base pairs in the molecule.

## 4.3.1 Biological Sequences

To complete the analysis of the effect of heterogeneity, and generally sequence-dependence, on the chaotic dynamics of the PBD model, we look at two biologically significant DNA sequences and compare their mLE values with those of random sequences with the same composition, i.e. exactly the same base pairs, with exactly the same alternation index, but otherwise arranged randomly within these constraints. The sequences chosen are an 86-base pair segment of the adenovirus-associated major late promoter (AdMLP), and a 129-base pair sequence from the *lac* operon, in both cases taken from around the transcription start site (TSS), the site along the strand where transcription begins [1]. These sequences are well-known transcriptional promoters, and have been studied in the context of bubble dynamics and DNA transcription activity [50, 100]. The precise sequence of the upper strand of each promoter is given below.

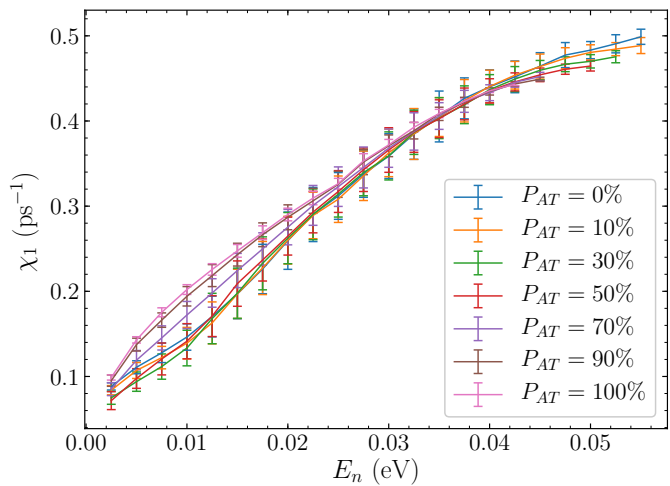

FIGURE 4.12: The average mLE $\langle \chi_1 \rangle$ as a function of energy density, for the ePBD model. Results are shown up to the melting point. This figure is the corresponding result to Fig. 4.5, but now for the ePBD model.

AdMLP:

**GCCACGTGACCAGGGGTCCCCGCCGGGGGGGGTATAAAAGGGGG
CGGACCTCTGTTCGTCCTCACTGTCTTCCGGATCGCTGTCCAG**

*lac* operon:

**GAAAGCGGGCAGTGAGCGCAACGCAATTAATGTGAGTTAGCTC
ACTCATTAGGCACCCCAGGCTTTACACTTTATGCTTCCGGCTC
GTATGTTGTGTGGAATTGTGAGCGGATAACAATTTCACACAGG**

We note that the AT percentage of the AdMLP is $P_{AT} = 33.7\%$, and of the *lac* operon is $P_{AT} = 51.9\%$. For each of these sequences, one hundred initial conditions were integrated, and for the random sequences the same procedure as before was followed, ten realisations with ten initial conditions each. In this case, the same number of base pairs was used in the random sequences as in the corresponding promoter.

The results of the mLE computations are shown in Figs. 4.11(a) and (b). We see that there is no significant difference in the $\chi_1$ values between the promoters and random sequences. All the variations that are present are within uncertainty of each other, and for the most part the lines are more or less identical. This suggests that there is no special chaotic behaviour inherent in the biological sequences. The chaotic dynamics are controlled by the heterogeneity, and particularly the alternation index, and while there is still some variation of the chaoticity of the sequences (as evidenced by the error bars), the biological significance of a sequence does not impact the value of the mLE.

## 4.4 Comparison with the Extended PBD Model

We will present a very short comparison between the results of mLE computations for the PBD model and the ePBD model. As we will see, the outcomes are very much the same for both models, which limits the usefulness of an exhaustive comparison between them.

Figure 4.12 shows the same results we saw in Fig. 4.5, but recomputed for the extended model. The overall behaviour is identical; we have the same three regions of distinct behaviour, and the same increase with energy density.

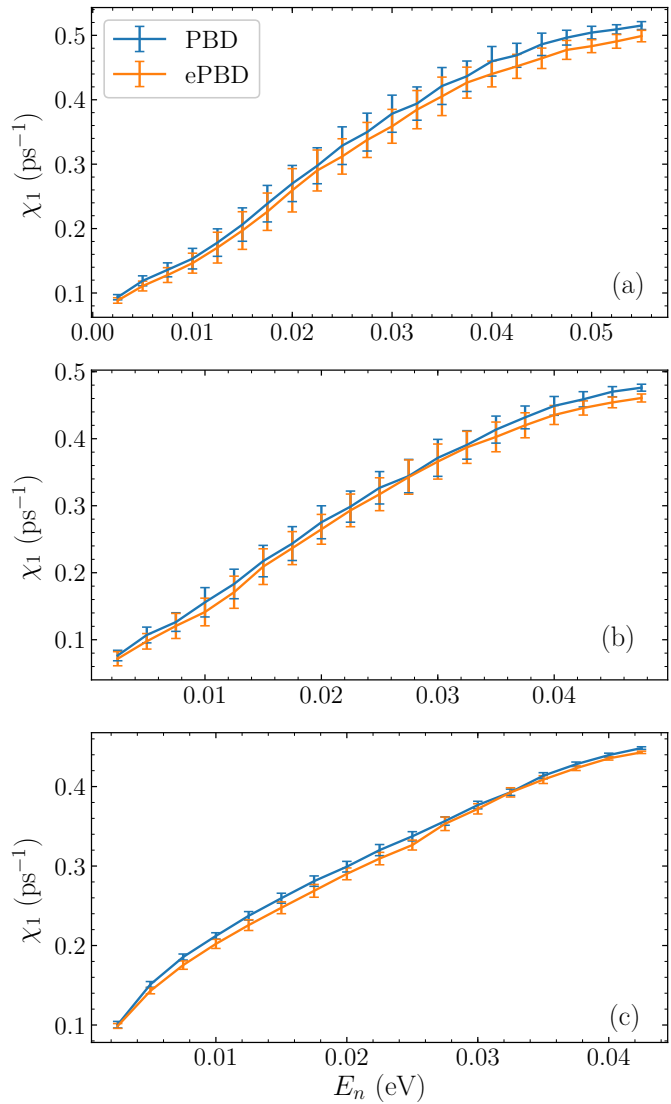

FIGURE 4.13: $\chi_1$ as a function of energy density $E_n$ for the PBD (blue) and ePBD (orange) models, with (a) $P_{AT} = 0\%$, (b) $P_{AT} = 50\%$, and (c) $P_{AT} = 100\%$. In all cases the values of $\chi_1$ for the two models are within uncertainty of each other.

Further comparison is easier to do on a case by case basis, in order to investigate whether or not there is any systematic difference. Figure 4.13(a) shows the estimation of $\chi_1$ for the PBD and ePBD models, across the energy spectrum for $P_{AT} = 100\%$. Here we see that the behaviours are very similar, but the ePBD model is slightly less chaotic. This suggests that despite the weaker coupling between base pairs due to the overall smaller values of the coupling constant $k$, the ePBD model predicts very slightly more stable behaviour than the PBD model. Since the two results are always within uncertainty of each other however, we cannot make any strong conclusions from this, with the general takeaway still being that the two models give more or less identical results.

The heterogeneous cases all exhibit similar behaviours; the $P_{AT} = 50\%$ result is shown as a representative case. Figure 4.13(b) shows that once again the ePBD model gives a marginally less chaotic behaviour than the PBD model, but the lines once again follow the same pattern, and remain within uncertainty of each other.

The pure GC case is shown in Fig. 4.13(c), once again displaying the same tendencies. Here however the difference, particularly at larger energies, is slightly greater than in the pure AT case, suggesting that the sequence-specific coupling constant has a more significant variation from the PBD value. This is the only case where some points are actually different even up to uncertainty; thus we could say that near the melting transition, the ePBD model gives a lower estimation of the chaoticity than the PBD model, for the pure GC case.

Overall, while the impression is clear that the exact nature of the sequence-dependence does not have a strong effect on the chaoticity of the DNA molecule, it is also apparent from these results that the ePBD model yields slightly less chaotic behaviour than the PBD model, in all cases showing smaller mLE values.

## 4.5   Deviation Vector Distributions

The final element we will mention in this study of the chaotic dynamics of DNA are the deviation vector distributions (DVDs). Here we want to investigate any link between the physically observable dynamics – displacements, and particularly the formation of bubbles – and the presence of strong nonlinearity. For instance, would regions of large displacement correspond to regions of significant nonlinearity? Or would they perhaps be devoid of chaotic behaviour, since the distant base pairs are only interacting relatively linearly? So our investigation of the DVD follows along these lines: Observing the displacements and the DVD in different contexts, investigating how different energies affect the dynamics, and how base pair sequences affect the dynamics.

### 4.5.1   DVD and Displacement at Low Energies

To investigate the general behaviour of the DVD in the low-energy regime, it is instructive to consider both individual cases, and averages over multiple runs. Figure 4.14 shows the time evolution of the DVD spectrum along with the displacement $y_n$, for a particular case with $P_{AT} = 30\%$ at $E_n = 0.005$eV. The displacement shows the natural behaviour – larger displacements between AT base pairs, especially in extended regions of AT base pairs. Note that the bar to the right of the contour plot shows the distribution of the AT and GC base pairs – white denoting AT base pairs, and black GC. So for instance, in

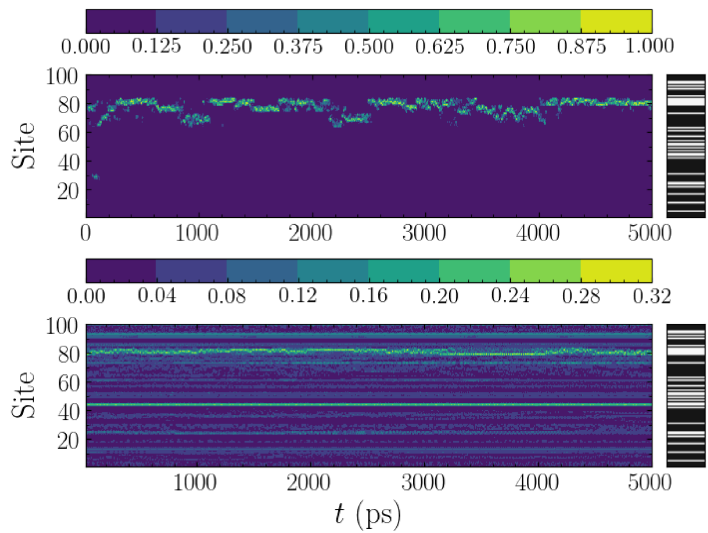

FIGURE 4.14: A single case of the evolution of the DVD (top) and the displacement (bottom), with $P_{AT} = 30\%$, at $E_n = 0.005$eV. The lighter areas correspond to higher values (as per the colour bar). The black and white bar on the right shows the AT (white) and GC (black) base pair distribution.

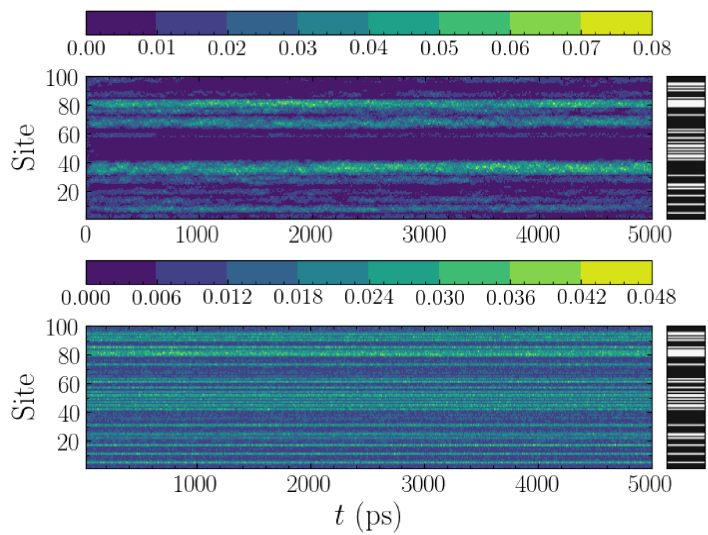

FIGURE 4.15: The average over one hundred initial conditions of the evolution of the DVD (top) and the displacement (bottom) for the same disorder realisation with $P_{AT} = 30\%$, $E_n = 0.005$eV. The light green areas correspond to higher values (as per the colour bar on top of each panel). The black and white bar on the right shows the AT (white) and GC (black) base pair distribution of the particular disorder realisation. A single case of these runs is shown in Fig. 4.14

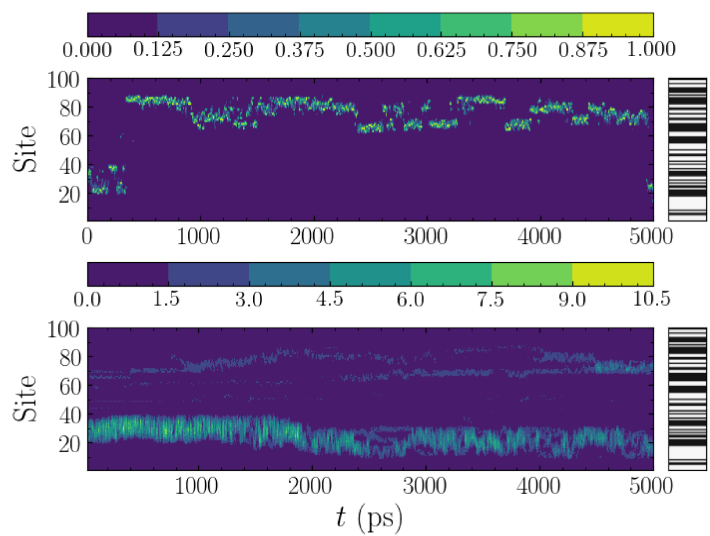

FIGURE 4.16: A single case of the evolution of the DVD (top) and the displacement (bottom), with $P_{AT} = 50\%$, at $E_n = 0.05$eV. The light green areas correspond to higher values (as per the colour bar above each panel). The black and white bar on the right shows the AT (white) and GC (black) base pair distribution for the particular disorder realisation.

Fig. 4.14 we see that there is a homogeneous island of AT base pairs near site 80. It is also in this region that the DVD generally concentrates, for this particular initial condition. The DVD however does not show the same clear tendency to stick to AT base pairs, despite the impression given by the displacement that the dynamics are mostly occurring in these regions. It jumps around, apparently randomly, but sticking to regions of homogeneity, regions occupied by either consecutive AT or GC base pairs. Here the simulations are performed up to a time of 5000ps, but in all the results presented there is no significant change if the system is allowd to evolve further.

To see if this is a general trend, we can perform many simulations with the same realisation but different initial conditions, and see if the DVD does in fact concentrate in the homogeneous islands. This result is shown in Fig. 4.15, for the same realisation as Fig. 4.14. We see that the DVD quite clearly focusses in these islands, regardless of the base pair type. Not only does it concentrate in these regions, but it very specifically stops at the edge of the islands, with an abrupt cutoff. This is particularly abrupt in the AT island near site 80, where there is almost no bleeding into neighbouring sites, even though it is bounded above by only a single GC base pair. That base pair alone is sufficient to inhibit the nonlinearity in that region. There is a slightly less sharp cutoff in the GC islands, but there is nonetheless an extremely clear tendency for the DVD to stick to these regions.

This is the general trend for the DVD at low energies/temperatures – what small amount of nonlinearity there is, is concentrated in the homogeneous islands, clusters of AT or GC base pairs.

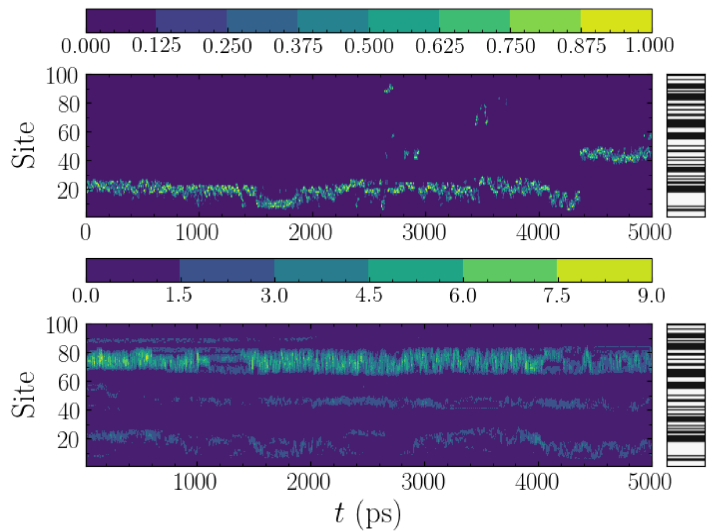

FIGURE 4.17: Another single case of the evolution of the DVD (top) and the displacement (bottom), with $P_{AT} = 50\%$, at $E_n = 0.05$eV, and the same realisation as Fig. 4.16. The light green areas correspond to higher values (as per the colour bar above each panel). The black and white bar on the right shows the AT (white) and GC (black) base pair distribution for the particular disorder realisation.

## 4.5.2 DVD and Displacement at Higher Energies

The behaviour described in Section 4.5.1 does not carry over to higher energies however. Considering a relatively high temperature system, we see that the DVD tends to follow regions of moderate displacement, but generally avoiding consistent, large bubbles. Figure 4.16 shows the DVD and displacement for a single case with $P_{AT} = 50\%$ at $E_n = 0.5$eV, corresponding to approximately $T = 330$K. This illustrates the described behaviour; while there is a clear region of significant, consistent opening near sites 10-40, the DVD follows the smaller displacements between sites 60-90. It is certainly not clear whether there is a particular pattern to this movement of the DVD, or what criteria (if any) there are for its choice of which displaced regions to follow.

To demonstrate this point, consider Fig. 4.17. Here we have the same base pair realisation, but a different initial condition for the momenta and deviation vector. Now we get a similar behaviour from the DVD, following the less consistent region of displacement, despite this being precisely the region it avoided in Fig. 4.16. Again it avoids the region of particularly high displacement, which is now in the region of sites 60-80. So it appears that the position of the DVD at high temperatures is not somehow determined by certain regions or sequences of base pairs, as seems to be the case at low temperatures, but it rather depends on the particular initial condition, and the regions of the DNA strand that are consistently or inconsistently displaced. Further, the jumping around of the DVD where it leaves a certain region abruptly, does not have a clear apparent reason. For instance in Fig. 4.17, we see the DVD jumping from around site 10 to site 40 shortly after 4000ps, while there is no discernible significance to the pattern of displacements.

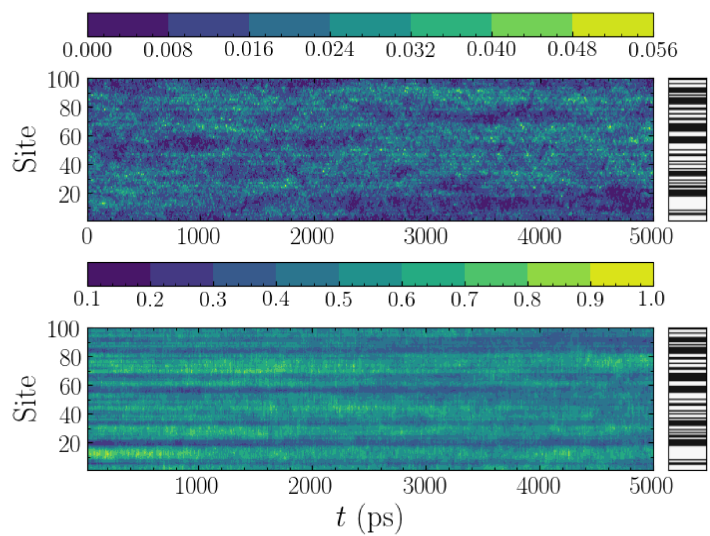

FIGURE 4.18: The average over one hundred initial conditions of the evolution of the DVD (top) and the displacement (bottom) for $P_{AT} = 50\%$, $E_n = 0.05$eV. The light green areas correspond to higher values (as per the colour bar above each panel). The black and white bar on the right shows the AT (white) and GC (black) base pair distribution for the particular disorder realisation. The realisation here is the same as shown in Figs. 4.16 and 4.17.

We confirm this impression by looking at the averaged DVD over 100 initial conditions (seen along with the displacements in Fig. 4.18). This average DVD does not show any clearly consistent patterns. The displacements once again form the expected pattern of larger displacements at AT sites, with more spreading and variation than at the very confining low temperatures. Apart from this slight trend, the DVD on average appears to be fairly uniformly distributed, not showing any strong preference for particular regions, although there is a slight tendency to avoid the region around sites 10-20, where there is also a corresponding tendency to have large, steady openings in the strand.

There is undeniably more scope for investigation of this behaviour, and particularly to investigate the significance of the jumps of the DVD, where it moves from one region to another. It remains possible that a link can be formed between these jumps and the formation of bubbles in the sequence. However, we now turn briefly to the consideration of known biological promoters.

### 4.5.3 DVD of Promoter Sequences

In terms of investigating properties of DNA strands, especially with an interest in the formation of bubbles, the natural avenue of study is to look at known biological sequences, and to see what links, if any, can be made between the biological significance and the dynamical observations. To this end, the two promoters whose mLE we investigated earlier in the Chapter (see Section 4.3.1), the AdMLP and the *lac* operon, are also investigated here. Since the primary region of physical interest is near physiological temperatures (around 310K), this is the temperature that we look at. Furthermore, since the aim here is specifically to understand the effect of the particular base pair sequence on the DVD, we consider only averages over many initial conditions, so as not to be overly distracted by the effect of a particular initial condition on the localisation of the DVD.

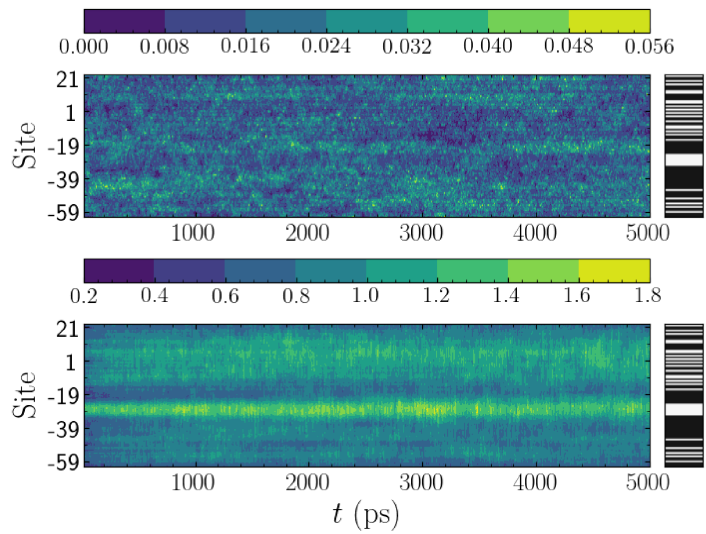

FIGURE 4.19: The average over one hundred initial conditions of the evolution of the DVD (top) and the displacement (bottom) for the AdMLP at $E_n = 0.044\text{eV}$. The light green areas correspond to higher values (as per the colour bar above each panel). The black and white bar on the right shows the AT (white) and GC (black) base pair distribution for the particular disorder realisation. The sites are labelled relative to the transcription start site (TSS), which is labelled as +1.

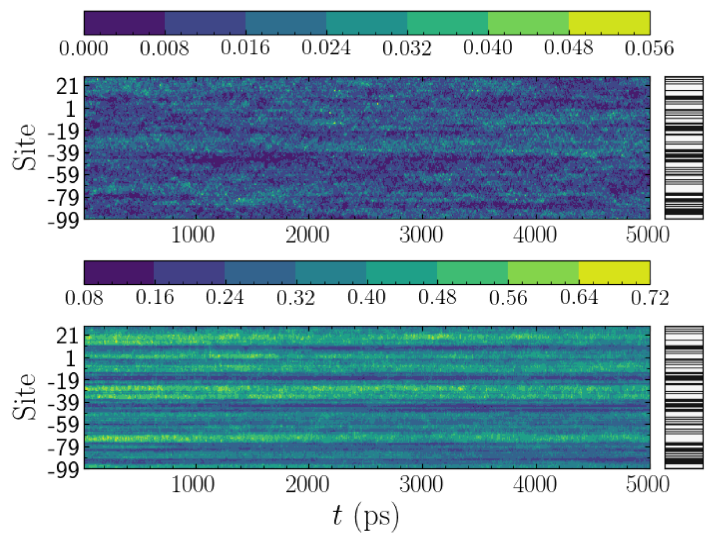

FIGURE 4.20: The average over one hundred initial conditions of the evolution of the DVD (top) and the displacement (bottom) for the *lac* operon at $E_n = 0.044\text{eV}$. The light green areas correspond to higher values (as per the colour bar above each panel). The black and white bar on the right shows the AT (white) and GC (black) base pair distribution for the particular disorder realisation. The sites are labelled relative to the transcription start site (TSS), which is labelled as +1.

Figure 4.19 shows the average DVD (upper panel) and displacement (lower panel) for the AdMLP, over a period of time. In general, this shows the characteristics found in Section 4.5.2, as these results are in the same energy regime. However, the dynamical behaviour of the AdMLP has been studied previously, and in particular the regions of high and low bubble probability [101], which gives us additional information that allows us to place the DVD results in context. Apart from the clear large displacements in the AT base pairs, particularly the TATA box around base pair −29 (a sequence of base pairs indicating the position of the TSS and the transcription direction [1]), we see that there is a region of large displacements centred around the TSS. Here the base pairs are labelled according to the biological convention, where the base pairs are numbered relative to the TSS, which is conventionally given the label 1. So for instance in the AdMLP, the $64^{th}$ site is the TSS, and so this is marked with 1. The downstream sites are labelled increasingly in the negative direction, relative to this site, so site 0 would be labelled −63. Upstream sites are labelled increasing in the positive direction so that site 84 is labelled with 21. The DVD generally avoids the TATA box, where we have consistent large displacements, which is in agreement with our earlier discussion of the DVD at higher temperatures. In addition to this however, we see that the DVD also avoids the region downstream of the TSS, which is known to be a low-probability region for large bubbles [101]. This again leads us to the conclusion that the DVD does not localise near long-lived, large, consistent bubbles, or in regions of little activity, but rather preferring the less predictable fluctuations in the DNA strand.

The same results for the *lac* operon, shown in Fig. 4.20, are still instructive. Once again, we know the bubble probability profile for this sequence, allowing us to compare this with the DVD [100]. Here we observe that the main areas avoided by the DVD appear to be around the TSS, and near sites −59 and below −39. These once again correspond to fairly high bubble probabilities (in the TSS case) and relatively low bubble probabilities (the downstream segment). Broadly these results support the understanding that the DVD avoids the extreme cases, with the DVD not concentrating at regions of large amplitude bubbles, or at regions of very small displacements.

It is worth noting at this point that the interpretation of the DVD is that the nonlinearity of the system is strongest at sites where the DVD is concentrated [23, 34]. This means that the DVD results here imply that the system is least chaotic in regions where the openings are systematically large, or where they are systematically small. Intuitively this matches with our previous analysis of the two near-linear borders of the system – the case of large displacements, where the potential effectively linearises, and the case of very little movement at low temperatures. The regions of large long-lived bubbles will exhibit approximately linear behaviour, and the regions of small displacements will be relatively less chaotic than other sites where the nonlinearity is being more strongly drawn out by the larger displacements.

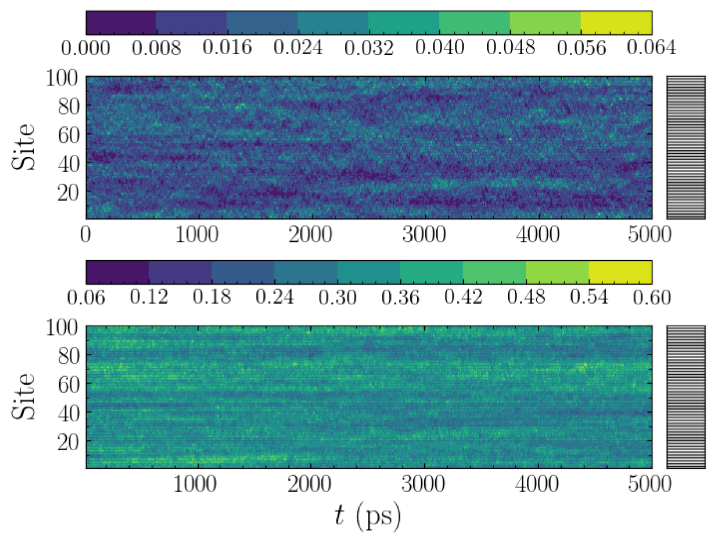

FIGURE 4.21: The average over one hundred initial conditions of the evolution of the DVD (top) and the displacement (bottom) for $P_{AT} = 50\%$, $E_n = 0.042$eV, with alternating base pairs. The light green areas correspond to higher values (as per the colour bar above each panel). The black and white bar on the right shows the AT (white) and GC (black) base pair distribution for the particular disorder realisation.

## 4.5.4   Effect of Base Pair Sequence on the DVD

Here we will digress from the theme of biological significance, and briefly consider some sequences that we would expect to be interesting from a purely dynamical standpoint. The aim is to gain a better understanding of how the DVD behaves in this system, and whether or not the intuition that more dynamic regions (i.e. more frequent openings, and thus AT-dominant regions) exhibit a higher propensity to host the deviation vector. To this end we start our investigation with precisely this point.

First, let us look at a sequence of 50 AT base pairs and 50 GC base pairs, completely mixed. That is to say $\alpha = 100$, with every base pair neighboured by two opposite type pairs. Here we would expect that since the system has no distinguishing features, on average the dynamics should be uniformly spread around the lattice. And indeed, as Fig. 4.21 shows, when averaging over many initial conditions the displacement and DVD both spread evenly through the lattice. There are nevertheless some notable points – when looking at the figures, one observes a striping in both displacement and DVD. This is expected in the displacements, where the AT base pairs should exhibit larger displacements. It is less clear what to expect from the DVD however, since we know that it requires nearby excitations, but prefers not to be localised at the site of the greatest excitation. The stripes turn out to favour the GC base pairs – around 55% of the DVD is present at sites with GC base pairs. So it seems that on average, while the DVD does spread everywhere corresponding to the universal displacements, it tends to localise slightly more at the GC sites, next to the highest displacements occurring in the AT base pairs.

If we now consider a sequence of alternating AT and GC base pairs, but replace the centre with a 10-base pair island of AT base pairs (Fig. 4.22), and GC base pairs (Fig. 4.23) respectively, we can find if there is something more to be seen about the effect of AT/GC arrangement on the DVD. In order to make the comparison as fair as possible, the exact same initial conditions were used in all three of these cases shown here (Figs.4.21, 4.22 and 4.23), but with the different base pair realisation.

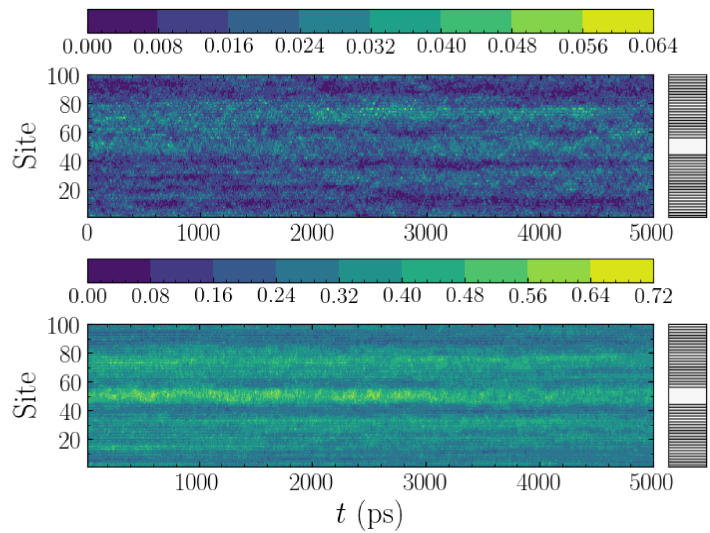

FIGURE 4.22: The average over one hundred initial conditions of the evolution of the DVD (top) and the displacement (bottom) for $P_{AT} = 55\%$, $E_n = 0.042$eV, with a specific arrangement of the base pairs. The light green areas correspond to higher values (as per the colour bar above each panel). The black and white bar on the right shows the AT (white) and GC (black) base pair distribution for the particular disorder realisation. The DVD is largely present everywhere in this case.

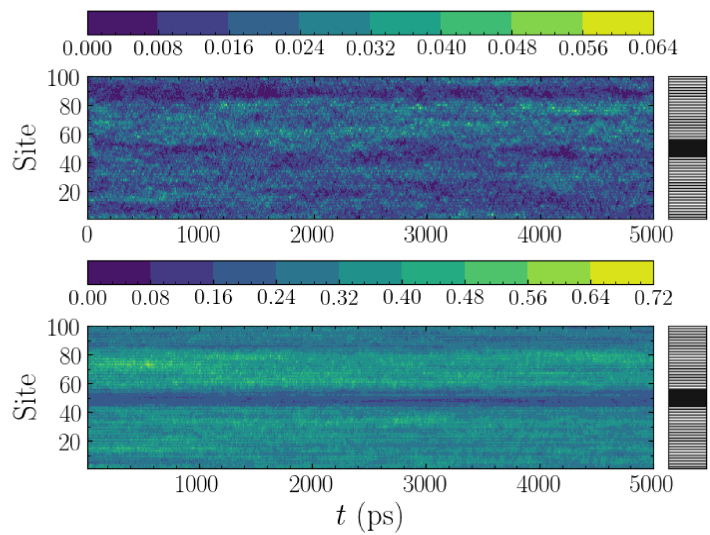

FIGURE 4.23: The average over one hundred initial conditions of the evolution of the DVD (top) and the displacement (bottom) for $P_{AT} = 45\%$, $E_n = 0.042$eV, with a specific arrangement of the base pairs. The light green areas correspond to higher values (as per the colour bar). The black and white bar on the right shows the AT (white) and GC (black) base pair distribution for the particular disorder realisation. Compared to Fig. 4.22, we can see that the DVD somewhat avoids the GC island in the middle.

In Fig. 4.22 it is apparent that the deviation vector has no overwhelming preference for its location. The displacement once again shows what we expect, a tendency to focus at the AT base pairs, resulting in a channel of high displacement in the centre. The DVD now does follow this excited region, showing a moderate tendency to focus in this central island. Around 14% of the DVD is focussed in this 10-base pair region, demonstrating a slightly above-average concentration, but nothing extraordinary. What is interesting is that once again the striping of the DVD reveals that despite the 45–55 base pair split in favour of AT base pairs, overall 48% of the DVD is focussed at GC base pairs. This means that outside the central region, we more or less maintain the previous fraction of 55% of the DVD at GC base pairs ($48/86 \approx 56\%$).

Turning to the case with a GC island in the centre, Fig. 4.23, we see that the displacement is lower in the central island as expected, and the DVD slightly avoids this region. This avoidance is very slight, as we see that approximately 9% of the DVD is found in this region. So relative to the AT island, there is a distinct difference. But compared to the general case, it is only 1 percentage point below the expected value, suggesting that the DVD is not particularly avoiding this region. Once again we see that outside of this central region, around 56% of the remaining DVD is localised at GC sites.

Neither island produced particularly dramatic results, and the trend clearly remains that the DVD localises nearby to the largest displacements, resulting in a slight tendency to focus at sites with GC base pairs in a completely mixed system. The DVD does not appear to show a universal affinity or aversion to particular base pair types. In all the cases examined in this section, it is clearly dependent not only on the particular base pair realisation, but also on the initial condition. By averaging over initial conditions we can reveal the dependence on the realisation alone, since the effect of the initial conditions themselves is minimised by the averaging, showing the tendency of the DVD to avoid consistently highly displaced regions.

## 4.6   Conclusions

In this chapter, we investigated the chaotic dynamics of the PBD model of DNA, as well as the sequence-dependent ePBD model. Using numerical methods, we studied the effect of AT and GC composition on the chaoticity of DNA sequences through the estimation of the mLE. We computed the mLE for DNA strands with several AT percentages, as well as fixing the alternation index and studying how the mLE can be affected by the ordering of base pairs even if the total number of AT and GC base pairs are the same. Additionally, we computed the DVD for different DNA sequences, and using single cases as well as averaged data studied the localisation of chaoticity in various cases.

Simply investigating the chaoticity of the model as a function of energy, the computation of the mLE informs us that the chaoticity increases with energy, with a slowing in the increase near the melting temperature. The effect of the percentage of AT base pairs in the sequence on the chaoticity is significant, with the low-energy regime yielding more chaoticity from AT-rich sequences, before a mid-energy region between energy densities of 0.025eV and 0.035eV shows little effect from changing the AT content, and finally at higher energies the GC-rich sequences become more chaotic.

The heterogeneity can be probed further, by use of the alternation index $\alpha$. Computing the mLE at different values of $\alpha$ we find that the more well-mixed a sequence is, the higher the value of $\alpha$, the less chaotic it is. Less heterogeneous sequences, with few alternations, provide more chaotic behaviour than the most likely values of $\alpha$. For sequences consisting primarily of one base pair type (for instance 90% AT base pairs), the alternation index has very little effect.

The addition of more sensitive sequence-dependence through the use of the extended PBD (ePBD) model yields no significant changes to the chaotic dynamics.

Investigation of the deviation vector distribution (DVD) yields that the nonlinearity in the system focusses near regions of large displacement, but not at those regions. In particular, the DVD avoids regions of consistently large bubbles, such as near the TATA box of biological promoters, as well as regions of very little fluctuational activity, where the displacements are consistently small. The DVD concentrates rather in regions of unpredictable and changing displacements. The DVD jumps around between these regions of changing displacement, often without apparent cause. However, at very low temperatures, the DVD concentrates in homogeneous islands in the sequence. This appears to be independent of the base pair that the island is comprised of, but the DVD will concentrate in islands of both AT and GC base pairs rather than spread through the more heterogeneous segments of the strand.

The effect of heterogeneity in DNA is clearly visible in the chaotic dynamics, and there are consequently several possibilities for future investigations. For instance, the study of different models, such as modifications of the PBD model to account for the twisting of the DNA strand explicitly [102], or other adaptations. There is also scope to study the chaotic dynamics through the computation of covariant Lyaponov vectors (CLVs) [103], which have the potential to provide valuable information about the nonlinearity of the models. Further investigation of the DVDs, possibly together with CLVs, is another direction that could prove fruitful. More insight into the relationship between bubbles and concentrated nonlinearity holds promise for the better understanding of DNA dynamics.

# Chapter 5

# Melting and Bubble Properties of DNA

The notion of thermally induced openings in DNA, or bubbles, has been introduced in both the introductory DNA chapter (Chapter 2) and mentioned in the context of the DVD in Chapter 4. We will now move into a thorough investigation of these bubbles, and try to build an understanding of bubble occurrences probabilities, as well as bubble lifetimes in typical DNA sequences. The biochemical significance of bubbles has been explored extensively in the literature, with results being found indicating a relationship between transcriptionally significant sites – the promoter TSS, as well as notable regulatory sites – and the spontaneous formation of bubbles [50, 54, 104, 105]. Against this background of biological interest, in this chapter we will present results pertaining to the formation of bubbles in DNA strands. This includes a relationship between the temperature and energy density, a distance threshold for considering base pairs open, and then distributions for bubble lengths in DNA and bubble lifetime probability distributions. A comparison of bubble lifetimes for a particular biological promoter with results obtained for general sequences is also presented, adding context to our findings.

The results of in this chapter are based on the research presented in [106].

## 5.1 Background

As an introduction to the study of bubbles, we can look at a number of essential aspects of bubbles that have warranted attention in the past. The biological importance of transcriptional sites has meant that many projects have been devoted to studying the distribution of bubble occurrence across the DNA sequence [50, 52–54, 75, 100, 101, 105, 107–111]. These investigations look at different features of thermally induced openings (i.e. bubbles) in DNA molecules, and compare dynamically active sites with experimentally determined transcription sites. One of the most significant suggestions to arise from studying the probability of formation and lifetimes of bubbles at the TSS of known promoters, and comparing this to known nonpromoter regions, is that DNA dynamically directs its own transcription [50, 54]. By performing numerical simulations at physiological temperatures, the probabilities and lifetimes of bubbles can be estimated, and from these distributions particular conclusions can be drawn – for instance, that "soft" regions in promoters, i.e. regions with a high propensity for thermal openings, correspond to transcription binding sites [101]. While the majority of studies (includ-ing the works cited here) perform simulations and track the appearance and lifetimes of bubbles directly, there are also works using methods such as autocorrelation functions [52], which have revealed further properties of base pair openings. As an important note with regard to computations with the PBD or similar models, simulations can be performed either at precise temperatures through Langevin dynamics, e.g. [50], or at approximate temperatures through constant-energy simulations, e.g. [101]. The use of Langevin dynamics has the added influence of simulating the behaviour of DNA in a solution, imitating the stochastic interaction with neighbouring particles and the effects of the larger system on the DNA molecule (see for instance [112] and references therein).

As a result of these works, the relationship between bubble formation and DNA transcription is strong enough to suggest that the sequence of base pairs alone is not enough to provide the relevant information for transcription. The (randomly occurring) thermal openings are important in the transcription process as they provide points along the molecule that are much more likely to attract the RNA polymerase which initiates the transcription [100, 101]. Consequently, there is strong motivation purely from this transcriptional point of view to pursue investigations into the properties of DNA bubbles, in a variety of transcriptionally significant sequences.

There have also been studies of the distribution of bubble lengths, the probabilities of longer or shorter bubbles forming in different sequences [60, 113]. These have found that in general the likelihood of long bubbles forming in a DNA molecule are extremely small, with the probability of bubble occurrence decreasing with increasing bubble length according to a power law modified exponential.

The significance of openings in DNA also goes beyond the transcriptional aspect. For instance, in studies of charge propagation along DNA strands, the presence of bubbles has been found to significantly impact the transport of the charge [93, 114–118]. The localisation of charge in DNA molecules, also called charge trapping, has been linked to these fluctuations in the strand, and in particular regions with a high propensity for bubble occurrence have a strong impact on charge trapping [94, 119]. Controlling charge transport has applications in different branches of bioelectronics, as the use of DNA-like wires for practical engineering considerations is a promising avenue for new developments [120].

Research has not been restricted to stationary openings in DNA either, with mobile discrete breathers providing a number of interesting aspects regarding the dynamics and functioning of DNA [121]. These mobile discrete breathers have also been linked to the trapping of charges in DNA, with recent results suggesting that the presence of these breathers can cause targetted movement of charge along the strand [120].

Additionally, results from dynamical openings in DNA have been used to study more properties of DNA such as free energy landscapes through modifying the PBD model [108], and investigating the process of transcription by identifying protein binding sites through their free energy characteristics [111]. The correspondence of these results with experimental findings provides promising signs for the possibility of useful predictions about promoters that are currently not well studied.

## 5.2    A General Threshold for Open Base Pairs

While the study of bubbles in DNA is clearly worthy of attention, when examining them from a molecular dynamics perspective there is the obvious problem of choosing a threshold for considering a base pair to be in a bubble. Since there is not a particular standard for this value, there are often wildly varying criteria applied, even within the same project (for some examples of results using different thresholds see [50, 107, 113] and references therein). Frequently the choice of a particular threshold value is made to emphasise a certain point – for instance the presence of uncharacteristically large amplitude bubbles at significant sites [54]. In other cases the implemented threshold value does not play a major role in the actual outcomes, affecting the exact numerical details but not the qualitative result [60].

Thus the proposition of a threshold which can be justified as a general measure of whether a base pair is open or closed would be useful. While in some cases the value will still be varied to investigate a specific property, a general criterion would enable more standardised results for cases where the exact threshold value is somewhat arbitrary.

Returning to the significance of the denaturation curves and melting temperatures, a logical criterion is to consider the definition of the melting temperature for DNA strands, which provides a reference point for open base pairs. Since melting is said to have occurred when precisely half the base pairs are open, this gives us a relationship between what we want (criterion for an open base pair) and something well established (the melting temperature). The approach now seems fairly straightforward – perform simulations, and retrieve data for the displacements of each base pair, and investigate this data near the melting temperature. Then by some process, settle on a threshold which gives precisely 50% of base pairs open at the melting temperature. These melting temperatures in Kelvin are taken from Kalosakas and Ares [60], where they have been calculated for DNA polymers of various AT and GC constituents and are given in (4.1), as

$$T_m = 325 + 0.4 P_{GC}.$$

In principle the same arbitrary-sequence computations could be applied to particular sequences where the melting temperature is known, if the boundary conditions can be correctly managed. However, if the aim is to provide a general threshold, there may be some variance in individual cases, especially in the ePBD model where the sequence dependent stacking interaction provides a significant difference in melting temperatures between similar molecules. We also note that for consistency and comparability with previously published results [60], in this chapter the AT/GC content of DNA sequences is quantified by the $GC$ percentage $P_{GC}$ rather than the AT percentage that has been used in prior chapters.

For our primary concern of general sequences, let us start with trialling a threshold value of 1Å, which has been used in previous studies as a useful value [60, 113]. In all these simulations, we have used 100 setups with random initial conditions, and in the case of heterogeneous sequences, random base pair realisations, with 1000 base pairs in the sequence. Periodic boundary conditions are used to avoid edge effects. We run the simulations for 10ns to thermalise, and then record data (displacement of each base pair) every 1ns for the next 10ns. This data can then be analysed using the desired threshold, in this case 1Å, to count the number of base pairs open beyond the threshold. The results shown in Fig. 5.1 give the fraction of open base pairs $f_o$, dividing the total number of recorded open base pairs by the total number of recorded base pairs.

Figure 5.1(a) shows the fraction of open base pairs $f_o$ in the DNA molecule as the temperature $T$ changes for five GC percentages, with a threshold of 1Å for considering the base pair to be open. Note that the temperature is once again computed as the

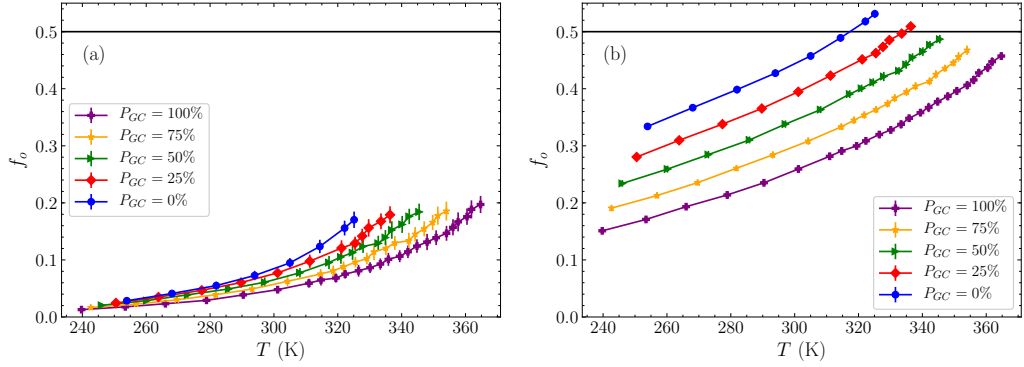

FIGURE 5.1: The fraction of open base pairs $f_o$ as a function of temperature $T$ for the PBD model given for five GC percentages: $P_{GC} = 0\%$ (blue circles), $P_{GC} = 25\%$ (red diamonds), $P_{GC} = 50\%$ (green triangles), $P_{GC} = 75\%$ (yellow stars) and $P_{GC} = 100\%$ (purple crosses). (a) A commonly used threshold of 1Å for considering a base pair to be open, predicts far below 50% of base pairs open at the melting temperature. The solid black line shows the 50% mark. The errorbars are given by the standard deviation from averaging the fraction of open base pairs across simulations. (b) Using a smaller threshold of 0.2Å produces results which are close for the AT-rich sequences, with nearly half the base pairs open at the melting temperature, but it still fails significantly as GC content increases, giving $f_o$ well below 50%.

rescaled mean kinetic energy, $T = 2\langle K_n \rangle / k_B$. We can immediately see that this threshold value is far too high to predict half the strand to be separated, as in all cases only about 20% of the base pairs are marked open by the time the melting temperature is reached. So we know that this is a fairly high threshold, and for our current purpose it is not a good choice. It does allow us to understand results that use this threshold better though – for instance, analysis performed with this (or even larger thresholds) will reflect properties of very widely separated bubbles, not considering the contributions from "slightly" open bubbles.

Moving past this threshold, in Fig. 5.1(b) we consider a threshold value of 0.2Å. This is a much better choice, giving us approximately 50% of all base pairs open at the melting temperature. It is not perfect though, and displays a clear trend of marking too many AT base pairs open and not enough GC base pairs. This is consistent with the weaker bonding between AT base pairs, resulting in a higher propensity to separate. This phenomenon has been documented in the study of bubble lengths of different DNA molecules [60], where AT base pairs have been seen to be more likely to open than GC base pairs. We thus need to consider some sort of heterogeneous approach to the threshold, in order to account for the different behaviours.

Having taken note of the region that this threshold falls in, we can now consider the properties of our model to see if there is a justification for any particular choice of threshold, and particularly any reasonable way of choosing one threshold for GC base pairs and one for AT base pairs. The natural place to look is at the on-site Morse potential, which primarily governs the distances between the bases. Referring all the

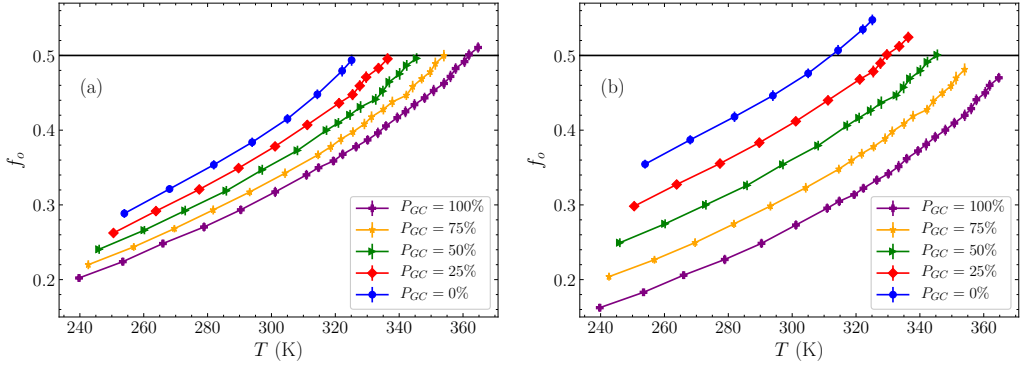

FIGURE 5.2: The fraction of open base pairs $f_o$ as a function of temperature $T$, for five GC percentages: $P_{GC} = 0\%$ (blue circles), $P_{GC} = 25\%$ (red diamonds), $P_{GC} = 50\%$ (green triangles), $P_{GC} = 75\%$ (yellow stars) and $P_{GC} = 100\%$ (purple crosses). (a) A heterogeneous threshold, with values $y_{AT}^{thr} = 0.24$Å for an AT base pair and $y_{GC}^{thr} = 0.15$Å for a GC base pair from inverting the characteristic value $a$ from the Morse potential gives a very accurate representation of 50% of the base pairs open at the melting temperature. (b) Using an average threshold in between the two values, $y = 0.185$Å. This works well for the intermediate cases around $P_{GC} = 50\%$, but fails on either extreme. Note that the 50% mark is shown by the solid black line im both panels.

way back to (2.4), we recall that the Morse potential has the form

$$V_n = D_n(e^{-a_n y_n} - 1)^2,$$

where the constants $D_n$ and $a_n$ depend on the base pair type at the particular site. This suggests a natural length scale based on the parameter $a_n$ – this provides the characteristic width of the Morse potential well, in units of inverse length. Consequently, there is good reason to investigate the choice of threshold $y_{AT}^{thr} = 1/a_{AT}$ and $y_{GC}^{thr} = 1/a_{GC}$. Calculating these quantities, we get $y_{AT}^{thr} \approx 0.24$Å and $y_{GC}^{thr} \approx 0.15$Å. Figure 5.2(a) shows the result of using these two thresholds for AT and GC base pairs respectively in the sequence. Here we find an almost exact presentation of 50% of the base pairs being open at the melting temperature. The only small blemish is the pure GC case (purple crosses), where around 51% of the base pairs are open, while for all other GC percentages almost exactly 50% of base pairs are open. This closeness very strongly motivates for this choice of threshold as a general measure of when a base pair is open or not.

For cases where only a rough measure is needed, a homogeneous averaged threshold value of $y_{AT}^{thr}$ and $y_{GC}^{thr}$, $y = 0.185$Å could be used, as shown in Fig. 5.2(b). This choice is evidently good for sequences with $P_{GC} = 50\%$, but becomes much less accurate for other GC percentages, up to 5% off for the pure AT case. So while this is not as accurate as the discriminatory threshold, it nonetheless provides useful information about the state of the strand.

## 5.2.1   Melting Temperatures for the ePBD Model

We are able to immediately make use of our new threshold in the investigation of the properties of the ePBD model of DNA. Recall that the only difference between the PBD and ePBD models is the sequence dependence built into the ePBD model (given by the values shown in Table 2.1), by varying the coupling strength depending on the exact ordering of the bases in the strand [75]. In light of this, and particularly due to the fact that the Morse potential [from (2.2)] remains unchanged between the two models, with the parameters chosen to reproduce physical results, the threshold we have found for the PBD model can be used for the ePBD model in exactly the same way.

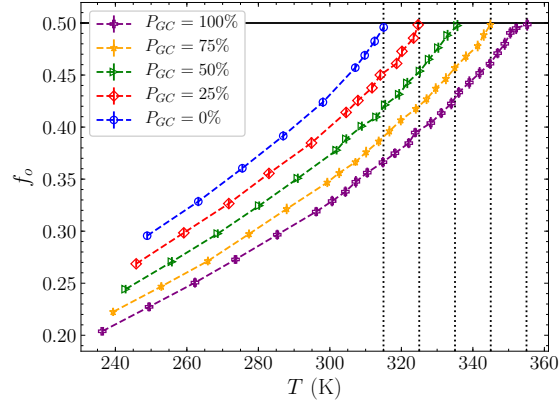

FIGURE 5.3: The fraction $f_o$ of open base pairs at a given temperature $T$ for the ePBD model, when the distinct threshold values $y_{GC}^{thr} = 1/6.9 \approx 0.15\text{Å}$ and $y_{AT}^{thr} = 1/4.2 \approx 0.24\text{Å}$ are used to define the opening of a GC and an AT base pair respectively. The solid line indicates 50% open base pairs. We see that using these thresholds, we can obtain approximate melting temperatures of $T_m = (315 + 0.4 P_{GC})\text{K}$, which retains the experimentally observed linear relationship. The dotted lines show the proposed melting temperatures for each GC percentage.

In order to find a general formula for melting temperatures of DNA molecules with the PBD model, a fairly complex procedure had to be followed involving fittings of bubble profiles [60]. The advantage of being able to carry the threshold values over from the PBD model circumvents this lengthy process in the ePBD model, as we can in fact directly find the melting temperatures by reversing the process applied above – instead of finding the threshold to match known melting temperatures, we can use the known threshold to locate the correct melting temperature. Figure 5.3 shows the result of following this approach by plotting the fraction of open base pairs $f_o$ against the temperature $T$ for the ePBD model. For all five GC percentages shown, we see that the thresholds predict very clear melting temperatures which follow the known linear relationship with the GC content $P_{GC}$ [122]. The obtained melting temperatures $T_M$ are shown by the dashed vertical lines in Fig. 5.3, with the horizontal solid black line marking 50% open base pairs. The melting temperature $T_M$ of DNA sequences as a function of the GC content $P_{GC}$ in the ePBD model is thus given by

$$T_M = (315 + 0.4 P_{GC})\text{K}. \tag{5.1}$$

The precise matching of this relation to the numerically observed melting points in Fig. 5.3 is convincingly close, and we are able to use this equation to estimate the melting temperatures of DNA molecules of various GC percentages. Since the ePBD model is designed to be more sensitive to the exact base pair sequencing, there is a lot of variability for every GC percentage. For instance, pure GC sequences are known to have melting temperatures varying by more than 25K, and this variance is captured by the ePBD model [75]. So we should be careful when applying this relation to specific cases, but for all of our investigations here, we are concerned with average behaviours of many realisations, meaning that this relation is certainly applicable.

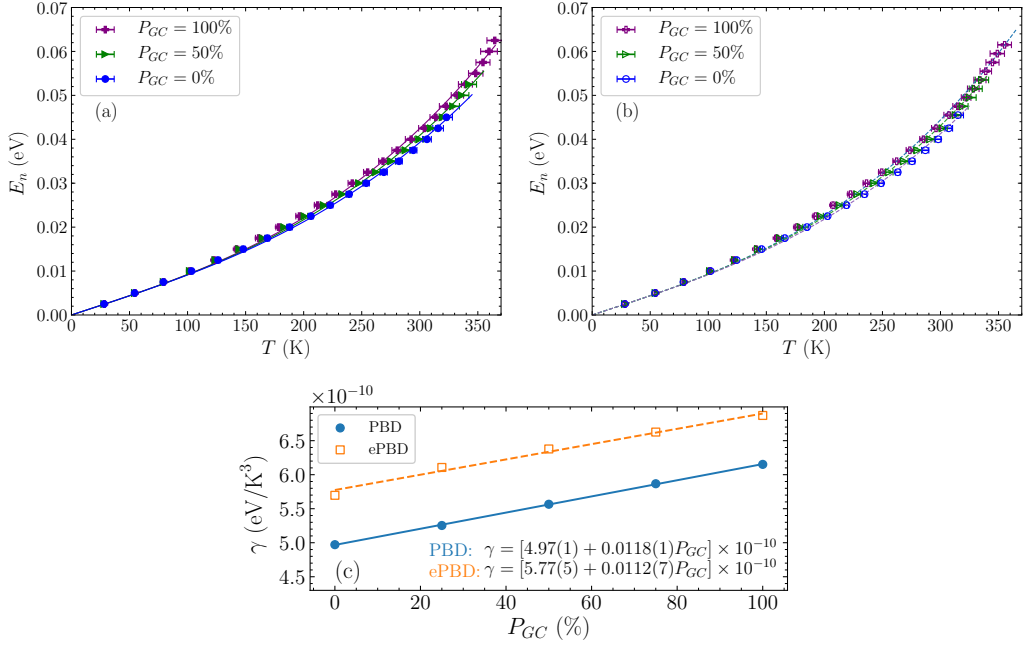

FIGURE 5.4: The energy density–temperature relationship for (a) the PBD model, and (b) the ePBD model for three GC percentages, $P_{GC} = 0\%$ (blue circles), $P_{GC} = 50\%$ (green triangles), and $P_{GC} = 100\%$ (purple crosses). The solid and dashed curves respectively indicate the fitting to (5.2) for each GC percentage shown. (c) The variation of the parameter $\gamma$ of (5.2) with GC content, for the PBD model (blue circles) and the ePBD model (empty orange squares). The fitted linear equations are marked by solid (PBD) and dashed (ePBD) lines, with the equations displayed in the panel. The error of the last significant digit of the obtained parameters is shown in parentheses after each fitted value, e.g. 4.97(1) corresponds to $4.97 \pm 0.01$.

## 5.2.2 Energy-Temperature Relationship

With the melting temperature (5.1) for the ePBD model, we can briefly interrogate the relationship between energy density $E_n$ and temperature $T$ for the PBD and ePBD models, now that we know the relevant bounds for the ePBD model. To this end, we performed simulations at several energy density values and computed the average temperature across these runs. For each GC percentage, at each energy density, 100 different random initial conditions were run, also with random base pair realisations in heterogeneous cases, using periodic sequences of 1000 base pairs. Each simulation was thermalised for 10ns, then data recorded for the next 1ns every 0.01ns.

The results of these simulations are shown in Fig. 5.4(a), for the PBD model and in Fig. 5.4(b) for the ePBD model. Only three GC percentages are shown, so as not to obscure the results, $P_{GC} = 0\%$ (blue circles), $P_{GC} = 50\%$ (green triangles), and $P_{GC} =$ 100% (purple crosses). GC-rich sequences [such as the pure GC case, purple crosses in Fig. 5.4(a)] tend to have a higher energy density at a given temperature, particularly at higher temperatures. In all cases we see that at low energies, there is an evident linear relationship between energy density and temperature, with the GC content making very little difference to the results. As the temperature increases however, a nonlinear trend emerges, and for temperatures above about 150K, the linear approximation is certainly insufficient. In order to model this change, we have fitted the data with a linear term proportional to the Boltzmann constant $k_B$, which is accurate for low energies, but also added a cubic term to more accurately follow the higher temperature behaviour,

$$E_n = k_B T + \gamma T^3. \tag{5.2}$$

The constant $\gamma$ is free to be fitted for the two models.

The fitted curves are shown in Fig. 5.4(a) and (b) by solid and dashed lines respectively. We see that in all cases (5.2) provides a very good fit to the data, despite the simplicity of the functional form. The values of the fitted constant $\gamma$ are shown in Fig. 5.4(c) as the GC content is varied, for both models. The linear increase of $\gamma$ with the GC percentage allows the fitting of a straight line to the data, resulting in a simple relation which is included in the figure. There is an overall tendency for $\gamma$ to be larger in the ePBD model, although the changes with GC content are very similar for the two models.

Equation (5.2) then gives us a very useful way of referring back and forth between $E_n$ and $T$ for the two models, combining (5.2) with the values of $\gamma$ given in Fig. 5.4(c) even providing us with a direct way of computing $E_n$ for a given $T$ and GC percentage. For investigations in the microcanonical ensemble, working with constant energy rather than temperature, it can frequently be useful to have a link between the two quantities, as often $T$ is a more physically useful quantity to consider. Of particular importance to our work here, where we will look largely at physiological temperatures, is that we can clearly tell from the outset that for different GC percentages we need to use different $E_n$ values in order to achieve the same $T$ values. So in order to compare results at the same temperatures, we can make use of this relationship (5.2) to find the appropriate energy value corresponding to $T = 310$K in each case.

## 5.3 Distributions of Bubble Lengths

With these initial results all cleared, we can investigate some properties of bubbles in DNA sequences using our heterogeneous threshold. The first aspect we will look at is the probability distribution for bubbles of various lengths to occur in a DNA strand at $T = 310$K, for strands with various base pair compositions. This investigation will help us to understand the likelihood of the occurrence of particularly long bubbles, and also the effect of the GC content on the appearance of bubbles. Results regarding these distributions have been previously reported for the physiological temperature regime [113], as well as for temperatures in the range of 270-350K [60]. These investigation made use of Monte-Carlo methods, and used a single threshold value of 1Å (which, as we saw in Fig. 5.1, does not necessarily capture small bubbles correctly), rather than our use of molecular dynamics and heterogeneous thresholds here.

For this study, we performed 8 000 simulations of DNA sequences with $N = 1\,000$ randomly chosen base pairs according to the desired GC percentage, each run with random initial conditions. Again, a 10ns thermalisation time was used, and then data were recorded for a further 10ns, taken every 0.1ns. At each recording step, for each base pair the information was stored as to whether it was open or closed, based on the threshold related to its base pair type. Thus, for each run data was recorded for a total of 1 000 base pairs at 100 steps, resulting in 100 000 data points (open/closed base pair) per run, which were analysed to find the lengths of all bubbles occurring. Combining all these runs together, we had a total of 800 000 000 recorded data points, providing strong statistics even up to the extreme ends of the probability distributions. Analysis of the full data set was compared with using only sections of the data to check that our results are not affected by a lack of data, and we found that our results were virtually identical even when only using 10% of the total data. This gives us a good deal of confidence in the data, and allows us to trust the tails of the distribution more strongly.

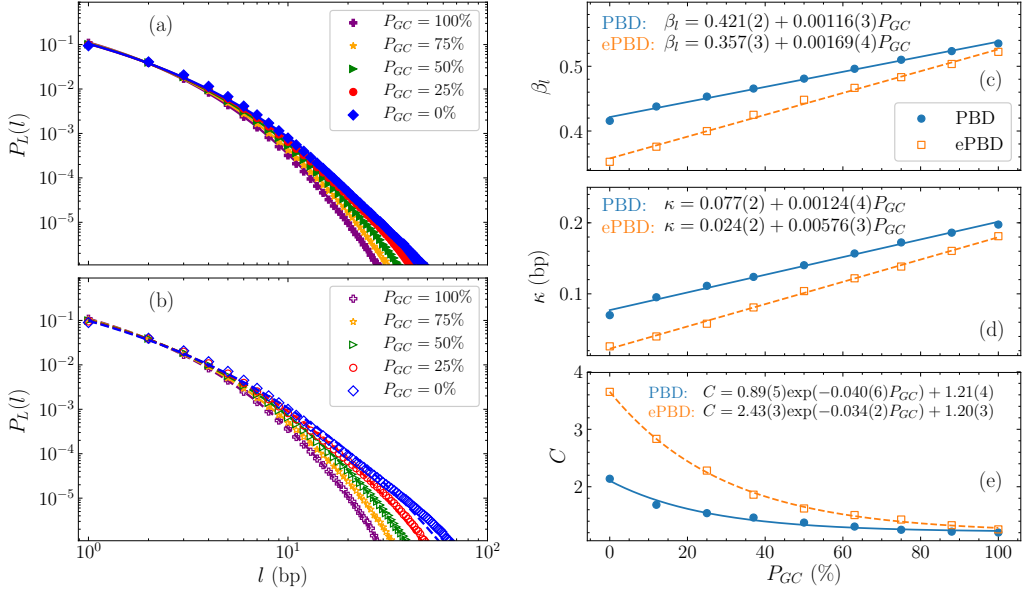

FIGURE 5.5: The distribution $P_L$ of bubble lengths in the PBD (a) and ePBD (b) models for five GC percentages, namely $P_{GC} = 0\%$ (blue circles), $P_{GC} = 25\%$ (red diamonds), $P_{GC} = 50\%$ (green triangles), $P_{GC} = 75\%$ (yellow stars) and $P_{GC} = 100\%$ (purple crosses), at 310K, plotted in log-log scale. The solid (dashed) curves correspond to fittings to (5.3). (c)–(e): The fitting parameters of (5.3) to the results of (a) and (b), the stretching parameter $\beta_l$, the characteristic length $\varkappa$, and the prefactor $C$ respectively as a function of the GC content $P_{GC}$. The PBD parameters are shown by solid blue circles, and the ePBD parameters by empty orange squares. For $\beta_l$ and $\varkappa$, the variation of the parameters with GC content has been fitted with a straight line, and the equation of the lines for the two models are displayed in each panel. The constant $C$ is fitted with a simple exponential function, and this equation is also presented in the figure.

The results of this analysis are shown in Fig. 5.5(a) and (b), showing the probability distribution $P_L(l)$ of bubble lengths for the PBD and ePBD models respectively. Some initial points are immediately apparent – for instance, the heterogeneity of the sequences appears to be counterbalanced by the base-pair dependent threshold for short bubbles, as bubbles with $l < 5$ base pairs (bp) do not show a significant variation in probability with GC content. However, as the lengths increase, the base pair composition becomes significant once again, as sequences with fewer GC base pairs demonstrate a markedly higher probability for long bubbles ($l > 10$bp) to appear, despite the larger threshold value for these base pairs. This behaviour is consistent across GC percentages, as we can see that the pure GC [purple crosses in Figs. 5.5(a) and (b)] and pure AT (blue diamonds) cases are the least and most likely to exhibit long bubbles, with the intermediate percentages showing a continuous trend of more long bubbles being present when the GC content is lower.

These overall behaviours are observed in both the PBD model [Fig. 5.5(a)] and the ePBD model [Fig. 5.5(b)], but there are some small differences between the two. Primarily, the ePBD model exhibits a larger probability for bubbles to appear at almost all lengths, and particularly for long bubbles. This effect is strongest in the AT-dominant strands (e.g. the $P_{GC} = 25\%$ case, red circles in Figs. 5.5(a) and (b), and the pure AT case, blue diamonds), where the tail of the distribution stretches noticeably longer than in the PBD model. The pure GC case [purple crosses in Figs. 5.5(a) and (b)] appears to be affected very little by the choice of model, with no discernible differences between the two plots. This behaviour results in a much more significant spread of results for the ePBD model, with the distributions distinctly more separated than in the PBD case.

We also note that these general trends have been seen previously, using a uniform threshold value [60], where particularly for large bubbles the behaviour is fairly similar. One distinct point of difference however is the behaviour for short bubbles, where in our case the base pair dependent threshold results in the probabilities of bubble occurrence being almost identical for all GC percentages.

These obtained $P_L(l)$ distributions can be fitted fairly accurately by a stretched exponential, of the form

$$P_L(l) = C \exp\left[-(l/\varkappa)^{\beta_l}\right], \tag{5.3}$$

with $C$, $\varkappa$ and $\beta_l$ parameters free to be fitted. This follows the data accurately even up to the tails in Figs. 5.5(a) and (b). These fits are shown for the PBD model in Fig. 5.5(a) as solid curves, and in Fig. 5.5(b) by dashed curves for the ePBD model, confirming that they provide a very good approximation of the numerical data.

The dependence of the fitting parameters of (5.3) on the GC content $P_{GC}$ is illustrated in Fig. 5.5(c)–(e). In Fig. 5.5(c) we see a linear increase in the stretching exponent $\beta_l$ as the GC content is increased. This holds for both the PBD model [blue circles in Fig. 5.5(c)] and the ePBD model [empty orange squares in Fig. 5.5(c)], albeit with different relations for the two models. The linear increase has in turn been fitted with a straight line, and the equations of these lines are given in the panel. For the characteristic length $\varkappa$, we also see a linear increase with the GC percentage, in Fig. 5.5(d). This linear function has also been fitted, and the equations are shown in the panel. For the prefactor $C$, the change with GC content is no longer linear, but rather exponential [Fig. 5.5(e)]. The exponential decrease can be fitted accurately with a simple exponential function for both models, as shown in Fig. 5.5.

In general, the values of the fitting parameters $\beta_l$, $\varkappa$ and $C$ confirm the impression from Figs. 5.5(a) and (b) that the difference between the PBD and ePBD models is more pronounced for lower GC content. For instance, the stretching exponent $\beta_l$ [Fig. 5.5(c)] and the prefactor $C$ [Fig. 5.5(e)] both are almost identical for pure GC sequences, while $\varkappa$ [Fig. 5.5(d)] is very close. At the other end of the spectrum though, pure AT sequences ($P_{GC} = 0\%$) show a very strong dependence on the model, with all the parameters exhibiting very different values, up to almost a factor of 2, for this case.

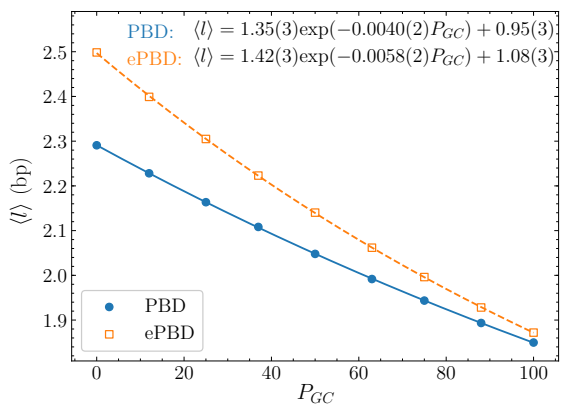

FIGURE 5.6: The average bubble length $\langle l \rangle$ (5.4) as a function of GC percentage $P_{GC}$, for the PBD model (blue circles) and the ePBD model (empty orange squares). The data are fitted with an exponential function, whose equations is shown in the figure. The solid blue and dashed orange curves show these fits for the PBD and ePBD models respectively.

An additional quantity we can calculate from the bubble length distribution $P_L(l)$ is the average bubble length $\langle l \rangle$, which can be calculated according to the equation (after [113])

$$\langle l \rangle = \frac{\sum_l l P_L(l)}{\sum_l P_L(l)}. \tag{5.4}$$

Here $P_L(l)$ is the probability of a bubble of length $l$ occurring [Figs. 5.5(a) and (b)]. Figure 5.6 shows the computed $\langle l \rangle$ values for both models. We see a clear decrease in the average bubble length $\langle l \rangle$ as the GC content increases (also previously observed for a constant threshold in [113]), for both models. The ePBD model (orange squares in Fig. 5.6) shows generally longer average bubble lengths, although once again the two models more or less converge for pure GC sequences. A large range of $\langle l \rangle$ values is also seen with the ePBD model, as a result of the greater sensitivity inherent in the model. The variation of the average bubble length with GC content can be fitted accurately with an exponential function, where the solid blue (dashed orange) curves show this fit for the PBD (ePBD) model in Fig. 5.6. The equations of the curves are embedded in the figure.

## 5.4 Distributions of Bubble Lifetimes

Another, less studied, aspect of denaturation bubbles is their lifetime. In a typical DNA double strand, how long would one expect a bubble to persist? In order to address this question, we can once again perform MD simulations, and by using detailed records of the base pair displacements, find distributions of the bubble lifetimes. The aim here is to find (at physiological temperature) the probability of a particular bubble lifetime given a) the GC percentage $P_{GC}$ of the sequence, and b) the length $l$ of the bubble. We already know from Section 5.3 that long bubbles form extremely rarely, but here we will go one step further, and see what we can learn about the lifetimes of these bubbles.

Firstly, let us outline the algorithm used to track and record bubbles, as this is a fairly non-trivial task. We note that this is much more complex than the analysis to produce the bubble length distributions, which does not require the tracking of bubbles through time. For a single simulation (one run of a single sequence with random initial condition), the following steps are followed to create the distribution of bubble lifetimes

1. For the given initial condition, perform the MD simulation to create, at each time step, a record of whether the base pair is open or closed, for each base pair in the sequence. In our simulations, the open/closed data is defined using the thresholds defined in Section 5.2.

2. For every time step, move through the sequence and record, at the start site of each bubble, the length of each bubble that occurs.

3. Check every recorded bubble (site and length) against the record from the previous time step.

   - If there is a bubble starting where previously there was no bubble recorded at that site, begin a record of that bubble – a tuple of (length, lifetime).

   - If a bubble starting at this site survives with the same length, then increment its lifetime by one time step.

   - If a bubble starting at this site survives but changes length, then end the record of that bubble, and begin a new tuple at that site with the new length.

   - If there is no bubble present where one was at the previous time step, then end the record and do not create a new one.

   - If there is no bubble present, and no bubble was previously present, then do nothing.

4. Finally, record the list of tuples of (length, lifetime) at each site.

5. Create the distribution of these (length, lifetime) data points, for each length.

This procedure allows us to study the bubble lifetime distribution for bubbles of varying lengths, and effectively combine many different runs to create a coherent picture of the general behaviour. As in all our investigations, we perform this analysis based on the GC content $P_{GC}$ of the sequences, by using many runs of random base pair realisations with the same numbers of GC and AT base pairs.

So we can now begin the investigation by creating bubble lifetime distributions for bubbles of varying lengths. For each GC percentage $P_{GC}$ studied, we have used 1 000 different realisations with random initial conditions, integrating a sequence of 100 base pairs with periodic boundary conditions. A thermalisation time of 10ns was again used to ensure equilibration of the system, and then data were recorded every 0.01ps for the next nanosecond. This data recording whether each base pair is open or closed was analysed using the algorithm described above, and we created normalised distributions of the observed bubble lifetimes, for lengths of up to $l = 10$ base pairs. We note that in all cases we have more than 200 000 recorded bubbles (with shorter bubbles having several million records), which gives a minimal likelihood for statistical anomalies to occur. Results could be presented for longer bubbles, but as noted in Section 5.3, these long bubbles are exponentially rare, and consequently the amount of data for these bubbles is significantly lower than for smaller bubbles.

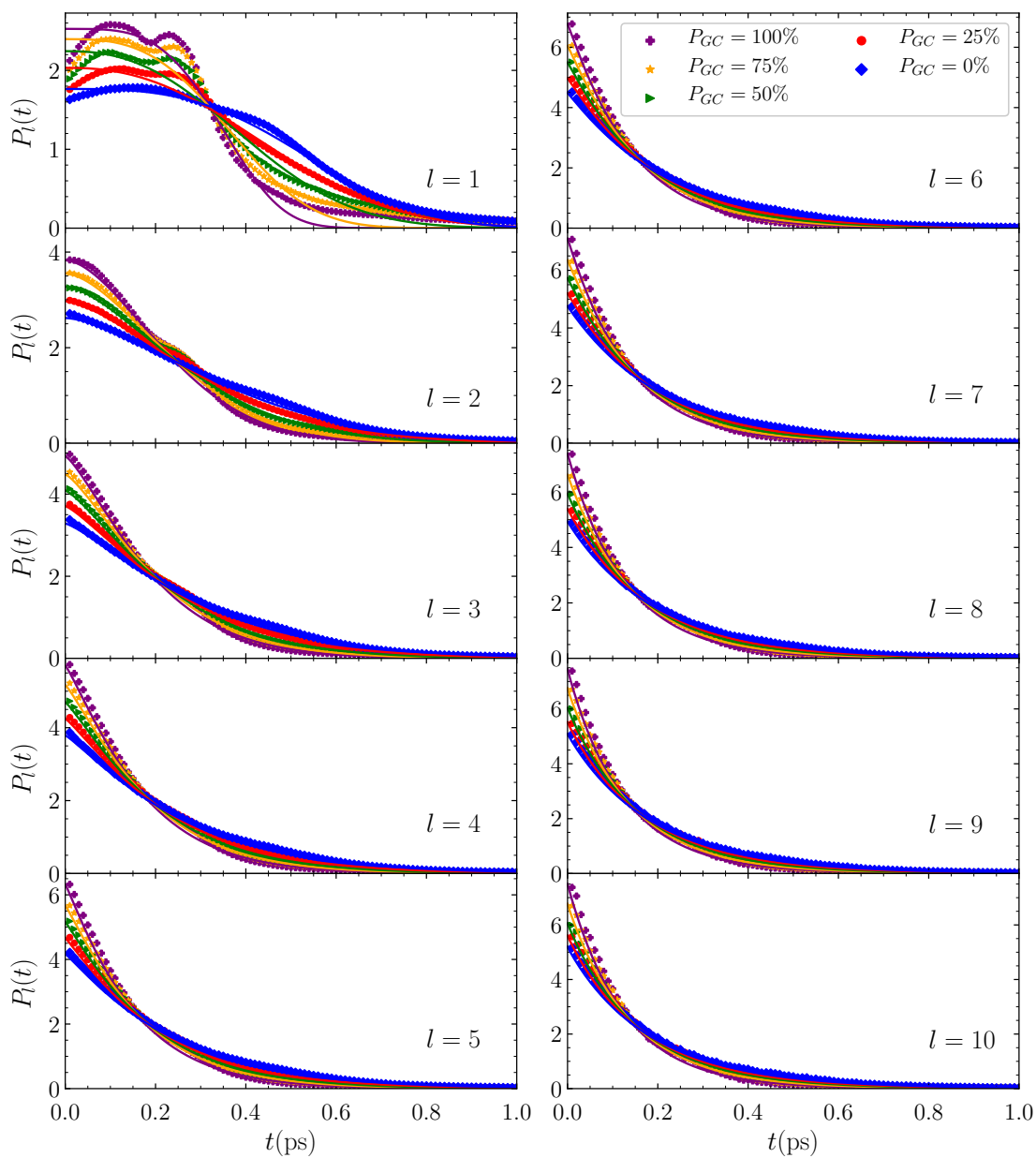

FIGURE 5.7: Bubble lifetime distributions $P_l(t)$ in the PBD model for five GC percentages, $P_{GC} = 0\%$ (blue circles), $P_{GC} = 25\%$ (red diamonds), $P_{GC} = 50\%$ (green triangles), $P_{GC} = 75\%$ (yellow stars) and $P_{GC} = 100\%$ (purple crosses). Results are shown for each bubble length up to $l = 10$, and each panel is labelled according to the relevant length. The solid curves show fittings to the stretched exponential (5.5).

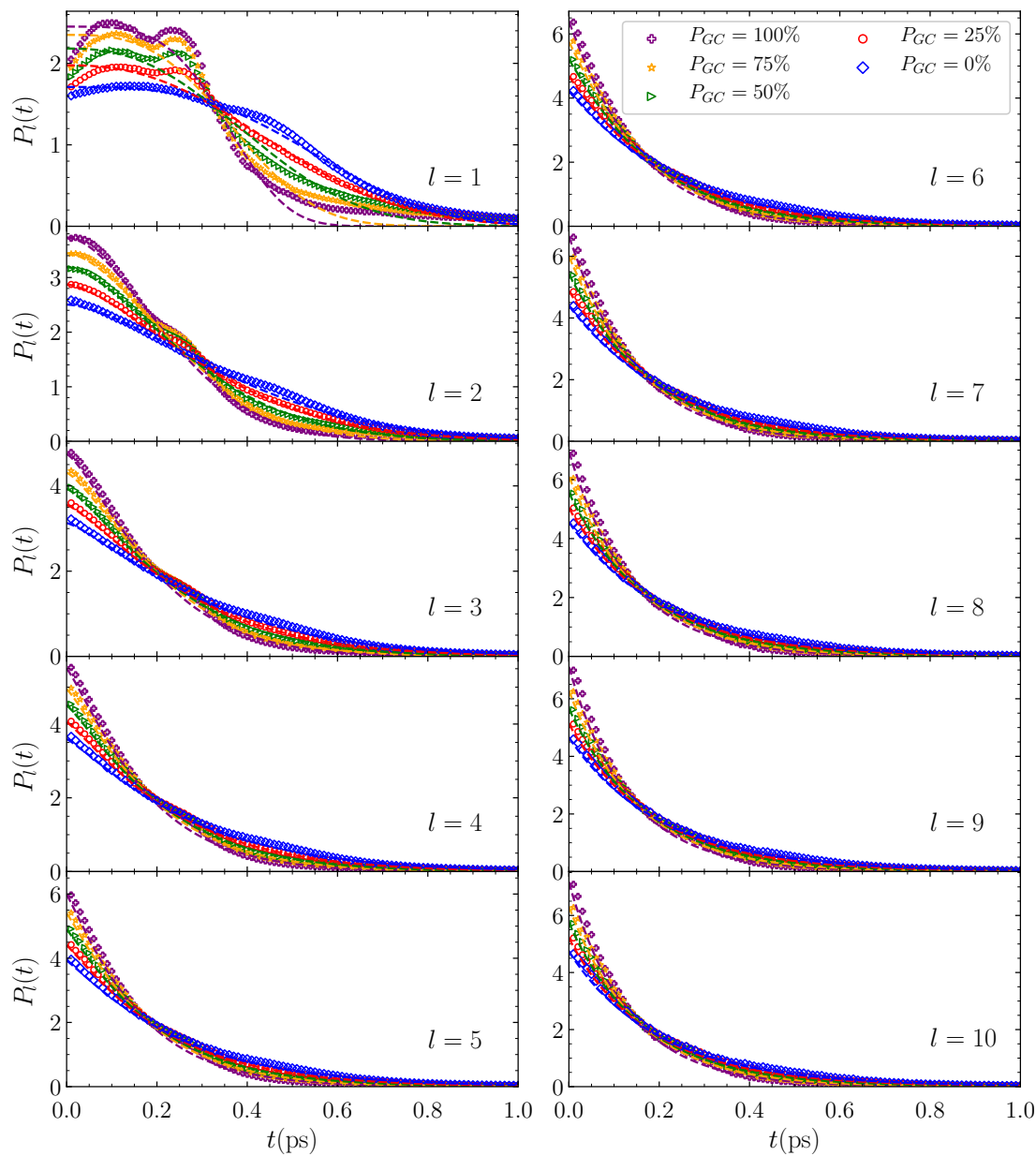

FIGURE 5.8: Similar to Fig. 5.7, but showing bubble lifetime distributions $P_l(t)$ for the ePBD model, for the same five GC percentages, $P_{GC} = 0\%$ (blue circles), $P_{GC} = 25\%$ (red diamonds), $P_{GC} = 50\%$ (green triangles), $P_{GC} = 75\%$ (yellow stars) and $P_{GC} = 100\%$ (purple crosses). The same stretched exponential (5.5) is fitted, shown by dashed curves.

The resulting bubble lifetime distributions $P_l(t)$ are shown in Figs. 5.7 and 5.8, for the PBD and ePBD model respectively. In each panel of the figure the distributions are shown for a particular bubble length $l$ (as denoted by the label on the panel), for five GC percentages, $P_{GC} = 0\%$ (blue circles), $P_{GC} = 25\%$ (red diamonds), $P_{GC} = 50\%$ (green triangles), $P_{GC} = 75\%$ (yellow stars) and $P_{GC} = 100\%$ (purple crosses). Immediately the result for "bubbles" of length $l = 1$ stands out (first panel of Figs. 5.7 and 5.8) – there appears to be a kind of double-peaked structure, with an initial peak around 0.1ps and a second following later at either 0.025ps (for GC-rich sequences) or 0.45ps (for AT-rich sequences). The peaks are also much broader in the pure AT sequences [blue diamonds in Fig. 5.7]. This structure may be the result of some interplay between the characteristic times of $l = 1$ bubbles and transient single-base-pair openings caused by the temporary opening and closing of larger bubbles, which can leave an isolated base pair open for some time before rejoining the larger bubble. As the length $l$ increases, this double peak feature does remain, although less clearly. For $l = 2$ in Fig. 5.7, the first peak is dramatically less pronounced, while the second peak is still visible but also much smaller in height. Past $l = 3$, we have much smoother distributions, with the peaks becoming less evident. The distributions also demonstrate a fairly clear tendency to exhibit fewer and fewer long-lived bubbles, although the last few cases $l = 7, 8, 9, 10$ in Fig. 5.7 are extremely similar visually.

The differences between the PBD results in Fig. 5.7 and the ePBD results displayed in Fig. 5.8 are not particularly strong. We find the same behaviour for single base pair openings ($l = 1$ in Fig. 5.8), with the peaks occurring in the same places as the PBD model. Thereafter a smoothing out occurs, and we end up with more or less exponential distributions again for long bubbles.

In order to quantify these distributions using an analytical expression, we can use a stretched exponential function of the form

$$P_l(t) = A \exp\left(-(t/\tau)^\beta\right), \qquad (5.5)$$

having as fitting parameters $A$, $\tau$, and $\beta$. This provides a very good approximation of the true distribution in all cases apart from the anomalous $l = 1$ openings. The fitting of the distributions to this function are shown in Figs. 5.7 and 5.8 by solid and dashed lines respectively. In both these figures we see that for the $l = 1$ case the stretched exponential does not follow the detailed structure of the distribution, but it does capture the essential behaviour of the decreasing probability, providing a useful estimation of the distribution nonetheless. For $l = 2$ and above, the distribution fits very closely, and gives us a very useful tool to understand the general likelihood of bubbles of particular lengths and lifetimes occurring.

The fitting parameters for the stretched exponential functions (5.5) are shown in Fig. 5.9 for the PBD and ePBD models. First considering the stretching exponent $\beta$ [Fig. 5.9(a)], we see that as bubble lengths increase $\beta$ rapidly approaches 1, corresponding to a standard exponential function in (5.5). This confirms the early impression from the distributions themselves [Figs. 5.7 and 5.8] that for long bubbles we had a consistent, exponential, behaviour. The very large value of $\beta$ for $l = 1$ is also explained by the fact that the $P_l(t)$ distribution for $l = 1$ is so distinct from the others. Other notable features are that as a consequence of the convergence towards 1, the GC content has little effect on the value of $\beta$ beyond the first two points, i.e. for $l \geq 3$, and that the two models have extremely close values for $\beta$ in all cases.

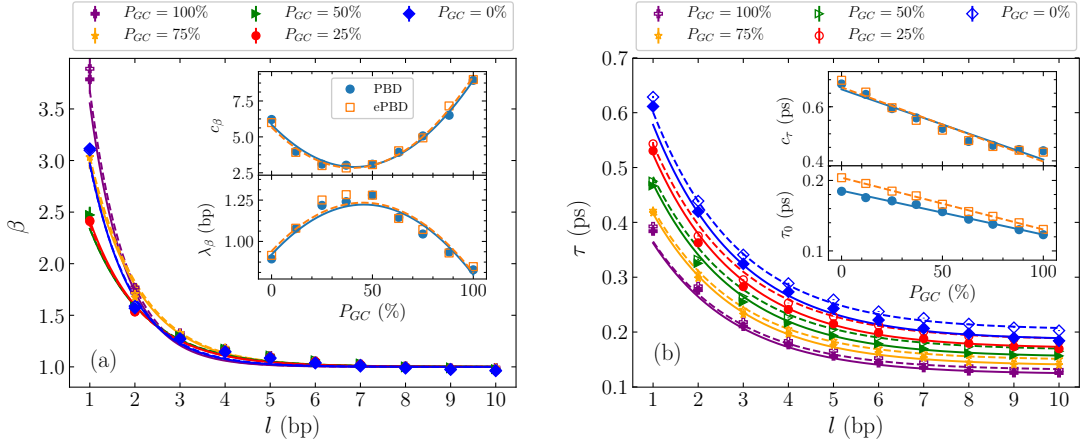

FIGURE 5.9: The parameters of (5.5) fitted to the bubble lifetime distributions of Figs. 5.7 and 5.8. (a) The stretching factor $\beta$ as a function of the bubble length $l$ for five GC percentages, $P_{GC} = 0\%$ (blue circles), $P_{GC} = 25\%$ (red diamonds), $P_{GC} = 50\%$ (green triangles), $P_{GC} = 75\%$ (yellow stars) and $P_{GC} = 100\%$ (purple crosses). Solid symbols correspond to the PBD model and empty symbols to the ePBD model. The exponential decrease in $\beta$ with increasing bubble length is fitted with a simple exponential, (5.6), which is shown by solid and dashed curves for the PBD and ePBD models respectively. The corresponding parameters $c_\beta$ and $\lambda_\beta$ of the fitting are shown by the insets as functions of the $P_{GC}$ values, with a quadratic fitting (solid blue curve for PBD, dashed orange for ePBD). (b) The characteristic time $\tau$, as a function of the bubble length $l$. Once again an exponential decrease is evident for both models, fitted with (5.7). The insets show the variation of the parameters $c_\tau$ and $\tau_0$ of (5.5) with $P_{GC}$, this time fitted by straight lines.

The characteristic time $\tau$ [Fig. 5.9(b)] also exhibits an exponential decrease with bubble length $l$, although nothing like so drastic as for $\beta$. Here we see a more steady decrease towards a constant asymptotic value, which is dependent on both the model and the GC percentage $P_{GC}$. The ePBD model (depicted by empty symbols in Fig. 5.9) has a consistent tendency to produce a larger $\tau$ value for all bubble lengths than the PBD model. The effect of the model is apparently stronger for AT-rich sequences, with the difference between the PBD and ePBD value of $\tau$ being most significant for the pure AT case [blue diamonds in Fig. 5.9(b)].

Since both these quantities vary exponentially with $l$, we are able to perform a fitting with simple exponential functions of the form

$$\beta = c_\beta \exp(-l/\lambda_\beta) + 1, \tag{5.6}$$

$$\tau = c_\tau \exp(-l/\lambda_\tau) + \tau_0. \tag{5.7}$$

These fits are shown in Figs. 5.9(a) and (b) by solid curves for the PBD model and dashed curves for the ePBD model, and the insets of these figures show the dependence of the parameters $c_\beta, \lambda_\beta, c_\tau$, and $\tau_0$ on the GC percentage. First addressing $\beta$, we see from the inset of Fig. 5.9(a) that the values of $c_\beta$ and $\lambda_\beta$ are very similar for the two models, with no consistent difference between the two. Furthermore, we can perform the simplest fitting that captures the correct behaviour of these values, which in this case is a quadratic function. These fittings are shown by the solid blue and dashed orange curves in the inset of Fig. 5.9(a). The fact that the fitted curves are almost identical also suggests that the effect of the model on the value of $\beta$ is minimal, and we can use the same values of $c_\beta$ and $\lambda_\beta$ for both cases while preserving accuracy. The computed equations for the parameters are $c_\beta = 0.0017(1)(P_{GC})^2 - 0.14(1)P_{GC} + 5.9(2)$, and $\lambda_\beta = -0.00015(2)(P_{GC})^2 + 0.014(2)P_{GC} + 0.93(4)$.

For the fittings to the values of $\tau$ [(5.7) and Fig. 5.9(b)], we found that a value of $\lambda_\tau = 1.9$ is maintained through all GC percentages for both models. In the inset of Fig. 5.9(b), we see that the parameters $c_\tau$ and $\tau_0$ of (5.5) both decrease linearly with increasing GC percentage $P_{GC}$. Fitting these values with straight lines, we see that for $c_\tau$ the fitted lines are effectively identical for both models, while there is a distinct difference between the models in the values of $\tau_0$ [see inset of Fig. 5.9(b)]. The computed equations for the parameters are given by $c_\tau = 0.66(1) - 0.0026(2)P_{GC}$, and then for the PBD model $\tau_0 = 0.19(1) - 0.0006(2)P_{GC}$ and for the ePBD model $\tau_0 = 0.20(1) - 0.0007(2)P_{GC}$.

Finally, the prefactor $A$ in (5.5) is considered to be free in our fitting procedure, but it is fundamentally defined by the normalisation of the distribution, which it maintains to an accuracy of around 5%. Note also that the lines of best fit shown in Figs. 5.9(a) and (b) are plotted using the parameter values estimated using the quadratic and linear fittings (shown in the insets of the figure), confirming that these fitted equations still produce an accurate representation of the true situation.

With these parameters for (5.5), we are now able to predict the probability of a bubble occurring with a particular length and lifetime, given only the GC content of the DNA sequence (and of course choosing a model). One point that is worth emphasising here is how extremely rare long-lived large bubbles are in the general DNA sequences, regardless of the AT/GC composition. The combination of the near-exponential rarity of long bubbles in general [see Figs. 5.5(a) and (b)] with the exponential decrease in probability of long-lived bubbles [Figs. 5.7 and 5.8] means that the chances of finding such a bubble by pure chance are "doubly exponentially small". This provides a useful backdrop to studies performed on bubble profiles in biological promoters, highlighting the fact that the presence of large long-lived bubbles at transcriptionally active sites is distinctly anomalous (see e.g. [50, 54])

A further property of the bubble lifetime distributions that we can investigate is the average lifetime; exploring how long the average bubble lives at different lengths and in strands of varying GC content. We can estimate the mean bubble lifetime $\langle t \rangle_l$ directly from the distributions, by using the formula

$$\langle t \rangle_l = \sum_{j=1}^{N_b} t_j P_l(t_j) \delta t, \tag{5.8}$$

where here the sum is over all the $N_b = 500$ bins of the histogrammed lifetime data, with width of $\delta t = 0.01$ps. Since even with our good statistics we have almost no bubbles with a lifetime longer than 5ps, we only consider the data up to this point. $t_j$ is the time at the centre of the $j^{\text{th}}$ bin, and $P_l(t_j)$ is the numerical probability of a bubble lifetime occurring within that bin from Figs. 5.7 and 5.8.

The result of performing this calculation for each bubble length is shown in Fig. 5.10 for a range of GC percentages, with the average bubble lifetime for the PBD model shown in Fig. 5.10(a), while the ePBD results are seen in Fig. 5.10(b). Similarly to the behaviour of $\tau$ [(5.7) and Fig. 5.9], we see an exponential decrease in the mean bubble lifetime $\langle t \rangle_l$ with increasing bubble length $l$. Sequences with more AT base pairs clearly exhibit a longer average lifetime, with a monotonic increase in $\langle t \rangle_l$ as the GC content $P_{GC}$ decreases at every bubble length. In addition, comparing the two models [Figs. 5.10(a) and (b)] we see that in general the ePBD model predicts longer average lifetimes, with a difference of around 0.02ps for each GC percentage. The difference between the two models is largest for the pure AT case [blue diamonds in Figs. 5.10(a) and (b)], and smallest in the pure GC case (purple crosses).

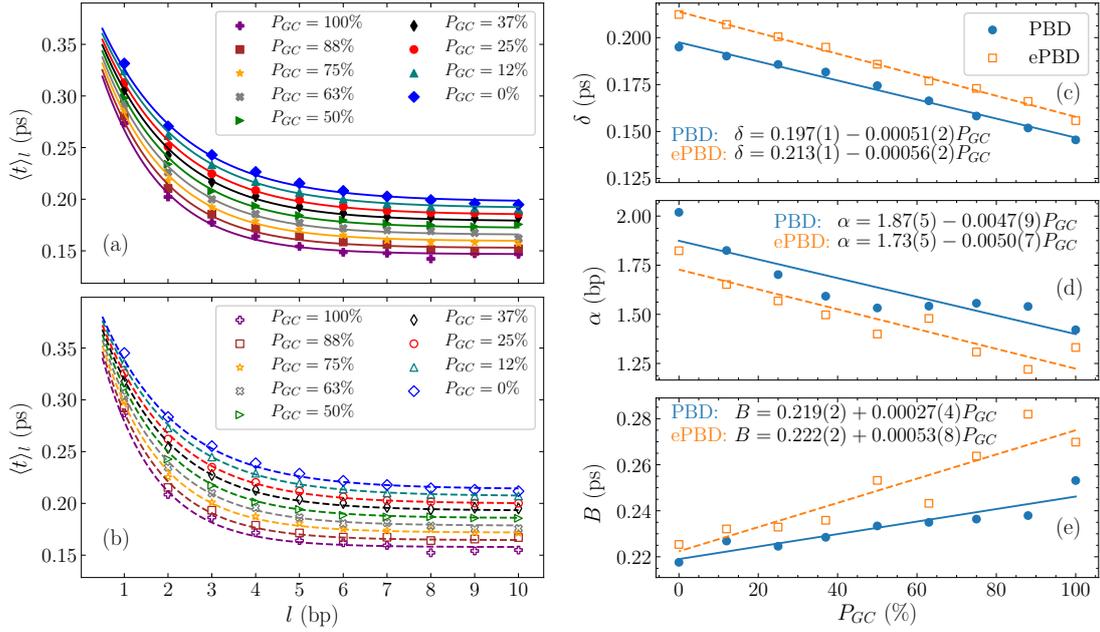

FIGURE 5.10: The numerical mean bubble lifetimes $\langle t \rangle_l$ for various GC percentage values $P_{GC}$ (5.8) as a function of the bubble length $l$, for (a) the PBD model, and (b) the ePBD model. Larger GC percentages correspond to shorter mean lifetimes, and shorter bubbles exhibit longer lifetimes. The decrease in mean lifetime with increasing bubble length is fitted by a simple exponential (5.9). These functions are shown by solid curves in (a) and dashed curves in (b). The parameters of this fit are shown in panels (c)–(e) as a function of the $P_{GC}$ value, the asymptotic limiting value $\delta$, the characteristic length $\alpha$, and the prefactor $B$ respectively. Results for the PBD model are shown by solid blue circles, and for the ePBD model by empty orange squares. All of these quantities vary more or less linearly with the GC percentage, and the fits are shown by solid blue lines (PBD) and dashed orange lines (ePBD), with the equations of the lines shown with respect to the bubble length $l$ in each panel.

The exponential decrease in mean lifetime $\langle t \rangle_l$ can be fitted accurately by the simple functional form

$$\langle t \rangle_l = B \exp(-l/\alpha) + \delta, \qquad (5.9)$$

with $B$, $\alpha$ and $\delta$ free fitted parameters. The fits of the data to this equation are shown by solid curves in Fig. 5.10(a) and dashed curves in Fig. 5.10(b), verifying that this form is an appropriate function to model the behaviour.

Figures 5.10(c)–(e) give the variation of the fitting parameters with GC percentage, the limiting constant $\delta$ in Fig. 5.10(c), the characteristic length $\alpha$ in Fig. 5.10(d), and then the prefactor $B$ in Fig. 5.10(e). In all cases we see a more or less linear change with GC percentage. For $\delta$ [Fig. 5.10(c)], we have a very clear linear decrease as the GC percentage increases, for both the PBD model (blue circles) and the ePBD model (empty orange squares). The ePBD model gives a consistently larger limiting value, which confirms our observation of the differences between the results in Figs. 5.10(a) and (b). The equation of the line of best fit is given in the panel. The characteristic length $\alpha$ [Fig. 5.10(d)] shows a less clearly linear behaviour, but nevertheless using a straight line fit gives a reasonable approximation. This time the PBD model exhibits higher values, with the slope once again fairly similar for the two models. Finally the prefactor $B$ [Fig. 5.10(e)] is also fitted with a linear function (the equation is also given in the panel), with the ePBD model giving higher values, as well as a steeper slope. The values of $B$ for small values of $P_{GC}$ are fairly similar, but the discrepancy is much larger for GC rich sequences.

## 5.4.1 Bubble Lifetimes in the AdMLP

A natural step after finding these general bubble lifetime distributions and properties is to compare them with results from particular biological promoters, in the same way as we explored whether the mLE is different for these promoters in Section 4.3.1. Here we will focus on the AdMLP, since it has been the focus of many previous studies (starting from [50]), and there is an established theme of large, long-lived bubbles forming at transcriptionally active sites [50]. Consequently, it will be instructive to see if there are differences in the bubble lifetime distributions inherent to the promoter sequence, as well as the average lifetimes.

A direct comparison of the lifetime distributions $P_l(t)$ are presented for the ePBD model in Fig. 5.11, showing the distributions for the AdMLP (empty circles in Fig. 5.11), as well as for sequences with exactly the same base pair composition (corresponding to $P_{GC} = 57/86 \approx 66\%$), but ordered randomly (solid curves in Fig. 5.11). From these results we clearly see that there are no visible major differences in the behaviour of the promoter. There are small indications, particularly in the early stages of the $l = 1$ distribution in Fig. 5.11, where the AdMLP exhibits a slightly stronger second peak around 0.25ps than the general sequences. For the longer bubbles however, there is no clear discrepancy. While these results are only presented for the ePBD model, very similar trends are seen in the PBD model, with the small differences even less apparent, likely due to the lack of the sequence-sensitivity of the PBD model.

The mean bubble lifetime calculated according to (5.8) can also be computed to compare the promoter with general sequences. This result is presented in Fig. 5.12 for both models. We see that using the PBD model, there is virtually no difference between the calculated values at all bubble lengths for the AdMLP (purple crosses in Fig. 5.12) and the randomly ordered sequences (yellow stars), as all points practically overlap. The ePBD model though does predict a slightly different behaviour, especially for large bubbles. For bubbles of length $l > 7$, the green triangles of the AdMLP are consistently above the red diamonds of the random sequences. These are small differences, but as the uncertainties of the mean value computation are very small (less than 1%), this adds more significance to the slightly different behaviours of the bubble lifetime distributions in Fig. 5.11. It is evident that while the observed trend was minor, the consistency is suffi-cient to affect the mean lifetime noticeably. It is also consistent with the previous studies that the AdMLP exhibits longer-lived bubbles, even if our results are not focussing on any particular regions of the promoter, where anomalously long-lived large bubbles have been found [50].

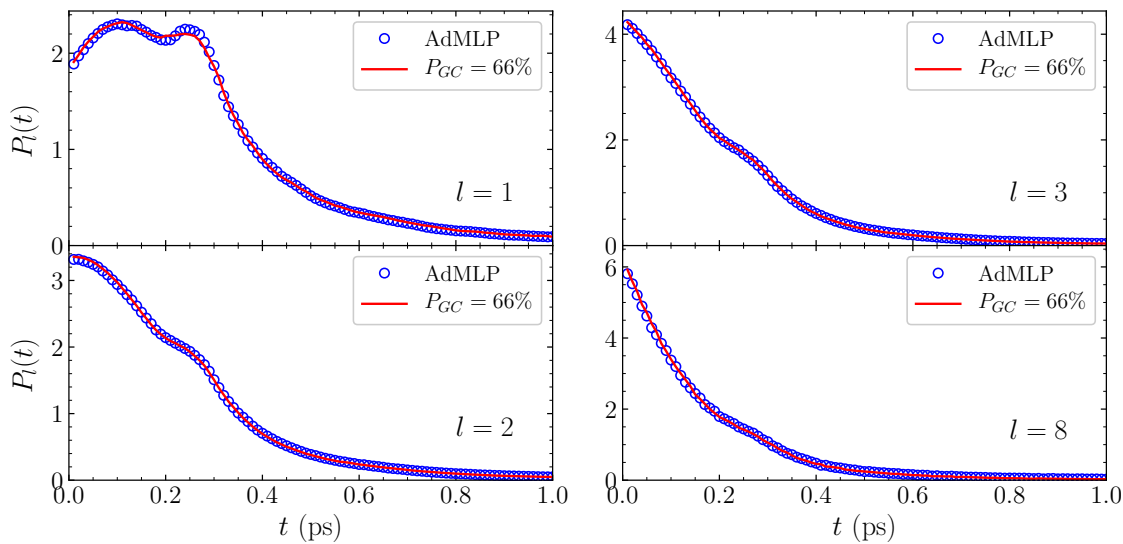

FIGURE 5.11: A comparison between the bubble lifetime distributions $P_l(t)$ of the AdMLP (shown by blue empty circles) and randomly ordered sequences comprised of the same base pairs (shown by the red solid line). Results are shown for four bubble lengths, $l = 1$, $l = 2$, $l = 3$ and $l = 8$ using the ePBD model (results for the PBD model are very similar and are not presented here). There is little discernible difference between the two cases, with very similar behaviour evident at all lengths and lifetimes.

These results are however very encouraging, as they suggest that the average distributions we have found can be meaningfully applied to specific sequences, without fear that in particular cases our results are not applicable. This allows the general results of this section to be used as a benchmark for the expected bubble lifetime behaviour of specific sequences, meaning that investigations into the transcriptionally significant regions of the sequence can be confidently compared to the general distribution in order to quantify how unusual the observed behaviours are. Using these bubble lifetime distributions is a promising future line of research to understand anomalous behaviour of functional DNA regions.

## 5.5 Conclusion

In this chapter we have studied properties of bubbles in DNA strands by introducing base pair dependent threshold values based on the characteristic length of the Morse potential, that satisfies the definition of the melting transition occurring when 50% of base pairs in the strand are open. Using these thresholds, we have investigated the energy-temperature relationship in DNA, distributions of bubble lengths and distributions of bubble lifetimes through extensive numerical simulations. In all aspects, both the PBD model and the sequence-dependent ePBD model were studied.

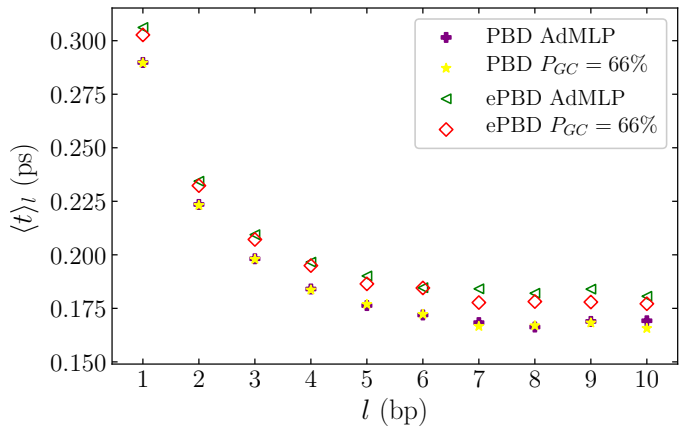

FIGURE 5.12: The mean bubble lifetime $\langle t \rangle_l$ (5.8) for the AdMLP and randomly generated sequences with the same base pairs in random arrangements. The results are shown for the PBD model (AdMLP: purple crosses, random sequences: yellow stars) and the ePBD model (AdMLP: green triangles, random sequences: red diamonds).

Using the derived threshold values, an estimation for the average melting temperature of the ePBD model was presented, which fits well with experimental understanding. A nonlinear relationship (5.2) was found between the temperature $T$ (calculated through the averaged kinetic energy) and the energy density $E_n$ of the system, where a cubic correction to the low-temperature relation $E_n = K_B T$ produces an accurate approximation. Similar results are found for the PBD and ePBD models.

The distributions of the lengths of bubbles in DNA was found (Fig. 5.5), which are approximated well by stretched exponential functions (5.3). Due to the considered heterogeneous thresholds, the probability for short bubbles to occur is similar for any base pair composition, but for long bubbles ($l > 4$bp), AT rich sequences exhibit substantially higher numbers of bubbles. In general the ePBD model predicts more bubbles than the PBD model, and this effect is again strongest in AT rich sequences.

Similarly, the distribution of bubble lifetimes was computed for bubbles of lengths up to ten base pairs (Fig. 5.7 and 5.8). The bubble lifetime distributions are also well fitted by stretched exponentials (5.5), although for long bubbles ($l \gtrsim 6$bp) the stretching is minimal and the function is essentially a pure exponential. The average bubble lifetimes (5.8) decrease with bubble length, and higher AT content leads to longer lived bubbles (Fig. 5.10). Again, the ePBD model predicts longer living bubbles for all lengths and GC percentages.

Finally, we compared the results of simulations using a well-known biological promoter, the AdMLP, to sequences with the same GC percentage but randomly arranged base pairs (Figs. 5.11 and 5.12). We found that the PBD model predicts virtually no differences between the two cases (AdMLP and random sequences), while the ePBD model predicts very slightly longer bubble lifetimes for the AdMLP than the random sequences. In both cases, the bubble lifetime distributions of the AdMLP are very similar to the random sequences.

Further investigations into the bubble lifetime distributions would be very interesting, particularly looking more at the distributions of bubble lengths and lifetimes at particular sites of biological promoters. It would also be instructive to investigate the ef-fect of the threshold values more closely; for instance to see if more widely open bubbles (with a larger threshold) exhibit longer lifetimes in general. Studying properties across the temperature spectrum is another option that could yield useful results, looking at near-melting effects or low temperature properties. For example, the existence of slow "glassy" dynamics suggested in [52] could be probed by detailed examination of bubble probabilities and lifetimes at lower temperatures.

# Chapter 6

# Dynamics of Graphene

This chapter represents a slight change of tack, moving away from the DNA models studied in the previous chapters, and presenting an investigation into the chaotic dynamics of a planar model of graphene. First we address the necessary background of graphene as a material as well as the computational machinery required to study a complex 2D system numerically, including scaling efficiencies for CPU and GPU parallel integration. Thereafter we look at results from the computation of the mLE for an "infinite" sheet of graphene, modelled as a graphene sheet with periodic boundary conditions to imitate the bulk behaviour of the material. Further, we will compute the mLE for finite width graphene nanoribbons (GNRs) with both zigzag and armchair edges. A short discussion of various simplifications of the model is also presented, briefly investigating the influence of the 2D hexagonal geometry on the chaoticity of the system.

The results in this chapter are based around the research presented in [123].

## 6.1 Structure of Graphene

### 6.1.1 Background

Since bursting onto the scene in 2004 [2], graphene has attracted a spectacular amount of attention due to its remarkable properties. Graphene in itself is a surprisingly simple material – just a single 2D layer of carbon atoms connected in a hexagonal lattice, covalently bonded together. As far as the bonding details go, each carbon atom forms an interatomic $\sigma$ bond (the strongest covalent bond, formed by the end-to-end overlapping of atomic orbitals) with each of its three neighbouring atoms, as well as a single $\pi$ bond (a weaker covalent bond, formed by lateral overlap of atomic orbitals) perpendicular to the plane [124]. The $\sigma$ bonds are the primary cause of the exceptional structural strength of graphene (and in fact the strength of this honeycomb structure in general) [124], while the $\pi$ bonds are the primary contributors to the electronic properties of graphene that present it as a candidate for so many technological innovations [125].

Among these applications are high energy supercapacitors [126], efficient transistors [127], electromechanical resonators [128] as well as advanced sensors [129]. These engineering applications are possible in large part due to the extraordinarily high electron mobility [130, 131] and the similarly exceptional thermal conductivity of graphene [132]. Further applications of graphene to structural engineering are highly promising due to its extremely high intrinsic strength, with a tensile strength measured at 130 GPa and a Young's modulus of 1TPa [133]. This strength translates into very high critical buckling strains under tension and compression [134], meaning that graphene has potential for extensive uses in structure development and reinforcement.

The dynamical aspects of graphene are of particular relevance to this work, and indeed MD simulations have been used extensively to investigate its properties. For instance, MD simulations have been used to study the interface interactions between graphene and different metals, studying the cohesive energy and bonding strength from dynamical considerations, which provides valuable information for the development of engineering applications [135]. MD simulations have also been useful in the further analysis of the thermal conductivity of graphene [136], and in particular to study the effects of the size of the sheet considered. A significant advantage of these numerical techniques is the ability to consider differently sized GNRs, thin strips of graphene, without having to painstakingly create an accurate sample in a laboratory. These investigations have revealed that the ribbon length plays a major role in the thermal conductivity, and also that applying strain to the ribbon can decrease the conductivity [136]. In addition to the conventional 2D forms of graphene, the material has further scope for physical applications in various conformations through controllable defect engineering, such as carbon nanotubes, extremely strong cylindrical forms of graphene [137], nanocages formed by careful chemical folding of the graphene sheet [138] and indeed a host of possibilities [139].

All these properties and avenues for further development present graphene as a significant part of future technologies. Understanding its stability and dynamical properties is consequently a very important part of the process of identifying effective use cases and searching the depths of its great potential.

## 6.1.2   A Dynamical Model of Graphene

Compared to the problem of modelling DNA, graphene presents a number of new difficulties. Foremost is that graphene is not only a two dimensional material, but the hexagonal lattice structure requires some detailed consideration to model accurately. Thus the potential governing the interatomic interactions needs to account for both the covalent bonding between atoms (not dissimilar to the bonding between the bases in DNA base pairs), but it also needs to firmly preserve the angles forming the hexagons. One simplification over DNA modelling is that in graphene all atoms have the same equilibrium distance, and disorder is only introduced through changing the masses of atoms (excluding the considerably more complex task of modelling defects at an MD scale, which is not considered here).

To achieve accurate modelling, a number of approaches have been used, such as MD simulations using empirical force fields [140] and using Monte Carlo methods with these force fields [141]. Tight-binding models have also been used with some success, particularly in studying the nonlinear stress response and elasticity of graphene [142]. Other methods used to study the stability and strength of graphene include continuum mechanics [143] and *ab initio* calculations [144]. All these methods have met with reasonable success, accurately reproducing many of the experimentally determined properties of graphene.

In this work, we consider an in-plane model derived from first principle density functional theory (DFT) calculations, where the empirically determined force fields are fitted with analytical expressions to produce a relatively simple atomistic Hamiltonian model of graphene [145]. This model splits the molecular potential into two parts: An interatomic potential to model the covalent bonds, and an angle potential to account for the effects of bending in the lattice, which maintains the hexagonal shape. Among the advantages of this approach is that accurate estimations of the empirical potentials can be found by careful exploration of the DFT calculations, and consequently the analytical expressions for the potential can be closely fitted, resulting in a very effective model.

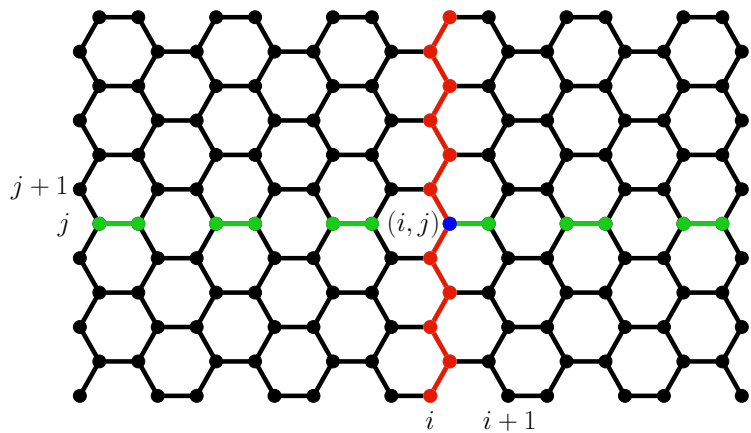

FIGURE 6.1: The hexagonal structure of graphene. Each of the atoms (represented by dots) is at equilibrium separated by $r_0 = 1.42$Å from each neighbour. The rows are labelled as shown by the bright green line (marking row $j$), while the zigzag columns are as shown by the red line (marking column $i$). The intersection of these is the point $(i, j)$, marked in blue.

Before progressing further with the discussion of the model, it is useful to first define the lattice we use. Figure 6.1 illustrates the honeycomb structure of graphene, and introduces the labelling convention for rows and columns in the model. One can think of an $N_x \times N_y$ graphene lattice as being constructed by taking $N_x$ zigzag columns with a length of $N_y$ atoms and placing them together into the honeycomb structure. This leads naturally to the $x$ (horizontal) index being given by the number of the zigzag column, counting from the left – as shown in Fig. 6.1, marked as $i$ and coloured in red. The $y$ (vertical) index is then given by the number of the atom along the zigzag column counting from the bottom, marked with $j$, coloured green. The origin of the indexing system is taken at the bottom left corner, with increasing indices up and to the right.

With this structure, we notice that there are two distinct cases for the possible neighbours of each atom. Some atoms, such as the blue atom $(i, j)$ in Fig. 6.1, have a neighbouring atom to the right as well as above and below. All atoms have a top and bottom neighbour, but depending on where along the zigzag chain a particular atom is, it can have either a left or a right neighbour. These two possibilities are illustrated in Fig. 6.2, where the case with a left neighbour is shown in Fig. 6.2(a) and the right neighbour in Fig. 6.2(b). So the blue atom in Fig. 6.1 is an example of case (b) in Fig. 6.2. Now recalling that the left bottom corner atom in Fig. 6.1 is indexed as $(0, 0)$, we see that this atom is an example of case (a), noting that we are using periodic boundary conditions here. Moving either up one atom or to the right by one atom (remember that the $x$ axis is indexed by zigzag columns), we move from case (a) to case (b). In general, by looking at the points in Fig. 6.1, we find that every point that has $i + j$ even has a left neighbour [case (a)], and every point with $i + j$ odd has a right neighbour [case (b)]. We will use this even/odd notation throughout to refer to the two cases.

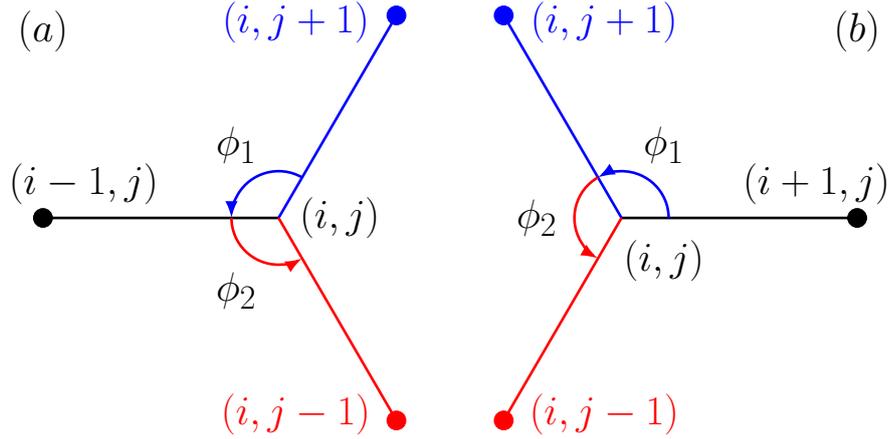

FIGURE 6.2: A schematic representation showing the two possible cases for atoms in the lattice – either having a neighbour to the left or to the right. The first case (a) occurs when the sum of the indices $i + j$ is even, and case (b) when $i + j$ is odd. Using our angle labelling convention, the angle $\phi_1$ in case (a) would be written as $_{i,j}\phi_{i,j+1}^{i-1,j}$.

The equilibrium interatomic distance in graphene is known to be $r_0 = 0.142$nm, or $1.42$Å [145], and the equilibrium angles necessary to create the hexagonal lattice are all $\phi_0 = 2\pi/3$ rad. These allow in principle the straightforward mathematical construction of the lattice through some trigonometry, with the only difficulty being accounting for the odd/even disparity. Adjusting for these two cases, we can write the position vectors for each atom in the lattice as

$$r_{i,j} = \begin{cases} \begin{pmatrix} (i-1)r_0\left[\cos\left(\frac{\phi_0}{2}\right)+1\right] \\ (j-1)r_0\sin\left(\frac{\phi_0}{2}\right) \end{pmatrix}, & i+j \text{ even} \\[2em] \begin{pmatrix} (i-1)r_0\left[\cos\left(\frac{\phi_0}{2}\right)+1\right]+r_0\cos\left(\frac{\phi_0}{2}\right) \\ (j-1)r_0\sin\left(\frac{\phi_0}{2}\right) \end{pmatrix}, & i+j \text{ odd.} \end{cases} \tag{6.1}$$

As one more required piece of labelling, we need a way of referring to each angle in the lattice. Each atom has three associated angles, formed between itself and its three nearest neighbours. Consider the angle labelled $\phi_1$ in Fig. 6.2(a). This angle is formed between the points $(i, j + 1)$, $(i − 1, j)$ and $(i, j)$, and centred at $(i, j)$. So we would like to unambiguously refer to this angle, while clearly maintaining the fact that it is centred at $(i, j)$. To this end we will define the slightly cumbersome but concise angle notation $_{i,j}\phi_{k,l}^{m,n}$ for the angle formed between the points $A = (m, n)$, $B = (i, j)$ and $C = (k, l)$, where moving around the central point in a counterclockwise direction the points are ordered $ABC$. Thus the angle $_{i,j}\phi_{k,l}^{m,n} = A\hat{B}C$, in the conventional geometric labelling. With this method, we label the angle $\phi_1$ from Fig. 6.2(a) with $_{i,j}\phi_{i,j+1}^{i-1,j}$. Note that our counterclockwise convention is satisfied here; starting from $(i, j)$ and moving counterclockwise through the points, the second point is indeed $(i, j + 1)$, and the third point $(i - 1, j)$. Following this through, the angle $\phi_2$ from Fig. 6.2(a) will be labelled $_{i,j}\phi_{i-1,j}^{i,j-1}$, and from Fig. 6.2(b) the angles $\phi_1 = {}_{i,j}\phi_{i+1,j}^{i,j+1}$ and $\phi_2 = {}_{i,j}\phi_{i,j+1}^{i,j-1}$.

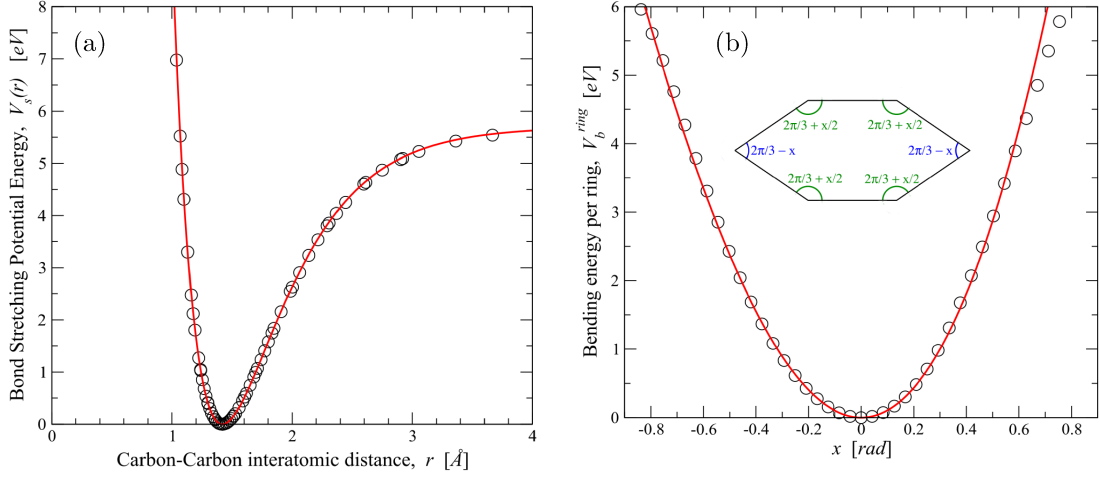

FIGURE 6.3: (a) The stretching potential energy plotted as a function of the distance between carbon atoms in a graphene lattice. The circles show the values from first principles DFT calculations, and the solid curve is the fitting of the Morse potential (6.2) to the data. We see that the Morse potential is a good choice for modelling the potential energy, and the fit is very close to the numerical data. (b) The bending potential energy plotted as a function of the displaced angle $x$ in a graphene lattice. Here the solid curve is the fitting of the bending potential of (6.3). The inset shows the displaced angle $x$ with reference to each of the angles in a hexagonal cell of the lattice. Figure taken from [145].

Now that we have set out the lattice, we can define the potential used to model the in-plane dynamics. In this model, the natural units of distance are Ångströms, Å, and time is measured in ps. Thus all reported displacements are measured in Å, and velocities in Å/ps, unless specifically given in other units for a particular result.

The first component, the stretching force due to interatomic covalent bonds, is modelled by a Morse potential, similarly to how the PBD model of DNA accounts for the bonds between bases. This potential is in the form

$$V_s\left(r_{i,j}^{k,l}\right) = D\left(e^{-a(r_{i,j}^{k,l} - r_0)} - 1\right)^2,\qquad(6.2)$$

where $r_{i,j}^{k,l}$ is the distance between the points $(i, j)$ and $(k, l)$, $r_{i,j}^{k,l} = \|r_{k,l} - r_{i,j}\|$ with $\|\cdot\|$ denoting the usual Euclidean norm of the vector, and the three parameters $D$, $a$ and $r_0$ are found through fitting the analytical expression to the DFT calculations. This fit is shown in Fig. 6.3(a), which is reproduced from [145]. We can see that the fitted expression gives an excellent representation of the potential energy, allowing us to use this formula with confidence. Additionally, the fitting procedure produces the known equilibrium distance $r_0 = 1.42$Å, as it should. The Morse well depth is found to be $D = 5.7$eV from the fittings, and the characteristic width scaling $a = 1.96$Å$^{-1}$.

The second part of the potential, accounting for the forces arising from bending in the lattice, is given by a quadratic potential well for the angle, with a cubic correction [145]:

$$V_b\left(_{i,j}\phi_{k,l}^{m,n}\right) = \frac{d}{2}\left[_{i,j}\phi_{k,l}^{m,n} - \phi_0\right]^2 - \frac{d'}{3}\left[_{i,j}\phi_{k,l}^{m,n} - \phi_0\right]^3, \qquad (6.3)$$

This expression has the distinct advantage of being a simple polynomial, but comes at the cost of being expressed in terms of the angles themselves, rather than a convenient cartesian form. Nevertheless, the parameters $d$ and $d'$ can be found through the fitting of DFT calculations, and as seen in Fig. 6.3(b) this form of the potential accurately captures the behaviour of the bending response. The optimised values are $d = 7.0\text{eV/rad}^2$ and $d' = 4.7\text{eV/rad}^3$.

So we now have expressions for each contribution to the overall potential, and need only to combine all these together to form the complete Hamiltonian. First considering the Morse potential components, we can see that summing over every point and adding up the contribution of all three bonds connected to the point will result in an overcounting by a factor of two – every bond will have its energy counted by the atoms on either end. The second consideration we need to make with this sum is to decide for each atom whether the Morse potential needs to be added for the neighbouring atom on the right or on the left. So for all points we will have contributions from the bonds going up and down, but we have to discriminate on a case by case basis depending on the parity of $i + j$ for the left or right bonds.

For the bending potential, we have to make the same even/odd distinction, since the potential of (6.3) needs to be evaluated with the correct three angles for each point. As such, the final potential will be split into two sums: One over all atoms with $i + j$ even, and one for $i + j$ odd.

With all this in mind, we can now write the Hamiltonian by adding in the kinetic energy, and then the two sums over the potential terms, ensuring to pass the correct angles as arguments to the bending potential $V_b$. Thus the expression of the Hamiltonian is

$$\begin{aligned}
H = &\sum_{i,j} \frac{m_{i,j}}{2}\left[\left(\dot{x}_{i,j}\right)^2 + \left(\dot{y}_{i,j}\right)^2\right] \\
&+ \sum_{i+j \text{ even}} \left\{\frac{1}{2}\left[V_s\left(r_{i,j}^{i,j+1}\right) + V_s\left(r_{i,j}^{i,j-1}\right) + V_s\left(r_{i,j}^{i-1,j}\right)\right]\right. \\
&\qquad\qquad \left. + V_b\left(_{i,j}\phi_{i,j+1}^{i-1,j}\right) + V_b\left(_{i,j}\phi_{i-1,j}^{i,j-1}\right) + V_b\left(_{i,j}\phi_{i,j-1}^{i,j+1}\right)\right\} \\
&+ \sum_{i+j \text{ odd}} \left\{\frac{1}{2}\left[V_s\left(r_{i,j}^{i,j+1}\right) + V_s\left(r_{i,j}^{i,j-1}\right) + V_s\left(r_{i,j}^{i+1,j}\right)\right]\right. \\
&\qquad\qquad \left. + V_b\left(_{i,j}\phi_{i+1,j}^{i,j+1}\right) + V_b\left(_{i,j}\phi_{i,j-1}^{i+1,j}\right) + V_b\left(_{i,j}\phi_{i,j+1}^{i,j-1}\right)\right\},
\end{aligned}$$

$$(6.4)$$

where we have written in terms of the velocities $\dot{x}$ and $\dot{y}$, with $p_x = m\dot{x}$ and $p_y = m\dot{y}$. Note that the mass $m$ is indexed as well, to allow for doping with othe carbon isotopes such as $^{13}\text{C}$ atoms. Unless otherwise stated, we use $^{12}\text{C}$ atoms, with the mass taken as 12u, but where we use $^{13}\text{C}$ atoms the mass is 13u. In the two sums, we have included the appropriate stretching terms – up and down neighbours in both cases, and then the left neighbour when $i + j$ is even, and the right neighbour when $i + j$ is odd. The bending potential terms are computed for the three angles that occur in each case, and here there is no double counting.

Having set up the Hamiltonian, we are now in principle ready to compute the equations of motion, and potentially even the variational equations, and proceed with numerical simulations. While $H$ (6.4) has a fairly involved form, it is at least possible to express it concisely, and the equations of motion take the usual form

$$m_{i,j}\ddot{x}_{i,j} = -\frac{\partial H}{\partial x_{i,j}}, \qquad m_{i,j}\ddot{y}_{i,j} = -\frac{\partial H}{\partial y_{i,j}}. \qquad (6.5)$$

The differentiation of the Morse potential $V_s$ (6.2) is straightforward enough, and these terms can be computed easily. While the bending potential $V_b$ (6.3) poses more difficulties due to the explicit angular dependence, the equations of motion can nevertheless be computed exactly, and these have been presented in [146].

## 6.2   Computational Considerations

Simulating the graphene lattice dynamics is a rather computationally challenging task for two main reasons:

1. The complexity of the equations to integrate. The equations of motion themselves involve performing a large number of operations, and either using the TM or the 2PM approaches will exacerbate the integration time requirements.

2. The relatively large number of atoms to be integrated in a 2D lattice (at least several thousand in most cases for reasonable results).

The integration is carried out using the fourth-order ABA864 symplectic integration method [22], which has been found to be one of the most efficient integrators for lattice integration [147].

In order to tackle the complexity of the task, we first consider the choice between the 2PM and the TM method for computing the mLE. For most models, the variational equations are an efficient way of estimating the mLE (see discussion in Section 1.3). In this case though, the variational equations are extremely complex (see [146]), and in fact the 2PM method is both simpler to implement and more efficient. As such we made use of the 2PM for all mLE computations, following the method outlined in Section 1.3.1. This leaves us only integrating the equations of motion (albeit twice), but does require a smaller time step in order to maintain the requisite accuracy for the small deviation.

Since in general for the 2PM, we require a deviation that is somehow small, but does not go beyond machine precision, as discussed in Section 1.3.1 a lower limit on the norm of the deviation vector has been suggested to be about $10^{-8}$ [37]. In our application, the limit of $10^{-8}$ is in fact somewhat smaller than can be accurately computed due to the large number of sites in the 2D lattice. For lattices of several thousand atoms such as we will want to use, the deviation vector will have on the order of ten thousand components (since it has components in the $x$, $y$, $p_x$ and $p_y$ directions). So more realistically, for the components to remain larger than $10^{-16}$ in magnitude, a deviation vector norm of $10^{-6}$ is necessary, although for small lattices a smaller deviation vector can be used.

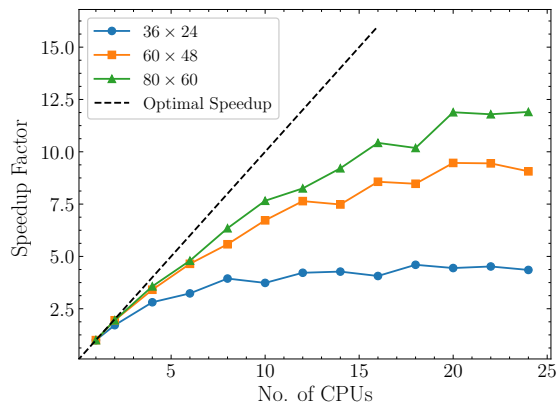

FIGURE 6.4: The total speedup factor as a function of number of CPU cores, simulating the graphene lattice up to a time of $10^4$ps, integrating two nearby initial conditions and computing the ftmLE. Three lattice sizes are shown here, $N_x = 36, N_y = 24$ ($N = N_x \cdot N_y = 864$) (blue circles), $N_x = 60, N_y = 48$ ($N = 2880$) (orange squares), and $N_x = 80, N_y = 60$ ($N = 4800$) (green triangles). The dashed straight line shows the perfect speedup rate. We see that the speedup drops off quite noticeably after 4 CPU cores for the 36×24 cases, and although for the larger lattice even up to 10 cores still provides a distinct improvement, after 4 cores the speedup becomes less efficient. The speedup past 12 CPUs is quite minimal in all cases, although for the 80×60 lattice there is still slow improvement.

It is also significant that even at the lowest energies considered in this investigation, the average fluctuations in atom positions and velocities are on the order of $10^{-2}$ (where positions are measured in Å and velocities in Å/ps). So taking the deviation norm of $10^{-6}$, if we have a lattice of more than a thousand atoms, as is typically the case, then the individual components of the deviation vector will be on the order of $10^{-8}$, which is very clearly smaller than the typical fluctuations, providing further assurance that this deviation is indeed small on the scale of the natural dynamics of the lattice.

The issue of a large number of atoms with complex equations of motion to be simulated is perfectly suited to parallel computing, since the lattice can be broken up into blocks of atoms that are integrated in parallel. This is possible due to the fact that changes in the position and momentum of a particular atom only affect its neighbours in the next iteration of the integration, allowing each integration step to be performed for each atom independently (see Section 1.2.2). By parallelising the integration, we can substantially speed up the integration process, especially as the lattice size increases.

In order to find a near-optimal number of CPU cores to use for the integration procedure, we performed scaling tests with varying numbers of CPU cores comparing the total wall time (real time elapsed, rather than CPU hours) taken to integrate the lattice up to a common final time. These tests were performed with three different lattice sizes, to get a better grasp of what is optimal for different sizes, especially since we expect larger lattices to be more amenable to parallelisation speedup. Figure 6.4 shows the speedup obtained (where a speedup factor of 2 corresponds to halving the compute time, and a factor of 3 to a reduction to one third, and so on) by using multiple cores to parallelise the integration, with lattices of $N_x \times N_y = 36 \times 24 = 864$ atoms (blue circles), $60 \times 48 = 2880$ atoms (orange squares) and $80 \times 60 = 4800$ atoms (green triangles). As we increase the number of cores used, there is at first a significant speedup observed in all cases. The addition of a second core is almost perfectly efficient, with a speedup factor of near 2. This keeps the point near the line of optimal speedup, which is when we have a speedup factor equal to the number of cores used. Going to 4 cores is not quite as efficient, but is nevertheless a strong improvement, with a speedup of close to perfect for the two larger lattices, with the $36 \times 24$ lattice (blue points) starting to show a decrease in efficiency. For this lattice, using more than 4 cores seems to be fairly wasteful, as the speedup drops off dramatically after this point. For the larger lattices however, there is still merit in going beyond 4 cores, even if the efficiency decreases. Up to around 10 cores, the speedup from each extra core is noticeable, and a speedup by a factor of 7 or 8 is definitely useful. Past this point though, while there is some speedup it is highly inefficient.

These results suggest that using 4 cores for most simulations will yield fairly efficient speedup, dramatically reducing the time needed to perform the computations, but not imposing a large overhead cost on the parallelisation. We are also then able to run 6 simulations simultaneously on each node.

Another aspect of the scaling to consider is a closer look at how the compute time scales with size, for a fixed number of CPU cores, and to compare this with using a GPU. As a first step, we will present some comparative results using CPU and GPU results, primarily to confirm that the GPU implementation is in fact producing correct results. Two main results are presented in Fig. 6.5(a) and (b), the relative energy error, and the estimation of the ftmLE. In both cases we see that the two computation methods produce near-identical results, with some small variations inevitable in the simulation of a chaotic system. Importantly, the energy remains conserved up to a small constant error. This is a strong indication that for both the CPU and GPU implementation the equations of motion are being accurately integrated. The estimation of the ftmLE is also identical up to the final time of $10^5$ps considered here, where the value has saturated to the final mLE. In both these measurements there are slight differences inevitable in the simulation of a chaotic system, but are effectively identical. These results allow us to continue with the GPU code in confidence that the results being produced are consistent with the CPU results which have previously been carefully tested.

In Fig. 6.5(c), we see the effect on the total simulation time of increasing the size of the lattice. Results are shown for integration with 4 and with 24 CPU cores (green triangles and orange squares respectively), as well as the GPU implementation (blue circles). From the results of Fig. 6.4, we have already seen that larger lattices benefit more from the use of many CPU cores. In the CPU results of Fig. 6.5(c), this is borne out by the much faster increase in compute time required when using only 4 CPUs (green triangle) when compared to the 24 CPU integration (orange squares). We also see that very quickly the 24 CPU parallelisation becomes faster than the 4 CPU case, requiring around a third of the time to integrate a lattice of 4096 atoms. The GPU parallelisation [blue circles in Fig. 6.5(c)] is clearly substantially faster when it comes to integrating very large lattices, and is even faster to integrate the 4096 atom lattice. For very small lattices it is hard to compare the times, but since the GPU takes the same time to integrate all lattices smaller than 5120, it is not at all efficient for integrating lattices of only a few thousand atoms.

It is however unfair to just compare the net compute time, for a number of reasons. Most obviously, the comparison between the 4 and 24 CPU cases ignores the fact that we can run 6 instances of the simulation at once when using only 4 CPUs, while we can only run a single instance with 24 CPUs. Similarly, the GPU is a much more intensive processor than a single CPU, and comparing even 24 CPUs in parallel with a single high-powered GPU is not necessarily a reasonable comparison. Relating CPU hours to GPU hours is thus a slightly tricky issue. We would then like to answer the question of "what is the most efficient way to integrate the lattice at different sizes?" by some definite measure. One logical such measure is to use the CPU hour - GPU hour relationship implemented by the computing cluster, which equates 1 GPU hour to 40 CPU hours [25]. Thus we can normalise the compute times to be in terms of CPU hours, which yields a much fairer estimation of the compute cost for performing many simulations. Of course if only one or two simulations need to be performed, then the results of Fig. 6.5(c) will apply more pertinently.

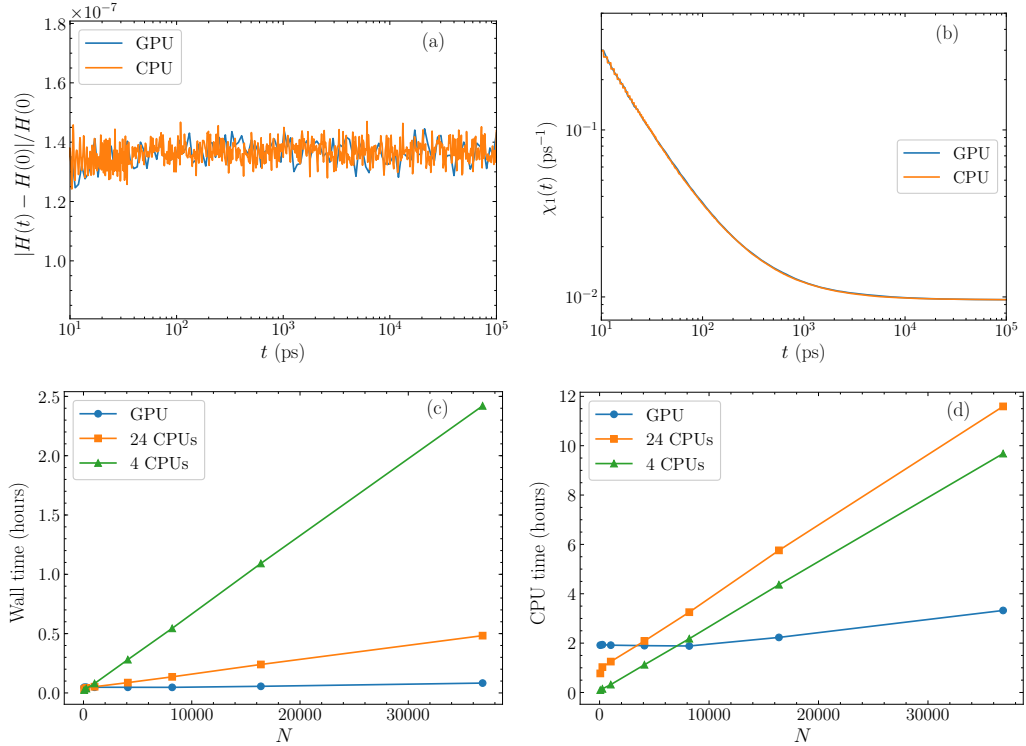

FIGURE 6.5: (a) The comparison of the time evolution of the relative energy error $|H(t) - H(0)|/H(0)$ for identical initial conditions in a lattice of $N_x \times N_y = 64 \times 64 = 4096$ atoms, computed using the CPU and the GPU code. We see that in both cases the error is consistently conserved around $10^{-7}$, as required. (b) The time evolution of the ftmLE $\chi_1(t)$ for both computation methods, for the same initial condition. Again we see that they both produce the same results, confirming that the GPU mLE computation algorithm is also correctly implemented. (c) The total compute time required to integrate lattices of size $N$ atoms up to a final time of $t = 10^4$ps, using a GPU (blue circles), 24 CPU cores (orange squares) and 4 CPU cores (green triangles). The lines connect the points to guide the eye. From this plot it is clear that for lattices of more than 4000 atoms the GPU requires the least time to integrate the system, and the 24 CPUs are faster than 4 CPUs. (d) The rescaled CPU time (see text) required to integrate the system, for the same three cases as (c). Now it is clear that the 4 CPU integration is more efficient for small lattices, of fewer than 8000 atoms. After this point the GPU is more efficient despite each GPU hour being equivalent in cost to 40 CPU hours (see text). As the lattice size increases, the GPU has by far the smallest gradient, indicating its strong performance for very large lattices.

These rescaled results are shown in Fig. 6.5(d). With this more practical approach, the 4 CPU case demonstrates much more efficient performance for lattices of fewer than 8000 atoms. The 24 CPU parallelisation is never more efficient than either of the alternatives, illustrating the problem with simply scaling up CPU multithreading. Despite the increase by a factor of 40 though, the GPU still massively outperforms the CPU code for atoms of more than 10000 atoms. This is consistent with the intended design of the GPU, which is to be efficient in performing highly parallelisable computations for very large systems of equations, which is exactly what we have with lattices of tens of thousands of atoms. As such, for any very large lattice simulations the GPU code is undoubtably the best choice, but for small lattice computations which are generally sufficient for our purposes, using around 4 CPUs is the most efficient option.

From this section, we can take away several conclusions. For most lattice sizes, using 4 or 6 CPU cores yields an efficient speedup, providing a good balance between reducing run times and not wasting CPU time on inefficient parallelisation. Simulating the system using GPU parallelisation is very effective for large lattices, and would be an efficient choice for investigations using lattices of more than around 5000 atoms. Further optimisation of the GPU code would also allow for even better efficiency in the integration of large lattices; for instance careful profiling to find inefficient computation blocks, and further minimisation of GPU-CPU data transfers. For our application however, where we do not need to consider such very large lattices, the best choice of parallelisation method is to combine OpenMP multithreading with GNU Parallel to run several parallelised CPU integrations alongside each other.

## 6.2.1   Initial Results

Having set up the graphene model and outlined the numerical processes, we can proceed with some first results of simulating the system. While the focus of this investigation is the chaotic dynamics of the lattice, in order to confirm the validity of the used model and our implementation of the equations of motion, we will reproduce some known results for the dynamical behaviour of graphene.

In all these results, we impose periodic boundary conditions on all edges to eliminate edge effects. The simulations are run using $N_x = 60$ atoms in the $x$ direction and $N_y = 48$ atoms in the $y$ direction, but changing the lattice size has no effect on the considered quantities, when using periodic boundaries. For each energy density $E_n$ (total energy divided by the number of atoms), we use ten different setups with random initial conditions (random excitations of the momentum in both the $x$ and $y$ directions, rescaled to give the required total energy value), run for 5ns to thermalise, and then data collected every 10ps for the next 5ns.

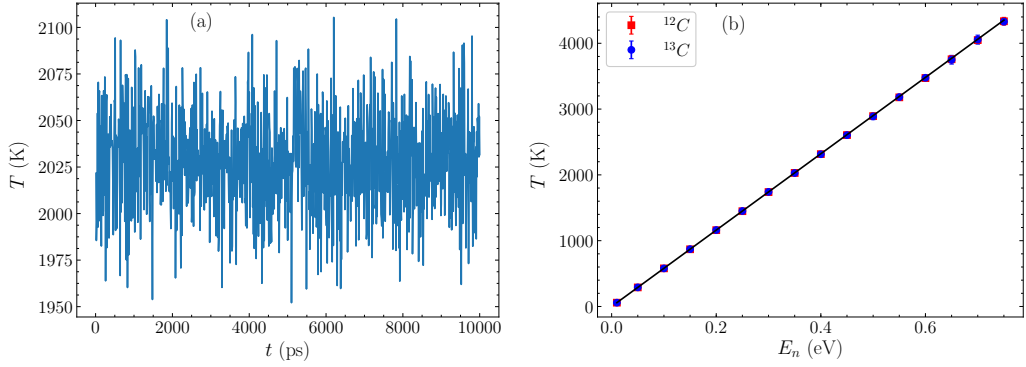

FIGURE 6.6: (a) A single case of the evolution of the lattice temperature in time, for energy density $E_n = 0.35\text{eV}$. The temperature quickly settles down to oscillating around a value of 2030K. (b) The relationship between temperature $T$ and energy density $E_n$ for the graphene model. The straight line shows the linear function of (6.6), confirming that the relationship holds not only at low temperatures, but even extending far beyond commonly experienced temperatures. The blue circles show results for the lattice using a mass of 12u ($^{12}$C), while the red squares show the results for a graphene sheet comprised entirely of $^{13}$C atoms. We see that for both masses the same energy-temperature relation is found, as the points overlap and are indistinguishable.

The first result we consider is the energy density-temperature relationship. Since we are still working in the microcanonical ensemble with constant energy, we estimate the temperature $T$ of the 2D system according to

$$T = \frac{K}{k_B N},$$ (6.6)

with $K$ being the total kinetic energy, the system having a total of $N$ atoms and $k_B$ being the Boltzmann constant.

Figure 6.6(a) shows the estimated temperature through time for a single run, at an energy density of $E_n = 0.35\text{eV}$. The energy spreads through the lattice very quickly, and the temperature oscillates fairly closely around a constant value. The initial thermalisation time of 5ns is clearly more than enough for the temperature computation, as we can see that even by 10ps (0.01ns) the temperature has settled to small oscillations. Averaging over the last fifty data points, we can find an average temperature measurement from this run.

By computing this for several initial conditions (ten in this case), and averaging over all the data points, we can numerically find the relationship between the energy density $E_n$ and temperature $T$. These results are shown by the blue circles in Fig. 6.6(b). There we see a purely linear relationship, extending far beyond the low-temperature region (as was seen in the DNA in Fig. 5.4). This straight line is in fact given by the low-temperature gradient for a 2D system of

$$E_n = 2k_B T,$$ (6.7)

where $k_B$ is the Boltzmann constant. The fact that no correction is needed to this low-temperature relation even when the temperatures reach extreme levels above 3000K bears testament to the stability of the graphene lattice. It is also very likely that the restriction of the model to in-plane motion adds to the stability of these high-temperature results, as the magnitude of out-of-plane motion is typically much greater at high temperatures.

In order to test the model for doping with $^{13}$C atoms, following the same procedure we also compute the temperature for a graphene lattice comprised entirely of $^{13}$C [red squares in Fig. 6.6(b)] as the most extreme case of this doping. We see that for this too the linear relation is maintained across the full energy density spectrum, with no discernible difference between the energy-temperature relation of the two isotopes.

Apart from the energy-temperature, we can look at the distributions of interatomic distances and atom speeds. For both of these quantities, we have a theoretical expectation for their distributions, which will allow us to support both the physicality and the accuracy of our computations. First, the distribution of the interatomic distances in the lattice should be normal [148]. So after equilibration of the lattice, we should find that the distances $r$ between neighbouring atoms are distributed according to

$$P(r) = \frac{1}{\sigma\sqrt{2\pi}} e^{-\frac{1}{2}\left(\frac{r-\mu}{\sigma}\right)^2}, \qquad (6.8)$$

where $\mu$ is the mean value (expected to be equal to $r_0 = 1.42$Å), and $\sigma$ is the standard deviation.

Figure 6.7(a) shows the distribution of distances $r$, using five different initial configurations and 500 snapshots from each run. The distributions are shown for five energy densities, both to check that the Gaussian form is maintained, and also to see how the shape of the curve changes. In each case, the symbols show the distributions from the simulations, and the solid curves show a fitting with (6.8). The normal distribution fits the data exactly for all energies, and correctly finds a mean value of $\mu = r_0 = 1.42$Å. The lowest energy, $E_n = 0.01$eV (blue circles), shows the narrowest distribution commensurate with the smaller displacements expected at low energies, and the fit yields a correspondingly small standard deviation. As the energy increases, the distribution flattens, but maintaining the mean value of $\mu = 1.42$Å.

For the particle speeds, we have a firm theoretical prediction, in the form of the Maxwell-Boltzmann distribution [148]. We expect that the speeds should be distributed according to the distribution

$$P(v) = \frac{mv}{k_B T} e^{-\frac{mv^2}{2k_B T}}, \qquad (6.9)$$

since we have a 2D lattice. This is of course fixed for a given temperature (or in light of the linear energy-temperature relation [(6.7) and Fig. 6.6(b)], a given energy density), meaning that we should find the atom speed distributions to follow (6.9) without any need for fitting. In Fig. 6.7(b) we see the theoretical prediction (solid curves) along with the numerical results obtained from our simulations (symbols) for several energy densities. In all cases, the speed distributions are exactly as they should be, following the Maxwell-Boltzmann distribution precisely. This serves as another confirmation of our temperature results, as fitting (6.9) to the distributions would be an alternative method of computing the temperature for a given energy density.

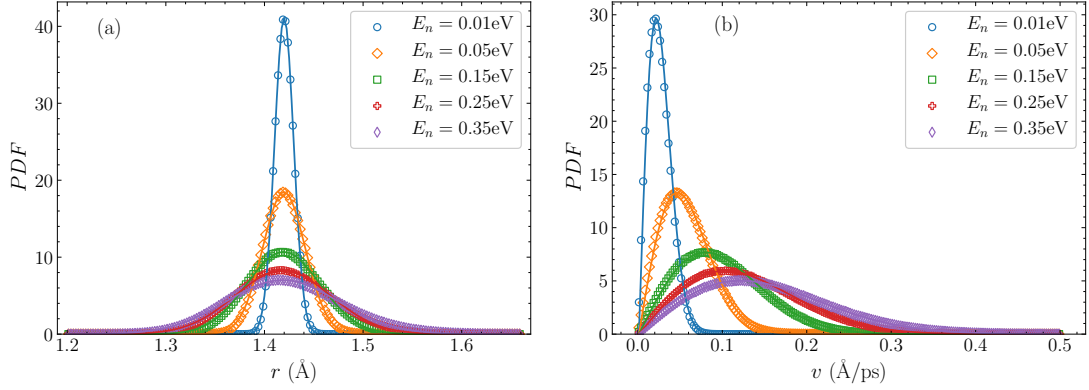

FIGURE 6.7: The distribution of (a) the distances $r$ between atoms, and (b) the particle speeds $v$ in the graphene lattice. For each case the results are shown at five energy densities, $E_n = 0.01$eV (blue), $E_n = 0.05$eV (orange), $E_n = 0.15$eV (green), $E_n = 0.25$eV (red) and $E_n = 0.35$eV (purple). We can see that the distances are normally distributed around the equilibrium distance $r_0 = 1.42$Å, with the Gaussian function of (6.8) (solid curves) fitting well in each case. As the temperature increases the distances spread more widely, consistent with the overall more active dynamics. The velocities also show the expected distribution, fitting closely to the Maxwell-Boltzmann distribution, (6.9). Like the distances, the velocity distribution is broader for higher temperatures.

## 6.3   Chaotic Dynamics of Graphene

With these computational methods and validated results, we are able to estimate the mLE of graphene lattices of different varieties, and investigate graphene's stability and chaotic dynamics. Here we will study graphene in two forms: A periodic sheet to approximate the dynamics of an infinite lattice, and finite width GNRs, with one pair of edges left free while the other two are joined with periodic boundary conditions.

### 6.3.1   Periodic Graphene Sheets

As our first task, we would like to understand the overall chaotic dynamics of graphene, without considering boundary effects or finite size issues. To this end, we will simulate an infinite graphene lattice by taking a reasonably large graphene lattice and applying periodic boundary conditions to all the sides. The resultant "torus" will allow for even spreading of energy, and should provide an indication of the behaviour of an infinite lattice.

The immediate question we need to answer is what constitutes a reasonably large lattice. In order to quantify this, we performed several simulations with varying lattice sizes, and computed the mLE at each size. Once we reach a point where the mLE no longer changes with lattice size, this will give a strong suggestion that we have reached a sufficiently large lattice size that the effects due to the fact that the sheet is not truly infinite are negligible.

Figure 6.8(a) shows the evolution of the ftmLE at an energy density of $E_n = 0.5$eV, for several lattice sizes: $N_x \times N_y = 24 \times 16 = 384$ (solid blue line), $36 \times 24 = 864$ (dashed

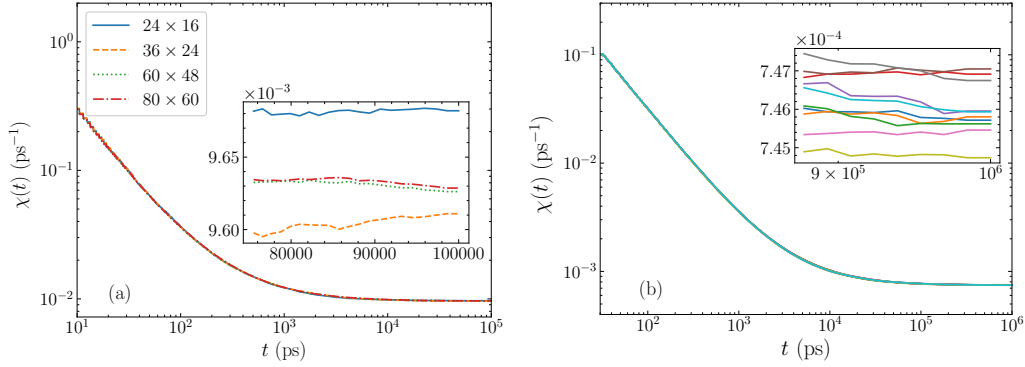

FIGURE 6.8: (a) The evolution of the ftmLE $\chi(t)$ for a periodic graphene lattice, for four lattice sizes, $N_x \times N_y = 24 \times 16 = 384$ (solid blue curve), $36 \times 24 = 864$ (dashed orange curve), $60 \times 48 = 2880$ (dotted green curve) and $80 \times 60 = 4800$ (dash-dotted red curve) at energy density $E_n = 0.5\text{eV}$. The inset shows the last few time steps of the evolution magnified. We see that for all lattices, the final mLE value is the same up to a precision of around $5 \times 10^{-5}$, but the $24 \times 16$ lattice systemically yields a slightly higher value. The $60 \times 48$ and $80 \times 60$ lattices however give extremely similar results. These results also illustrate that for $E_n = 0.5\text{eV}$, the ftmLE has saturated to its final value by a time of $10^5\text{ps}$. (b) The time evolution of the ftmLE $\chi(t)$ for a periodic graphene lattice, at an energy density of $E_n = 0.05\text{eV}$ for a lattice of size $N_x \times N_y = 60 \times 48 = 2880$ atoms. The integration of ten initial conditions are shown, exhibiting near-identical behaviour. We see that by a time of $10^6\text{ps}$ the value has saturated to a constant, as confirmed by the inset which shows a magnification of the last stage of the evolution.

orange line), $60 \times 48 = 2880$ (dotted green line) and $80 \times 60 = 4800$ (dash-dotted red line). It is clear in this plot that all four of these cases give extremely similar ftmLEs, with the evolution near identical at all times. The final value of the ftmLE saturates at the same value up to a spreading of $1 \times 10^{-4}\text{ps}^{-1}$, as seen in the inset. The blue curve of the smallest lattice size, $24 \times 16$, does exhibit a slightly larger final saturated value for the mLE, which we have also observed in other cases, suggesting that this size is likely too small to be reliable as a good estimator of the dynamics of the infinite lattice. The $36 \times 24$ lattice, given by the dashed orange curve, gives an mLE value very slightly below the two larger lattices, but within a range of $2 \times 10^{-5}\text{ps}^{-1}$ in a measurement on the scale of $10^{-2}\text{ps}^{-1}$, which is negligible in the final analysis. However, the two largest lattices, $60 \times 48$ (green dotted curve) and $80 \times 60$ (red dash-dotted curve) are almost identical even in the magnified image presented in the inset. This inspires confidence that the effects of the finite lattice size have been neutralised by the time we have a lattice of $N_x \times N_y = 60 \times 48$ atoms, and most likely even before this. Since the mLE values no longer change even at a tiny scale past this point, we will use this lattice size of $60 \times 48$ atoms in all our periodic lattice computations.

With this chosen lattice size, we are able to build up a coherent picture of the relationship between the mLE and the energy density in much the same way as with the DNA model. Taking the final points of the ftmLE as an estimate of the actual mLE for that initial condition, we can average over multiple simulations to construct an average mLE for a given energy density.

To be sure that all simulations have completely converged to the final mLE value, we have integrated most cases up to a final time of $10^5$ ps [as in Fig. 6.8(a)], as this gives a clear saturation. For particularly small energies ($E_n = 0.01$ eV and $E_n = 0.05$ eV), we have run the simulations to a longer time of $10^6$ ps, as these cases take longer to thermalise and for the ftmLE to saturate. In Fig. 6.8(b), we see the evolution of the ftmLE up to a time of $10^6$ ps for $E_n = 0.05$ eV, for ten different randomly generated initial conditions. All cases give almost precisely the same results, and have clearly saturated by the final time. The inset in Fig. 6.8(b) shows the last stage of the ftmLE evolution for all the considered cases, clearly displaying that the ftmLE has ceased to decrease and has stabilised at the final value of the mLE. We also note that the choice of initial conditions for the lattice has no actual bearing on the mLE value, with different types of initial excitation trialled resulting in identical final ftmLE values.

To study the overall dynamics of the system now, we can take exactly the same approach as for the DNA (see Section 4.2), and collate average mLE values from multiple runs to build up a picture of the energy dependence of the system's chaoticity. To this end, we will perform the same computations shown in Fig. 6.8(b) for multiple energy densities $E_n$, and average over the ten runs to find an average mLE value for each $E_n$ value. We will use the final value of the ftmLE, averaged over the ten runs, with an errorbar given by the standard deviation of these data as an estimation of the system's mLE $X_1$.

Figure 6.9 presents the average mLE $X_1$ across the energy spectrum for the periodic graphene lattice. The computation of the mLE was performed for both a lattice of purely $^{12}$C atoms, shown by red squares in Fig. 6.9 and $^{13}$C atoms, shown by blue circles. We see that in both cases the mLE increases monotonically with the energy density, with the $^{13}$C lattice showing slightly more stable behaviour, arising from the higher mass of each atom. Unlike the DNA, here we see no changing in the behaviour of the mLE, the increase with energy (or equivalently temperature, since the relationship between the two quantities is linear) is smoothly monotonic. This implies that there is no change in the structural stability in the graphene shell in the range of energy densities considered here, which cover temperatures up to around 4000K. The extraordinary stability of graphene is once again illustrated here, with even extreme conditions unable to provoke any irregularity in its dynamics (although investigating the effects of out-of-plane atomic motion would be interesting here).

At low energies the mLE is more or less linearly dependent on the energy density $E_n$, curving slightly upwards as it moves past $E_n \approx 0.1$ eV. The numerical data of Fig. 6.9(a) can be fitted accurately with a simple quadratic function,

$$X_1(E_n) = \beta E_n + \gamma E_n^2. \tag{6.10}$$

where $\beta$ and $\gamma$ are free constants to be determined. Performing this fitting for $^{12}$C we find $\beta = 0.01447 \pm 0.00005$ ps$^{-1}$eV$^{-1}$, and $\gamma = 0.00951 \pm 0.00008$ ps$^{-1}$eV$^{-2}$. Similarly, for $^{13}$C we find that $\beta = 0.01389 \pm 0.00004$ ps$^{-1}$eV$^{-1}$, and $\gamma = 0.00921 \pm 0.00007$ ps$^{-1}$eV$^{-2}$. Due to the similarity of the mLE values, these fitted values are fairly close together, once again demonstrating the minor effect that $^{13}$C doping has on the chaotic dynamics and stability of the lattice. The small variance in each coefficient also reflects the closeness of the quadratic fit.

Considering the Lyapunov time $T_L = 1/X_1$ of the system (see Section 1.3), also pro-

vides some useful insight into the process of chaotisation in graphene. Estimating $T_L$ by inverting the mLE values in Fig. 6.9(a) will give us the time (in picoseconds) by which the system has reached a chaotic state. What we are really interested in however is how this time relates to the natural time scales of the lattice. This will allow us to understand the chaotisation time in the specific context of the graphene sheet.

In the lattice, one clear time scale is specified by the frequency of the normal modes, and in particular the highest frequency of the normal modes. The normal modes of the model used here have been computed through phonon dispersion relations [149], and we can use the highest frequency thus obtained, the transverse optical (TO) mode for the time scale of the system. Measured in units of cm$^{-1}$, using wavenumber notation, the frequency of this mode is around $\omega_W \approx 2100 \text{cm}^{-1}$. In order to convert this to an actual frequency in Hz, we rescale $\omega = 2\pi 100 c \omega_W$, where $c$ is the speed of light. With this rescaling, we find a frequency of $\omega \approx 396 \text{THz}$, and a characteristic time $\tau \approx 2.5 \times 10^{-3} \text{ps}$ for each oscillation of the system from inverting this frequency.

So now, instead of just presenting the Lyapunov time $T_L$, we are able to present the rescaled Lyapunov time

$$\widetilde{T}_L = T_L/\tau, \tag{6.11}$$

which gives us the number of oscillations of the highest frequency mode before chaos sets in. Figure 6.9(b) shows $\widetilde{T}_L$ as a function of the energy density $E_n$, demonstrating the very large number of normal mode oscillations needed before chaotisation occurs. At the lowest energies, more than $10^5$ oscillations are required before the system becomes fully chaotic, with the room temperature case of $E_n = 0.05\text{eV}$ taking around $5 \times 10^5$ oscillations. As the energy increases, chaotisation happens more and more quickly, but even at the extreme energies of $E_n \approx 0.6\text{eV}$ more than $10^4$ oscillations are needed. From the results of Fig. 6.9(b) we can say that in graphene, the onset of chaos is very slow, which is consistent with the material's stable behaviour.

### Linearisation of the Model

As a short aside, we will briefly consider the possibilities for simplifying the model, in order to better understand the source of chaos in graphene. Considering the potential functions of eqs. (6.2) and (6.3), there is evidently nonlinearity inherent in both of them. The question is then what the effects are of the 2D hexagonal geometry itself, as well as the actual functional form of the potentials. For instance, if we kill the bending potential $V_b$ (6.3) completely and linearise the Morse potential to a harmonic coupling, how does the mLE behave? And what effect does the quadratic part of the bending potential $V_b$ have; since it is quadratic in the angles, it is certainly nonlinear in the coordinates $x$ and $y$, but it is not clear how this will interact with the interatomic coupling.

In order to investigate these effects, we performed simulations with three variations of the full nonlinear potential. These three cases are as follows:

1. Only the harmonic interatomic coupling:

$$V_s = k(r - r_0)^2. \tag{6.12}$$

2. Only a harmonic interatomic coupling with a quadratic angular potential:

$$V_s = k(r - r_0)^2, V_b = k'(\phi - \phi_0)^2. \tag{6.13}$$

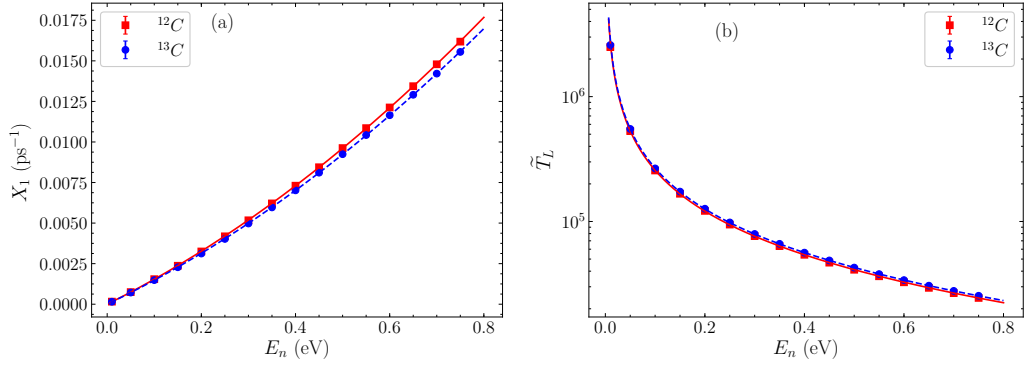

FIGURE 6.9: (a) The average mLE $X_1$ as a function of the energy density $E_n$, for the periodic graphene lattice. Results are shown for a lattice composed solely of $^{12}$C atoms (red squares) and $^{13}$C atoms (blue circles). The error-bars are given by the standard deviation of the mLE from each run, but since the values are extremely close the errors are not visble on the scale of the plot. The solid red and dashed blue curves show a fit to (6.10) for $^{12}$C and $^{13}$C respectively. (b) The rescaled Lyapunov time $\widetilde{T}_L$ (6.11) shown in a semilogarithmic plot on the $y$ axis, counting the number of oscillations of the fastest normal mode of the lattice before chaotisation occurs. The fitted curves are given by (6.10) rescaled like the data.

3. A full linearisation of the equations of motion, by first order approximation with a Taylor expansion. To obtain the true linearisation, we consider the equations of motion of the Hamiltonian (6.4) in vector form. If the positions of the atoms in the lattice are given by a vector

$$\boldsymbol{u} = \left( x_{1,1}, x_{1,2}, \ldots, x_{N_x,1} \ldots, x_{N_x,N_y}, y_{1,1}, y_{1,2}, \ldots, y_{N_x,1} \ldots, y_{N_x,N_y} \right),$$

then the equations of motion are given by the accelerations of the atoms,

$$m \ddot{u}_i = -\frac{\partial H}{\partial u_i},$$

which can be compressed into a vector form as

$$m \ddot{\boldsymbol{u}} = -\frac{\partial H}{\partial \boldsymbol{u}}. \tag{6.14}$$

We note that this expression corresponds to the equations of motion in (6.5). In this form, we are able to perform a Taylor expansion around the equilibrium position $\boldsymbol{u}_0$. Keeping only the term in first order of $\boldsymbol{u} - \boldsymbol{u}_0$, we have

$$m \ddot{\boldsymbol{u}} = -\frac{\partial H}{\partial \boldsymbol{u}}(\boldsymbol{u}_0) - \frac{\partial^2 H}{\partial \boldsymbol{u}^2}(\boldsymbol{u}_0)(\boldsymbol{u} - \boldsymbol{u}_0). \tag{6.15}$$

Evaluated at the equilibrium positions, all the values in the vector $\partial H / \partial u_i$ are zero, and the matrix of elastic coefficients is given by

$$Q_{i,j} = \frac{\partial^2 H}{\partial u_i \partial u_j}. \tag{6.16}$$

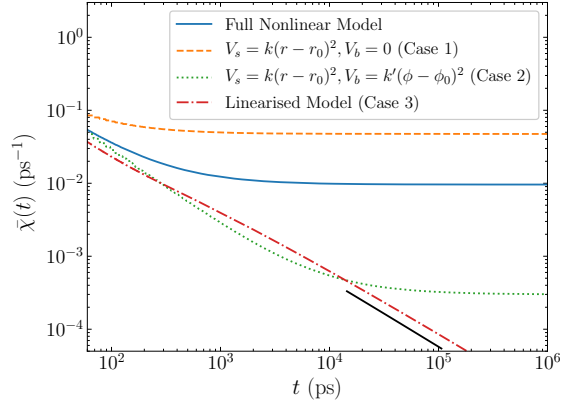

FIGURE 6.10: The ftmLE averaged over ten initial conditions, for four differ-
ent variations of the graphene model at $E_n = 0.5$eV. The solid blue curve
depicts the original nonlinear model. The dashed orange curve shows the
result using only a harmonic coupling between atoms, the dotted green
curve using a harmonic coupling and a quadratic angular potential, and
the red dash-dotted curve the full linearisation of the model. The solid
black line shows a slope of $\ln(t)/t$.

We thus have the final linearised equations of motion as

$$\ddot{\boldsymbol{u}} = -\frac{\boldsymbol{Q}}{m}(\boldsymbol{u} - \boldsymbol{u}_0). \tag{6.17}$$

For each of these cases, we can compute the system's mLE, using $k = D$ for the har-
monic potential and $k' = d/2$ for the angular potential. In Fig. 6.10, we see the time
evolution of an averaged ftmLE for the three modified versions of the model, as well as
the full model. Each curve corresponds to the average over ten simulations with random
initial conditions, at an energy density of $E_n = 0.5$eV. The ftmLE of the fully nonlinear
potential is shown as the solid blue curve in Fig. 6.10. Case 1 corresponding to only the
harmonic interatomic coupling is shown by the orange dashed curve, demonstrating a
markedly more chaotic behaviour than the full system. We find that the addition of a
quadratic angular potential (case 2 – green dotted curve) stabilises the system dramati-
cally, with the mLE of this potential being almost two orders of magnitude below the
full system. Finally, the result of the linearised equations of motion (case 3) is displayed
by the dash-dotted red line, demonstrating the expected continuous decrease towards
zero as in this case we have no chaos at all.

From these results, a number of interesting points arise. The simplest form of the
potential, the simple harmonic coupling (case 1), actually results in the most chaotic
behaviour. This is at first remarkable, as we expect that harmonic oscillators should
produce linear behaviour. However, due to the 2D geometry of the system, and possibly
exacerbated by the hexagonal structure, there is a large amount of motion transverse to
the direction of the oscillators. Including the full 2D range of motion in the potential,
without ignoring this transverse motion, results in a strongly nonlinear interaction.

The addition of the quadratic angular potential (case 2) results in a significantly more
stable system. At first this is counterintuitive as well, since this is the part of the poten-
tial we expected to be nonlinear. However, this potential $V_b$ acts as a potential well for each of the angles, which maintains the hexagonal shape much more strongly than simply allowing the atoms to move around constrained only by the springs (atom-atom interactions). The effect of this added rigidity is that the motion is restricted, and thus transverse displacements are a lot smaller and the system closer to a linear behaviour. The fact that case 2 is still nonlinear is due to both the small transverse motions as well as the inherently nonlinear behaviour of the angular potential.

Finally, the ftmLE of the completely linearised potential decreases towards zero as we expect, since in this case the system is actually linear. The slope of $\ln(t)/t$ is not as steep as the $1/t$ slope that is expected for a non-growing deviation vector, but this is just due to a linear growth of the deviation vector [29], which still meets the requirement for regular motion of a subexponential deviation growth (see Section 1.3).

### 6.3.2 Graphene Nanoribbons

Apart from the infinite sheet or bulk behaviour of graphene, we can also study the dynamics of a very physically relevant graphene structure – the GNR. A GNR is simply a narrow strip of graphene, with the transverse direction containing relatively few atoms, and the longitudinal direction containing a large number of atoms. In our simulations, we will approximate the very long ribbon using periodic boundary conditions once again in one direction, and the leave the ribbon edges free to move in the other direction. This of course leaves us both options as to which edges we leave free – the armchair or the zigzag edges. The choice of which edge is free describes the type of ribbon, and thus the labels "armchair" and "zigzag" GNRs refer to the free edge. Harking back to Fig. 6.1, we can see that by choosing the free edges on top and bottom, we would have an armchair GNR. If we instead had the left and right edges free, this would be a zigzag GNR.

In the same vein as with the periodic sheet, we can build up a plot of the average mLE of the GNRs by performing multiple runs at each energy density, and averaging the mLE values from each run to provide a single data point. Here however we have an additional parameter that we are concerned with, beyond the energy density. For the GNRs, we would like to study how the width affects the stability of the lattice, ranging from very narrow ribbons of only a few atoms to larger ribbons where the dynamics should start to approximate the infinite lattice. Based on the geometry of the hexagonal lattice (see Fig. 6.1), we can calculate the width of the nanoribbons using some trigonometry, yielding

$$W_A = \sqrt{3}\left(N_y - 1\right) r_0/2, \tag{6.18}$$

for the width of an armchair ribbon, while

$$W_Z = (3N_x/2 - 1) r_0, \tag{6.19}$$

for their zigzag counterparts. Here again $N_x$ refers to the number of atoms in the $x$ direction (the number of zigzag columns), and $N_y$ to the number of atoms in the vertical $y$ direction.

So now we can perform a series of computations, estimating the mLE for a) different energy densities, and b) different ribbon widths, for both GNR types. For these runs, we used periodic boundary conditions on the longitudinal direction, with a length of 60 atoms, as increasing the length beyond this was found to have no effect on the dynamics. The results of these runs are presented in Fig. 6.11, showing the mLE as a function of ribbon width for several energy densities: $E_n = 0.01$eV (blue circles), $E_n = 0.05$eV (orange diamonds), $E_n = 0.1$eV (green stars), $E_n = 0.15$eV (red triangles) and $E_n = 0.2$eV (purple crosses). Each data point is the average over ten runs with random initial conditions. Figure 6.11(a) shows the mLE for armchair ribbons. At all energy densities, there is a monotonic decrease towards the mLE of the periodic sheet (this value is shown by dashed lines for each energy density) as the width increases. For low energies, particularly the $E_n = 0.01$eV (blue points), the width has little perceptible effect on the mLE. For the larger energies, such as $E_n = 0.2$eV (purple crosses), the difference between the smallest and largest ribbons simulated is considerable, with the very narrowest ribbons exhibiting particularly chaotic behaviour. The results for zigzag GNRs are shown in Fig. 6.11(b), with generally similar trends to the armchair case. Here there is a slightly anomalous non-monotonic behaviour in the narrowest widths considered, but thereafter we see the same steady decrease towards the periodic sheet limiting value.

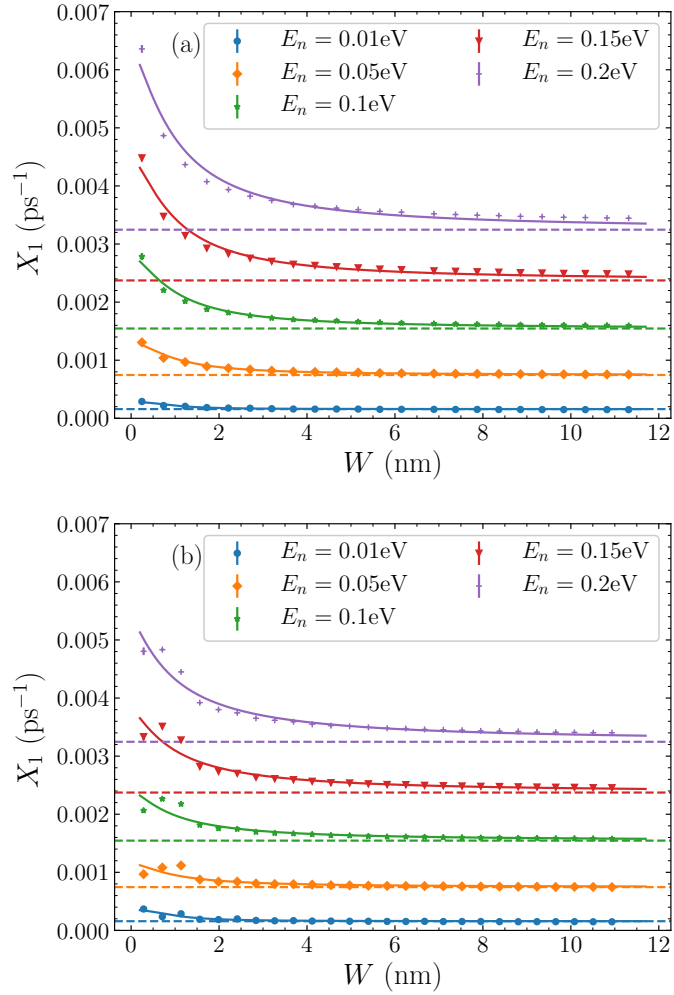

FIGURE 6.11: The average mLE of GNRs as a function of ribbon width $W$ in nm, for several energy densities. (a) Armchair ribbons, and (b) zigzag ribbons. The horizontal dashed lines indicate the value of the mLE of the periodic sheet at that energy density, which are shown in Fig. 6.9(a). The data for each $E_n$ value in both plots are fitted with (6.20), shown as solid curves.

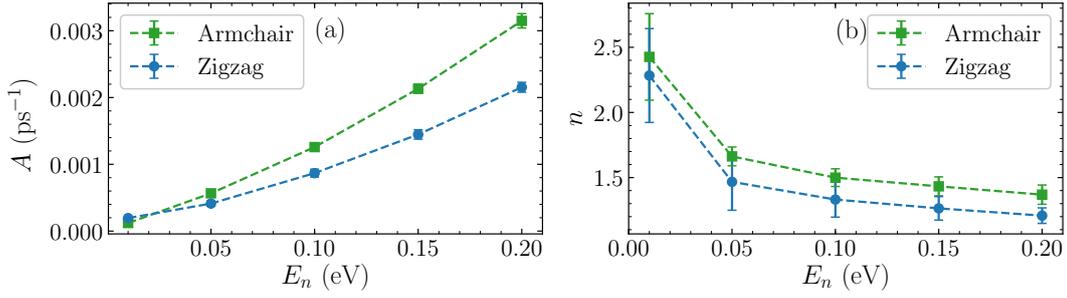

FIGURE 6.12: The parameters of (6.20) fitted to the data of Fig. 6.11, for arm-
chair (green squares) and zigzag (blue circles) nanoribbons across several
energy densities. (a) The coefficient $A$, and (b) the exponent $n$. The dashed
lines are to guide the eye.

For both GNR types, the mLE values can be fitted quite accurately with a decreasing
Hill function, with a constant limiting value added,

$$X_1(W) = \frac{A}{1 + W^n} + X_1^b. \tag{6.20}$$

This constant is simply given by the periodic sheet mLE value for the particular energy
density [see Fig. 6.9(a)]. As seen by the solid fitting curves in Fig. 6.11, this functional
form captures the rate of decrease effectively, allowing us to understand how the increas-
ing ribbon width affects the chaoticity of the lattice at different energies, and to compare
the behaviour of the two GNR types.

The computed value of the two parameters, $A$ and $n$, of the fitting are given in
Fig. 6.12. The scale coefficient $A$ is shown in Fig. 6.12(a), exhibiting a distinct increase
with energy density commensurate with the overall trend of increasing mLE values with
energy. This parameter $A$ is related to the range of values exhibited by the mLE, which
can be seen more clearly by rewriting (6.20) as $X_1(W) - X_1^b = A/(1 + W^n)$. Because of
this, the larger values of $A$ seen for the armchair GNRs are consistent with what we see in
Fig. 6.11, where the range of mLE values is larger for the armchair case [Fig. 6.12(a)] than
the zigzag ribbons [Fig.6.12(b)]. For instance, let us take the extreme case of $E_n = 0.2$eV
[purple crosses in Figs. 6.11(a) and (b)]. Because of the anomalous decrease in the mLE
for the smallest width zigzag ribbons [$W \approx 0.5$nm in Fig. 6.11(b)], the largest mLE
value is just below 0.005ps$^{-1}$, while the limiting value is approximately 0.0032ps$^{-1}$. This
difference of $\sim 0.0017$ps$^{-1}$ is substantially smaller than the corresponding difference of
around $\sim 0.0032$ps$^{-1}$ for the armchair case [Fig. 6.11(a)].

The exponent $n$ decreases with increasing energy density, showing that for higher
energies there is in fact a relatively slower decrease towards the limiting value. Once
again, the armchair ribbons have consistently larger values of $n$ compared to zigzag
GNRs, which demonstrates that the mLE of the armchair ribbons tends to decrease
more quickly towards the bulk value. Although this rate of decrease is systemically
faster in armchair GNRs, for both cases there is an eventual decrease to the asymptotic
value corresponding to the mLE of the infinite sheet.

## 6.4 Conclusion

In this chapter, in order to study the chaoticity of graphene we have used a Hamiltonian model which very effectively simulates the dynamics of the material while remaining relatively easy to use and computationally manageable. The unavoidable computational difficulties inherent in studying large two dimensional lattices, compounded by the distinctly non-trivial equations of motion arising from the geometrical complexity of the graphene model, were met using CPU multithreading and optimisation techniques, which allowed us to integrate a large number of atoms (typically on the order of 3000) in a reasonable time frame. A GPU implementation of our code was also written using CUDA, which drastically outperforms the CPU code for very large lattices, consisting of tens of thousands of atoms. Future work that involves integrating such large systems would definitely benefit from using the GPU implementation of the code rather than CPU parallelisation.

We have computed the mLE of various graphene structures using the 2PM, due to the extreme complexity of the variational equations. The mLE of an infinite graphene lattice was estimated by approximating this system through the use of periodic boundary conditions, and the mLE was found to increase quadratically with energy density [Fig.6.9(a)]. Replacing the $^{12}$C atoms in the lattice with $^{13}$C atoms results in a slightly more stable structure due to the higher masses, with a marginally smaller mLE value estimated. By inverting the mLE, Lyapunov times were also explored, and compared to the natural time scales of graphene lattices given by the highest frequency normal modes. While chaos sets in faster as energy density increases, the number of oscillations of the normal modes with the highest frequency required before the system is fully chaotic is over $10^4$ even up to temperatures of 4000K [Fig. 6.9(b)].

Some simple modifications of the model have been investigated, by approximating the nonlinear Morse potential for interatomic interactions with a simple harmonic coupling, as well as by using only a quadratic angular potential for the bending potential. We found that a hexagonal lattice with only harmonic interatomic coupling was more chaotic than even the full nonlinear system, while the addition of a quadratic angular potential to the harmonic coupling dramatically stabilises the system by discouraging motion transverse to the oscillators (Fig. 6.10). We have also confirmed that the full linearisation of the equations of motion using a Taylor approximation leads to regular behaviour.

The stability of GNRs was also studied using the mLE, with both edge types – armchair and zigzag – investigated. We found that the mLE decreases as the GNR width increases for both cases (Fig. 6.11), with both the range of mLE values and the rate of decrease being higher for armchair ribbons than zigzag, quantified by fitting to a decreasing Hill function (6.20). In both cases the mLE converges to the value of the infinite sheet asymptotically as the width increases.

There is plenty of scope to expand this work, and to use the model (6.4) for further investigations. One example would be studying the effect of stretching or compression of the lattice, as computing the mLE under these conditions could lead to insight about the stability of graphene. The addition of a term in the potential to account for out-of-plane deformations, as outlined in [149], would allow for a more robust analysis of high temperature regions, as the current model is likely overstating the stability of graphene in these extreme conditions.

Further investigations based on the results of this chapter would be to look at the effects of geometric anharmonism on the chaoticity of 2D (and potentially 3D) lattices. Our finding that the harmonic interatomic coupling yields strongly chaotic behaviour, and that the addition of a simple angular potential is capable of stabilising the system so much, raises questions about whether this is a particular property of the honeycomb lattice, and thus a contributing factor to graphene's extraordinary strength, or if it holds true in different 2D geometries. So performing estimations of the mLE and carrying out other stability computations across triangular, rectangular and hexagonal lattices could provide useful insights. Results obtained by using both harmonic and anharmonic couplings could also be compared, to give an understanding of when, counterintuitively, a nonlinear potential produces less chaotic results.

# Chapter 7

# Conclusions and Perspectives

This thesis has been an exploration of the use of methods of chaotic dynamics in biological and chemical systems. The aim was to provide both new insights into the systems themselves and a deeper understanding of how chaotic dynamics, and particularly the maximum Lyapunov Exponent (mLE), can be used to study physical models. We have explored the nonlinear behaviour of the Peyrard-Bishop-Dauxois (PBD) model of DNA, investigating the effects of heterogeneity and temperature on the chaoticity, quantified by the mLE. We have further probed the biological significance of DNA by studying the properties of bubbles in the double strand in detail. Beyond DNA, we have performed an extensive study of the dynamical stability of graphene, again using the mLE as a quantifier for both large graphene sheets and finite width graphene nanoribbons (GNRs). In combination, these studies have in turn led to a better grasp of the value that the mLE provides in the study of biochemical models. Here we summarise the salient findings from these research efforts, as well as pointing towards a number of exciting possible research avenues to be followed in the future.

We made use of symplectic integration methods as well as parallel computing techniques to numerically simulate the dynamics of the Hamiltonian systems, with an emphasis on computing the mLE as accurately and quickly as possible. The model of DNA which forms the core focus of this thesis, the PBD model, uses a relatively simple mesoscale potential approach to accurately approximate key dynamical features of DNA, particularly melting curves. This allows the efficient computation of macroscopic dynamical quantities, which is perfect for our application.

The investigation of DNA is presented in three distinct parts, each corresponding to a chapter of this thesis. In Chapter 3, we introduced the notion of the alternation index $\alpha$, which counts the number of times that the base pair type [adenine-thymine (AT) or guanine-cytosine (GC)] changes in the sequence. This value $\alpha$ quantifies the heterogeneity of a given sequence by describing how well-mixed the base pairs are, whether there is a homogeneous clumping of base pairs (small $\alpha$) or a heterogeneous mixture of base pairs in the strand (large $\alpha$). In order to understand which arrangements of base pairs are more or less likely, which would facilitate studying the effect of heterogeneity on the system's dynamics, we found a probability distribution for $\alpha$ given a particular number of AT and GC base pairs in a periodic DNA sequence. Although it was a challenging mathematical problem, we used Pólya counting theory and some tools from group theory to find this distribution, which enabled us to construct an algorithm for computing the number of possible arrangements of base pairs with a given number of alternations, and hence a complete distribution for all possible values of $\alpha$. While Monte Carlo simulations using around 20 000 runs yield a good approximation of the true distribution, the Pólya algorithm is more efficient even for short sequences, and is substantially faster for long sequences. These distributions now allow the investigation of high and low probability regions, as well as understanding how probable the extreme cases are. With an eye to future developments, while there is no obvious generalisation of this quantity to 2D lattices one could certainly find an analagous measurement to quantify the heterogeneity of a 2D lattice. The use of $\alpha$ and a 2D or 3D generalisation could be very informative for the study of heterogeneity in lattices.

The next phase of our investigation was to compute the mLE of the considered DNA models and study their chaotic dynamics (Chapter 4). Particularly, our interest lay in the heterogeneity; firstly the inherent heterogeneity arising from DNA sequences comprising of both AT and GC base pairs in various ratios (which can be quantified by considering the percentage of AT or GC base pairs in the strand), and secondly the heterogeneity within a set group of base pairs given by the clumping or mixing of base pairs (quantified by $\alpha$). The mLE was computed for sequences with varying AT and GC content, as well as across a spectrum of energy density (or equivalently, temperature) values. In all cases, the mLE increases with energy, as the DNA molecule destabilises. At low energies, a higher concentration of AT base pairs results in more chaotic behaviour, while GC-rich sequences are more chaotic at high energies. Moving past the melting point (and the physical validity of the model), we see that the mLE decreases to zero, behaving like a dynamical order parameter for the system. Looking at the effect of $\alpha$, we found that the more well-mixed heterogeneous sequences produce consistently more stable (i.e. less chaotic) dynamics. Particularly for sequences with a roughly equal number of AT and GC base pairs, more homogeneous molecules are substantially more chaotic than the typical sequences. Beyond the global information gleaned from the mLE, the local chaoticity was investigated by computing the deviation vector distribution (DVD) for a number of sequences. For all cases, the DVD tends to concentrate near regions of large displacement, avoiding areas of very little motion, but also not focussing at highly displaced sites. In particular, this means that generally the DVD avoids regions with a very high or very low probability of bubble occurrence. Results for the PBD model and the sequence-dependent ePBD model are found to be very similar with regard to their chaoticity.

The final piece of our tripartite study of DNA was a thorough investigation in Chapter 5 of the properties of thermally induced openings, called bubbles. As a starting point, we introduced physically motivated threshold values for how widely separated a base pair needs to be in order to be considered open, or part of a bubble. A DNA strand is defined to be melted when 50% of the base pairs are open, and using this criterion we could find very effective base pair-dependent thresholds, one for AT base pairs and one for GC. Using these thresholds, we found a remarkably precise melting temperature for the ePBD model, and established a relationship between the temperature and energy density for both the PBD and ePBD models. This relationship was fitted with a simple cubic polynomial. Following this, and by using a large number of molecular dynamics simulations to create reliable statistics, we found distributions for the lengths of bubbles in DNA sequences of varying base pair content at physiological temperatures. Using the base-pair-discrimating thresholds we introduced in our work means that probabilities for short bubbles (4 or fewer base pairs open) are similar for all se-quences regardless of AT/GC composition, but longer bubbles are more likely to occur in AT-rich sequences. The bubble length distribution was accurately fitted by a stretched exponential function, and the parameters provided in Section 5.3 (and also in [106]) allow researchers to estimate the probability of a bubble of a given length occurring in any desired sequence. Additionally, the ePBD model consistently predicts that more bubbles occur than in the PBD model, particularly for large bubbles in AT-rich sequences. The final, and most significant, result of our study was the determination of bubble lifetime distributions for the two models, again at physiological temperatures. For a variety of base pair compositions and bubble lengths, the lifetime distributions have been accurately computed. These distributions were also fitted with stretched exponential functions, which provide an excellent approximation for bubbles of two or more base pairs. Computing the mean bubble lifetime for each length reveals an exponential decrease in average lifetime as the bubble length increases, with longer lifetimes for sequences with more AT content. The same trends were observed for the PBD and ePBD models, with generally longer lifetimes predicted using the sequence-dependent ePBD version of the model. The general results found from simulating random sequences were compared to simulations of the well-studied AdMLP sequence. While the ePBD model predicted slightly longer lifetimes in the AdMLP, consistent with its known propensity for large long-lived bubbles, the distributions for random sequences and the AdMLP are very similar using both models, confirming that our results are applicable to specific promoters, which holds great promise for aiding future investigations.

In Chapter 6 we then considered a Hamiltonian model of graphene and studied the chaoticity of several graphene structures. Firstly, an infinite graphene sheet was modelled by applying periodic boundary conditions to all edges. There, in the same vein as the DNA, the mLE was computed for a spectrum of energy density values. In the graphene however, there is no phase transition, and the mLE continues to grow quadratically until the model loses meaning at extreme temperatures. Mild disorder was introduced in the graphene sheet by doping with $^{13}C$ atoms, but even considering a sheet comprised of purely $^{13}C$ atoms the mLE was only very slightly affected. The mLE was found to be very small (orders of magnitude smaller than the mLEs computed for the DNA) and consequently the related Lyapunov time (the inverted mLE) revealed that chaos takes a very long time to manifest in graphene when considered on the natural time scale of the system's fastest normal modes. Several simplifications of the nonlinear system were investigated, with the angular potential found to be a strong stabilising influence on the dynamics. Apart from this bulk behaviour, finite width GNRs were also studied, with the mLE computed as a function of ribbon width for several energies. For both zigzag and armchair ribbons the mLE was found to decrease as the width grows according to a decreasing Hill function towards the bulk value, with the armchair ribbons being slightly more chaotic at very narrow widths.

In all these investigations, and particularly the graphene simulations, computational optimisation played a fundamental role. The use of multithreading methods such as GNU Parallel and OpenMP, as well as optimisation techniques like code profiling, have enabled the pushing of numerical boundaries that have been previously impassable. The hundredfold improvement in computational time from the first implementation of the graphene equations of motion to the final version speaks to the value of devoting time to writing efficient code and optimising bottleneck routines. While this is not a major result of this thesis *per se*, computational optimisation is an increasingly important aspect of numerically studying dynamical systems. Particular aspects of the optimisation and parallelisation process were discussed in Chapter 1.

Looking over the results for both systems, and the behaviour of the mLE in each case, we can make some more general observations. Firstly, it is gratifying that the mLE of the very stable graphene model is so much smaller than that of the soft DNA. This purely dynamical quantification of the systems' stability matches neatly with graphene's known mechanical stability. Further work to consider a number of molecular compounds of varying mechanical strength (such as other proteins, or crystal lattices) with appropriate dynamical models and compute the mLE in each system could provide useful insight about this relationship. Studying the effects of different force models, such as the Lennard-Jones interatomic potential [150] or the anisotropic Gay-Berne potential [151], could also prove interesting. We also saw that in the DNA as the energy increases, openings in the strand cause local linearisation of the system due to the form of the intermolecular Morse potential. These linearisations lead to a slowing down of the increase in the mLE with energy, giving a flattening behaviour as melting is approached. Due to the complete lack of this behaviour in the 2D graphene sheet, the mLE continues to grow with no changing behaviour. In this way the mLE describes the overall dynamics of the systems effectively across the energy spectrum. As energy increases, there will be an inevitable decrease in stability due to stronger atomic fluctuations, but in the DNA system the mLE captures the process of denaturation which does not occur in the graphene lattice. Further investigation of this idea would be interesting – for instance, if a graphene sheet is placed under strain would there be some characteristic change in the behaviour of the mLE as the breaking point is approached? Or in other systems that exhibit local linearisations of the potentials, do we see behaviours in the mLE like that of the DNA? The capturing of this global dynamical behaviour through the mLE could be a very promising avenue of research, which could lead to better understanding of structural stability as well as dynamical stability in a variety of compounds.

Another strong possibility for future work is to look more closely at the localisation of strong nonlinearity. This is more relevant in the heterogeneous DNA system, where particular regions of the sequence are of biological interest, but could also apply to graphene and related models. The behaviour of the DVD in the DNA suggests that it could provide information about the formation of bubbles or other physical phenomena. Further study of the DVD, as well as local stretching of deviations and covariant Lyapunov vectors would be very interesting based on the results we have found so far.

The contribution to the overall stability of graphene from the angular potential is another potentially significant point, particularly linked to the appeareance of chaos in the purely harmonically coupled lattice. Since any 2D lattice with harmonic coupling that allows atoms full 2D range of motion will exhibit chaotic motion, it could be instructive to further investigate how this chaoticity depends on the geometry (e.g. hexagonal, triangular, rectangular or Kagome lattices), as well as how a potential well for the angles between atoms affects the stability in each case. Since there are many physical materials with varying lattice structures, this could in turn inform solid state experimentation.

In all numerical studies, there is always a limitation that the results can only be as accurate as the model studied allows by its faithful representation of the physical system. In our case, the models used have been rigorously tested and used extensively, giving strong evidence that our numerical simulations are applicable to the true systems. However, the exploration of other models, which may capture different dynamical aspects of the real systems more accurately, could always provide deeper understanding of these systems. For instance, DNA models that include twisting force terms [102, 152], different statistical mechanics approaches [153], or other ideas (see e.g. [154] and references therein) would be worth studying in more detail.

The logical next step regarding the complexity of the graphene model is to consider out of plane torsional effects, rather than confining all motion to a 2D plane. The use of a model including such terms (such as in [149]) would allow a more realistic investigation of higher-energy regions, as well as provide insight to the significance of this out of plane motion. While this addition would make the equations of motion much more complex, and the computational expense greater, it would be a very useful step towards a dynamical understanding of graphene.

In conclusion, we note that our research has provided new and significant knowledge about the dynamics and stability of DNA and graphene, along with additional evidence of the usefulness of nonlinear dynamics methods (like the mLE and the DVD) for studying complex physical systems. This work has laid the foundation for a number of exciting new prospects, which could be influential in a broad array of applied sciences.